\documentclass[secnumarabic,amssymb, nobibnotes, aps, prd]{revtex4-2}
\usepackage{graphicx}
\usepackage{amsmath}
\usepackage{subcaption}
\usepackage{xcolor}
\usepackage{adjustbox}
\usepackage{multirow}

\usepackage{amsmath, amssymb, booktabs, geometry}
\begin{document}
	
	\title{Thermodynamic Properties and Shadows of Black Holes in $f(R,T)$ Gravity }%
	\author{Bidyut Hazarika$^1$}
	
	\email{$rs_bidyuthazarika@dibru.ac.in$}
	
	\author{Prabwal Phukon$^{1,2}$}
	\email{prabwal@dibru.ac.in}
	
	\affiliation{$1.$Department of Physics, Dibrugarh University, Dibrugarh, Assam,786004.\\$2.$Theoretical Physics Division, Centre for Atmospheric Studies, Dibrugarh University, Dibrugarh, Assam,786004.}

	\maketitle

	\maketitle

	\section*{Abstract}
	In this paper, we explore two $f(R,T)$ gravity models and derive black hole solutions within these models. We focus on investigating how the $f(R,T)$ model influences the thermodynamic characteristics of black holes by studying their thermodynamic topology and thermodynamic geometry. We consider five specific values of the thermodynamic parameter $\omega$, which signify five different classes of black hole solutions in general relativity (GR). We observe significant changes in the local topological properties of these black holes compared to GR, depending on the model parameters. Notably, we identify an additional topological class $W=0$ for some values of $\omega$ that is absent in the GR framework.
We also study the thermodynamic geometry of the black hole using the Geometrothermodynamics (GTD) formalism. Our analysis demonstrates that the singular point, where the GTD scalar curvature diverges, corresponds exactly to the point where the heat capacity changes sign. Additionally, we constrain the model parameters of both models considered by utilizing black hole shadow data from the Sgr A* black hole, measured by the Event Horizon Telescope (EHT).
	\section{Introduction:}
	Since its inception in 1915, general relativity (GR) has remained the cornerstone of modern theoretical physics. GR is widely regarded as a successful theory of gravity, with experimental confirmations from phenomena such as the perihelion precession of Mercury, the deflection of light during the solar eclipse of 1919, the precise detection of gravitational waves by the Laser Interferometer Gravitational-Wave Observatory (LIGO) in 2015 \cite{ligo}, and the release of the first image of the supermassive black hole at the center of galaxy M87 by the Event Horizon Telescope (EHT) in 2019 \cite{m87a, m87b, m87c, m87d, m87e, m87f}. Despite these successes, however, GR faces significant challenges, such as the discovery of the universe's accelerating expansion \cite{reiss, perlmutter, spergel, astier} and the galaxy rotation curves, which imply the existence of unseen matter \cite{naselskii}, commonly referred to as dark matter.\\
	In response to the challenges faced by general relativity, modified theories of gravity have gained significant attention. With $f(R)$ gravity being one of the most studied to tackle issues related to dark energy and dark matter, inspired by this $f(R,T)$ gravity presents a novel approach that was first introduced in \cite{harko}.  In this framework, a functional dependence  of the action on the Ricci scalar $R$, and  the trace of the energy-momentum tensor $T$ is taken into account as follows \cite{harko}
	\begin{equation}
S=\frac{1}{16\pi}\int f(R,T)\sqrt{-g}d^4x + \int L_{m}\sqrt{-g}d^4x,
\end{equation}
where the $L_m$ part represents the matter Lagrangian density associated with any energy-momentum tensor. The coupling between matter and geometry in the framework of $f(R,T)$ gravity yields intriguing results, particularly in cosmology and the dynamics of massive particles. This theory also holds promise as an alternative to general relativity, with the potential to account for both dark energy and dark matter\cite{astier,reiss,dj}. Recently,a number of work has been done in different fields of physics using the $f(R,T)$ theory framework \cite{cosmo1,cosmo2,cosmo3,santos,sood,Mohan,Gohain,takol,pranjal}.\\
	
	The link between general relativity and thermodynamics was established nearly fifty years ago through the pioneering works of Bekenstein and Hawking  \cite{Bekenstein:1973ur,Hawking:1974rv,Hawking:1975vcx}. Bekenstein's introduction of the concept of black hole entropy and Hawking's discovery of black hole radiation fundamentally altered the way physicists perceive black holes, suggesting that they are not merely voids in spacetime but rather entities that adhere to thermodynamic principles\cite{Bardeen:1973gs}. Since then, numerous fascinating developments have emerged reinforcing this intricate connection. \cite{Wald:1979zz,bekenstein1980black,Wald:1999vt,Carlip:2014pma,Wall:2018ydq,Candelas:1977zz,Mahapatra:2011si,fop1}. One particularly noteworthy result in black hole mechanics is the phenomenon of black hole (BH) phase transitions\cite{Davies:1989ey,Hawking:1982dh,curir_rotating_1981,Curir1981,Pavon:1988in,Pavon:1991kh,OKaburaki,Cai:1996df,Cai:1998ep,Wei:2009zzf,Bhattacharya:2019awq,Kastor:2009wy,Dolan:2010ha,Dolan:2011xt,Dolan:2011jm,Dolan:2012jh,Kubiznak:2012wp,Kubiznak:2016qmn,Bhattacharya:2017nru,fop2}, first identified by Davies\cite{Davies:1989ey} . This critical insight revealed that a phase transition occurs at a specific point characterized by a discontinuity in the heat capacity.
The Hawking–Page phase transition, introduced in \cite{Hawking:1982dh}, is another type of phase transition characterized by a change in the sign of free energy. The transition of black holes from non-extremal to extremal states has been studied in several works \cite{curir_rotating_1981,Curir1981,Pavon:1988in,Pavon:1991kh,OKaburaki,Cai:1996df,Cai:1998ep,Wei:2009zzf,Bhattacharya:2019awq}. Additionally, the behavior of phase transitions that resemble the van der Waals type has been explored in various studies \cite{Kastor:2009wy,Dolan:2010ha,Dolan:2011xt,Dolan:2011jm,Dolan:2012jh,Kubiznak:2012wp,Kubiznak:2016qmn,Bhattacharya:2017nru}.\\

Recent advancements in black hole thermodynamics have underscored the importance of thermodynamic topology as a valuable tool for probing the intricate phase behaviour of black holes. Initially, topological techniques were utilized to analyze phenomena such as light rings and time-like circular orbits in black hole spacetimes \cite{PRL119-251102, PRL124-181101, PRD102-064039, PRD103-104031, PRD105-024049, PRD108-104041, 2401.05495, PRD107-064006, JCAP0723049, 2406.13270}. The application of topology in the context of black hole thermodynamics was pioneered in \cite{28}, taking inspiration from Duan's earlier contributions \cite{d1, d2} in the study of relativistic particle systems. Central to this approach is the idea of topological defects, represented by the zero points of a vector field, which correspond to the system's critical points. These zero points act as markers of phase transitions and can be characterized by their winding numbers, allowing black holes to be grouped into distinct topological classes based on their thermodynamic behaviour.This method has been widely adopted across various black hole models in the literature \cite{PRD105-104053, PLB835-137591, PRD107-046013, PRD107-106009, JHEP0623115, 2305.05595, 2305.05916, 2305.15674, 2305.15910, 2306.16117, PRD106-064059, PRD107-044026, PRD107-064015, 2212.04341, 2302.06201, 2304.14988, 64, 2309.00224, 2312.12784, 2402.18791, 2403.14730, 2404.02526}.
In this study, we utilize the topological framework outlined in \cite{29}, which is particularly effective for investigating black hole thermodynamics. Using the off-shell free energy method, black holes are modeled as topological defects within their thermodynamic structure. This approach sheds light on both the local and global topological features of black holes, where their topological charge and stability are described through winding numbers. The stability of a black hole can be deduced from the sign of its winding number. This methodology has been successfully applied to a wide range of black hole systems in various gravitational theories \cite{PRD107-064023, PRD107-024024, PRD107-084002, PRD107-084053, 2303.06814, 2303.13105, 2304.02889, 2306.13286, 2304.05695, 2306.05692, 2306.11212, EPJC83-365, 2306.02324, PRD108-084041, 2307.12873, 2309.14069, AP458-169486, 2310.09602, 2310.09907, 2310.15182, 2311.04050, 2311.11606, 2312.04325, 2312.06324, 2312.13577, 2312.12814, PS99-025003, 2401.16756, AP463-169617, PDU44-101437, 686, 2404.08243, 2405.02328, 685, 2405.20022, 2406.08793, AC48-100853, 682,nwu1}. In \cite{nwu2,nwu3,nwu4,nwu5} a novel scheme in the context of classifying the black holes topologically is being discussed. \\

Recent studies have emphasized the importance of thermodynamic geometry in understanding the rich phase structure of black hole systems \cite{Weinhold, Ruppeiner1, Ruppeiner2, Mrugala, Aman, Que1, Que2}. A crucial aspect of any thermodynamic framework is its inherent fluctuation theory, which establishes a connection between macroscopic properties and their microscopic origins. To decipher the implications of these thermodynamic fluctuations for microscopic characteristics, we utilize the thermodynamic Ricci curvature scalar 
$R$. The scalar $R$ serves as a thermodynamic invariant within the geometric framework of thermodynamics. Assuming the fundamental universality of thermodynamics, it is reasonable to anticipate that characteristics $R$ in ordinary thermodynamics,  may also apply within the black hole context. In ordinary thermodynamics, 
the magnitude of R represents the average volume occupied by groups of atoms, which is organized according to their interparticle interactions. Near critical points, this average volume corresponds to the correlation length of these interactions.\\
The choice of a robust metric in thermodynamic state space is crucial.In this study, we focus exclusively on two-dimensional thermodynamic metric geometries .Weinhold \cite{Weinhold} was the first to propose a Riemannian metric for thermodynamic systems. Followed by Ruppeiner \cite{Ruppeiner1, Ruppeiner2} who introduced a new metric in the late 1970s, which is defined as the negative Hessian of entropy with respect to other extensive variables. The Ruppeiner metric is given by \cite{Ruppeiner1, Ruppeiner2}  : 
\begin{equation} 
g_{ij}^{R} = -\partial_{i} \partial_{j} S(U, X), 
\end{equation} 
where the entropy $S(U, X)$ depends on the internal energy $U$ and other extensive variables $X$. This metric has proven useful in measuring the distance between equilibrium states, thereby allowing for a more detailed analysis of the system’s thermodynamic behaviour. A negative scalar curvature implies predominantly attractive interactions, a positive value suggests repulsive interactions and a flat geometry ($R_{\text{rupp}} = 0$) indicates no interaction. While both Weinhold and Ruppeiner geometries have provided valuable insights into the phase structures of different thermodynamic systems, they exhibit inconsistencies in some cases. For instance, regarding Kerr-AdS black holes \cite{Aman}, the Weinhold metric does not successfully predict phase transitions, which stands in contrast to findings from conventional black hole thermodynamics. Conversely, the Ruppeiner metric does indicate the presence of phase transitions, although this is contingent upon the selection of particular thermodynamic potentials.To address these shortcomings, a new framework known as geometrothermodynamics (GTD) was proposed by Quevedo \cite{Que1, Que2}. GTD unifies the properties of both the phase space and the space of equilibrium states, and unlike the Weinhold and Ruppeiner geometries, the GTD metric is Legendre invariant, meaning it does not depend on the choice of thermodynamic potential. In GTD, phase transitions inferred from the black hole’s heat capacity are directly linked to singularities in the scalar curvature of the GTD metric. A singularity in the GTD curvature $R_{\text{GTD}}$ aligns with the phase transitions obtained from the heat capacity. The general form of the GTD metric is \cite{Que1, Que2}: 
\begin{equation} 
g = \left(E^c \frac{\partial \Phi}{\partial E^c}\right) \left(\eta_{ab} \delta^{bc} \frac{\partial^2 \Phi}{\partial E^c \partial E^d} dE^a dE^d\right), 
\label{gtd}
\end{equation} 
where $\Phi$ is the thermodynamic potential, $E^a$ represents extensive thermodynamic variables (with $a = 1, 2, 3, \ldots$), $\eta_{ab} = \text{diag}(-1, 1, 1, \ldots)$, and $\delta^{bc} = \text{diag}(1, 1, 1, \ldots)$. This formalism successfully addresses the discrepancies found in earlier geometries and provides a more consistent approach to studying the thermodynamic properties of black holes.\\

The motivation behind this work is to investigate how $f(R,T)$ gravity influences the well-known Kiselev black hole solutions in general relativity.  We investigate black hole solutions in two distinct $f(R,T)$  models to explore how this modified gravity theory influences thermodynamic behavior, phase transitions, and topological classifications compared to general relativity (GR), with a particular emphasis on thermodynamic geometry and topology. A critical aspect of our study is the selection of model parameters, which significantly influence the thermodynamic behavior of the black holes under consideration. To illustrate the range of potential outcomes and to highlight the dependence of the thermodynamic properties on these parameters, we have employed arbitrary values throughout our analysis.But to establish a meaningful connection between our models and the recently observed data from black holes  we utilize the study of black hole shadows to impose constraints on the model parameters. By doing so, we aim to ensure that the parameters we consider are not only theoretically sound but also compatible with empirical observations.\\
Although a significant amount of work has been done in various fields of physics based on the framework of $f(R,T)$ gravity, comparatively fewer studies focus on black hole solutions. In \cite{santos}, Kiselev-type black hole solutions were evaluated in this framework for the first time. However, the study of phase transitions and thermodynamic properties of these black holes was not addressed. Our work introduces a novel perspective by analyzing the thermodynamic topology and thermodynamic geometry of these solutions, which has not been explored before.  In \cite{sood}, the phase transitions and photon orbits of Kiselev black holes in $f(R,T)$ gravity were discussed using "Model I," a model also considered in our manuscript. However, their analysis was performed in AdS space, while our work is based on flat spacetime, where the cosmological constant and the AdS boundary are absent.  Furthermore, we present a new $f(R,T)$ model of the type $f(R,T) = f_1(R) + f_2(R)f_3(T)$, referred to as "Model II" in our manuscript, and derive a new black hole solution that had not been explored previously. Additionally, our work focuses on constraining the $f(R,T)$ gravity model parameters using black hole shadow data for both models, which represents another unexplored aspect.  In summary, our manuscript addresses several novel aspects, including thermodynamic topology, thermodynamic geometry, and parameter constraints using shadow observations, thereby filling critical gaps in the existing literature on $f(R,T)$ gravity and black hole solutions.  

The black hole shadow has recently attracted significant attention, particularly due to the release of groundbreaking data and images of black holes at the centres of the M87 galaxy and Sgr A*. Recent studies has successfully constrained different parameters of modified theories of gravity using contemporary measurements of shadow radius data..The technique we used in this paper to constrain the model parameter is shown in \cite{s0} and adopted across various literature \cite{s01,s02,s03}. Numerous studies in recent literature \cite{s1,s2,s3,s4,s5,s6,s7,s8,s9,s11,s12,s13,s14,s16,s17,s18,s19,s20,s21,s22,s23,s24,s25} underscore the significance of black hole shadows in the context of constraining various gravitational theories and their associated parameters. Apart from shadow data,  in \cite{s15} black hole parameters are constrained with the precessing jet nozzle of M87*.\\
 
	This paper is organized as follows: In section \textbf{II}, we briefly review the field equations in $f(R,T)$ gravity and the energy-momentum tensor of the Kiselev black hole. In section \textbf{III}, we evaluate the black hole solution in Model I and study its basic thermodynamic properties in subsection \textbf{III.1}. In subsection \textbf{III.2}, we analyze the thermodynamic topology of the black hole solution, followed by the thermodynamic geometry in subsection \textbf{III.3}. In subsection \textbf{III.4}, we constrain the model parameters using black hole shadow data. In section \textbf{IV}, we evaluate a novel black hole solution and study the SEC and horizon structure of the black hole in Model II. The thermodynamic quantities, thermodynamic topology, and thermodynamic geometry are briefly discussed in subsection \textbf{IV.1}. The model parameters of the black holes are constrained in subsection \textbf{IV.2}. Finally, the conclusions are presented in section \textbf{V}.

\section{Revisiting field equations in $f(R,T)$ gravity and energy-momentum of the Kiselev black hole}
	We start with the action in the $f(R,T)$ gravity framework which is given by  \cite{harko} :
\begin{equation}
S=\frac{1}{16\pi}\int f(R,T)\sqrt{-g}d^4x + \int L_{m}\sqrt{-g}d^4x,
    \label{e1}
\end{equation}
where $f(R,T)$ is a function of the Ricci scalar, $R$, and of the trace $T$ of the  energy-momentum tensor of the
matter. $L_m$ represents the matter Lagrangian density.  The field equation is obtained by varying the action $S$ with respect to  the metric tensor as  \cite{harko} :
\begin{equation}
 f_{R}(R,T)R_{\mu\nu}-\frac{g_{\mu\nu}}{2}f(R,T)+
(g_{\mu\nu}\Box - \nabla_{\mu}\nabla_{\nu})f_{R}(R,T) = 8\pi T_{\mu\nu} - f_{T}(R,T)T_{\mu\nu}- f_{T}(R,T)\Theta_{\mu\nu}
   \label{fieldeq}
\end{equation}
where $f_R=\frac{d f(R,T)}{d R}$ and $f_T=\frac{d f(R,T)}{d T}$. The $\Theta_{\mu \nu}$ is associated with matter Lagrangian density and the energy-momentum tensor as follows  \cite{harko} 
\begin{equation}
\Theta_{\mu\nu} =-2T_{\mu\nu}+g_{\mu\nu}L_{m}-2g^{\eta\xi}\frac{\partial^2L_{m}}{\partial g^{\mu\nu}g^{\eta\xi}},
    \label{e4}
\end{equation}
	There are three classes of models proposed by \cite{harko} from which we can obtained different  theoretical model depending upon the functional form of $f.$ By varying the combinations of matter model $f(T)$ and the Ricci scalar models $f(R)$ a number of models can be obtained. These three classes are given as  \cite{harko} :
	$$\textbf{A.}f(R,T)=  f_1(R)+ f_2(T)$$
	$$\textbf{B.}f(R,T)=  R + 2 f(T)$$
	$$\textbf{C.} f(R,T)=  f_1(R)+ f_2(R)f_3(T)$$

The field equations in these models are influenced by the tensor $\tau_{\mu\nu}$, which depends on the properties of the matter field. Therefore, in $f(R,T)$ gravity, the nature of the matter source plays a crucial role.  Recently a number of works have been conducted regarding the development of concepts of these models in different aspects of cosmology and black hole physics\cite{h1,h2,h3,h4,h5,h6,h7,h8,h9,h10}.

In this paper, we consider two of these three particular three $f(R,T) $ model classes(A and C) and obtain the black hole solutions. We choose the  $T$ as the trace of the energy-momentum of the spherically symmetric Kiselev black hole which has the components of the energy-momentum tensor adequately connected to an anisotropic fluid.The Kiselev energy-momentum tensor is chosen because the $t$-component and the $r$-component are equal, while the $\theta$- and $\phi$-components are equivalent. This simplifies the tensor, making it easier to work with when evaluating the field equations. The components are given as \cite{santos}
 
\begin{align}
    T^{t}_{\:\:\:t}= T^{r}_{\:\:\:r}&=\rho(r), \label{t11}\\
   T^{\theta}_{\:\:\:\theta}= T^{\phi}_{\:\:\:\phi}&=-\frac{1}{2}\rho (3\omega+1),
   \label{t22}
\end{align}
where $\omega$ is the parameter of equation of state and $\rho$ is the energy density. The trace of the above tensor is calculated to be $T=\rho-3\omega \rho$. The matter Lagrangian density of the Kiselev black holes which is associated with the anisotropic fluid is given by \cite{santos}
$$L_{m}=(-1/3)(p_r+2p_t)$$
where $p_t=\frac{1}{2} \rho (3 \omega +1)$ and $p_r=-\rho$ are the transverse and radial pressure respectively 
 Using the expression of $L_m$, one can obtain the  $\Theta_{\mu\nu}$ as \cite{santos} :
\begin{equation}
    \Theta_{\mu\nu}=-2T_{\mu\nu}-\frac{1}{3}(p_r+2p_t)g_{\mu\nu}.
    \label{theta}
\end{equation}
In the following sections, we will utilize this information to derive the Kiselev black hole solution and explore some of its key properties.
	%%%%%%%%%%%%%%%%%%%%%%%%%%%%%%%%%%%%%%%%%%%%%%%%%%%%%%%%%%%%%%%%%%%%%%%%%%%%%%%%%%%%%%%%%%%%%%%%%%%%%%%%%%%%%%%%%%%%%%%%%%%%%%%%%%%%%%%%%%%%%%%%%%%%%%%%%%%%%%%%%%%%%%%%%%%%%
	\section{Model I : $f(R,T)=f_1(R)+f_2(T)$}
	We consider $f_1(R)=\alpha R$ and $f_2(T)= \beta T$ where $\alpha$ and $\beta$ are the two model parameters. Using eq.(\ref{theta}) in the field equation in eq.(\ref{fieldeq}) we get
	$$ \alpha  R_{\mu  \nu }-\frac{1}{2} \alpha  R g_{\mu  \nu }=\beta  \left(\frac{1}{3} g_{\mu  \nu } \left(p _r+2 p _t\right)+2 T_{\mu  \nu }\right)+\frac{1}{2} \beta  T g_{\mu  \nu }-\beta  T_{\mu  \nu }+8 \pi  T_{\mu  \nu }$$
	$$\alpha  G^\mu{}_ \nu =\frac {\beta T}{2}+\beta  \left(\rho  \omega +T^\mu{}_ \nu \right)+8 \pi  T^\mu{}_ \nu $$
	In the above equation, we have substituted
	\begin{equation}
	G^\mu{} _\nu =R^\mu{}_ \nu -\frac{1}{2} \delta^\mu{}_\nu R
	\label{gmunu}
\end{equation}	 
	Using  eq.(\ref{t11}) and eq.(\ref{t22}), finally the first field equation is obtained as
	\begin{equation}
	 G^r{}_r= G^t{}_t=\frac{1}{\alpha}(\beta  (\rho  \omega +\rho )+\frac{1}{2} \beta  (\rho -3 \rho  \omega )+8 \pi  \rho)
	 \label{fe1}
	\end{equation}
Similarly using and  eq.(\ref{t2}) the other field equation is found to be 
	$$\alpha  G^{\theta }{}_{\theta }=\beta  \left(\rho  \omega -\frac{3 \rho  \omega }{2}-\frac{\rho }{2}\right)+\frac{1}{2} \beta  (\rho -3 \rho  \omega )+\frac{1}{2} (-(8 \pi )) \rho  (3 \omega +1)$$
	\begin{equation}
	 G^{\theta }{}_{\theta }=\frac{1}{\alpha}\left(-\frac{1}{2} (\omega +1) (\beta  \rho )+\frac{1}{2} (1-3 \omega ) (\beta  \rho )+\pi  (-\rho ) (12 \omega +4)\right)
	 \label{fe2}
	\end{equation}
	For a spherically symmetric Kiselev black hole solution,  the line element is given by :
	\begin{equation}
	ds^2=N(r)dt^2-M(r) dr^2-r^2(d\theta^2+sin \theta^2 d\phi^2)
	\label{line}
	\end{equation}
	As it is evident from eq.(\ref{fe1}) and eq.(\ref{fe2}), there is a symmetry in the field equations as $G^t{}_t=G^r{}_r$. This symmetry results in the condition 
	$$M(r)=\frac{1}{N(r)}$$
	Next, apply this condition in eq.(\ref{gmunu})  and evaluate the following two equation
	\begin{align}
   G^{t}_{\:\:t}=G^{r}_{\:\:r}=& -\frac{1}{r}\frac{dN(r)}{dr} - \frac{N(r)}{r^2} + \frac{1}{r^2} \label{f1}\\
 G^{\theta}_{\:\:\theta}=G^{\phi}_{\:\:\phi}=& -\frac{1}{2}\frac{d^2N(r)}{dr^2}-\frac{1}{r}\frac{dN(r)}{dr} 
    \label{f2}.
\end{align}
equating these two equation with eq.(\ref{fe1}) and eq.(\ref{fe2}), we get
\begin{equation}
-\frac{1}{r}\frac{dN(r)}{dr} - \frac{N(r)}{r^2} + \frac{1}{r^2}=\frac{1}{\alpha}(\beta  (\rho  \omega +\rho )+\frac{1}{2} \beta  (\rho -3 \rho  \omega )+8 \pi  \rho)
\label{d1}
\end{equation}
and 
\begin{equation}
-\frac{1}{r}\frac{dN(r)}{dr} - \frac{N(r)}{r^2} + \frac{1}{r^2}=\frac{1}{\alpha}(-\frac{1}{2} (\omega +1) (\beta  \rho )+\frac{1}{2} (1-3 \omega ) (\beta  \rho )+\pi  (-\rho ) (12 \omega +4))
\label{d2}
\end{equation}
dividing eq.(\ref{d1}) by  eq.(\ref{d2}) we eliminate the energy density  term $\rho$ and the model parameter $\alpha$ and finally get the differential equation
\begin{equation}
-\frac{1}{r}\left(\frac{16 \pi -\beta  (\omega -3)}{8 (\beta  \omega +\pi  (6 \omega +2))}\right)\frac{d}{dr}\left(\frac{d}{dr}(r N(r))-1 \right)=\frac{1}{r^2} \left(\frac{d}{dr} (r N(r))-1 \right)
\end{equation}
Solving this equation yields the required black hole solution with the Lapse function
\begin{equation}
N(r)=1+\frac{c_1}{r}+c_2 r^{-\frac{8 (\beta  \omega +\pi  (6 \omega +2))}{16 \pi -\beta  (\omega -3)}}
\label{lapse1}
\end{equation}
where $c_1$ is obviously equals to $-2M$ with $M$ being the mass and  we have considered $c_2=1$ in this paper.
According to \cite{santos}, by setting $\omega = 0$ and the model parameter $\beta = 0$, we recover the simple Schwarzschild black hole solution. Similarly, if we substitute $\omega = \frac{1}{3}$ and $\beta = 0$, we obtain the Reissner-Nordström (RN) charged black hole, with $c_2$ representing the effective charge. Our black hole solution satisfies both of these conditions. In the following section, we will discuss the thermodynamic properties of this black hole solution.
	\subsection{Thermodynamical properties}
\begin{figure}[h]	
		\centering
		\begin{subfigure}{0.30\textwidth}
			\includegraphics[width=\linewidth]{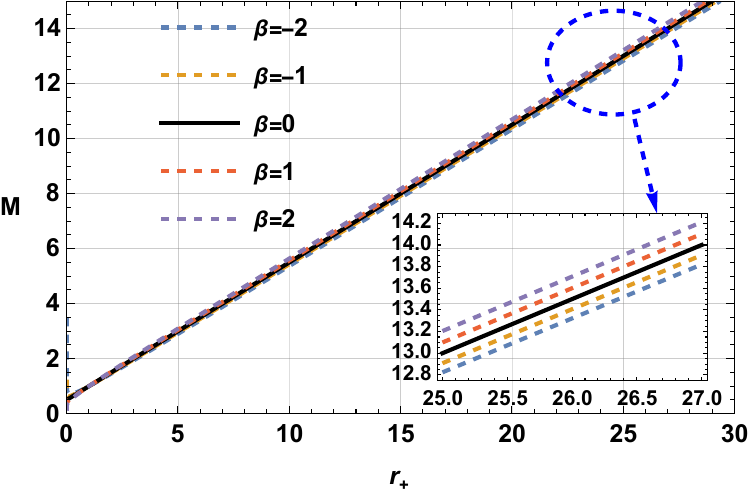}
			\caption{$\omega=0$}
			\label{1a}
		\end{subfigure}
		\begin{subfigure}{0.30\textwidth}
			\includegraphics[width=\linewidth]{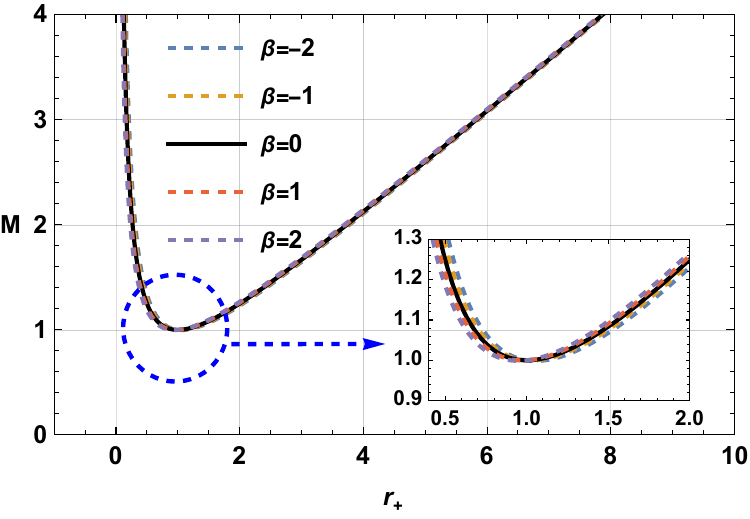}
			\caption{$\omega=\frac{1}{3}$}
			\label{1b}
		\end{subfigure}
		\begin{subfigure}{0.30\textwidth}
			\includegraphics[width=\linewidth]{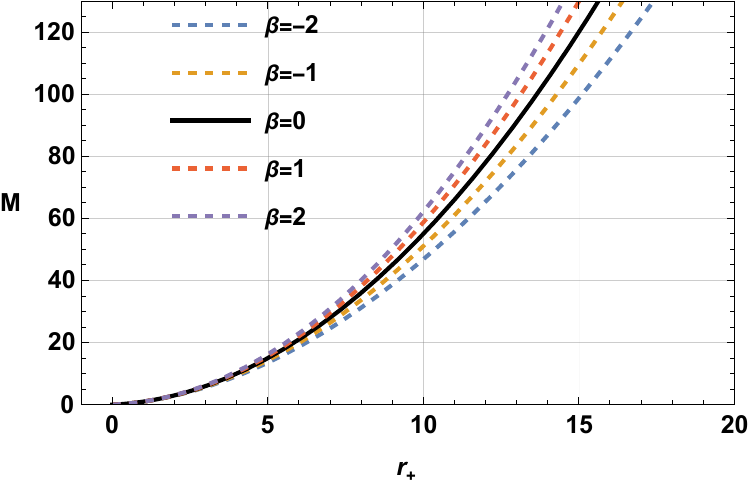}
			\caption{$\omega=-\frac{2}{3}$}
			\label{1c}
		\end{subfigure}
		\begin{subfigure}{0.30\textwidth}
			\includegraphics[width=\linewidth]{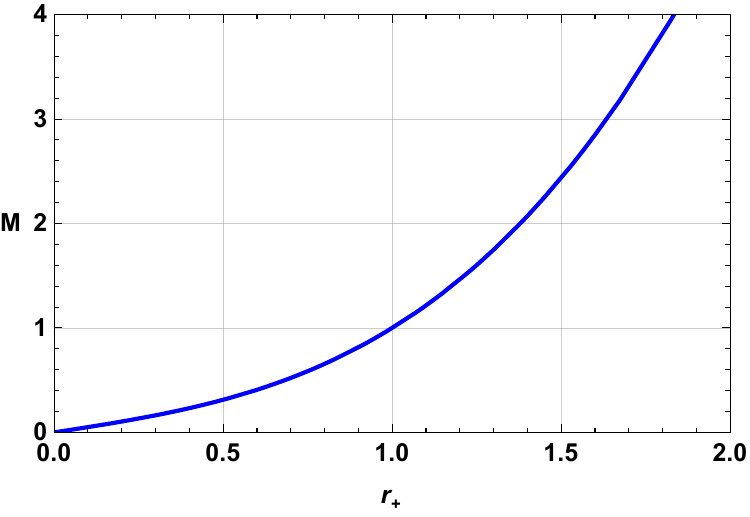}
			\caption{$\omega=-1$}
			\label{1d}
			\end{subfigure}
		\begin{subfigure}{0.30\textwidth}
		\includegraphics[width=\linewidth]{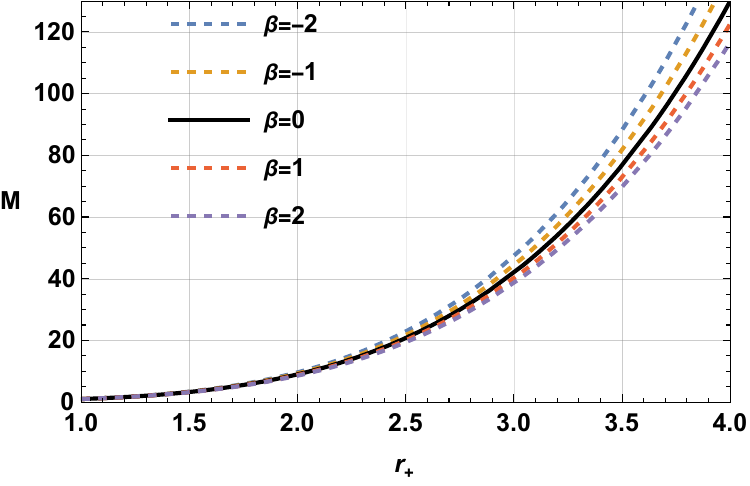}
		\caption{$\omega=-\frac{4}{3}$}
			\label{1f}
		\end{subfigure}
		\caption{ $M$ vs $r_+$ plot for different values of $\omega$. The impact of model  parameter $\beta$ is shown for a specific value of $\omega$}
		\label{fig1}
	\end{figure}

\begin{figure}[h]	
		\centering
		\begin{subfigure}{0.30\textwidth}
			\includegraphics[width=\linewidth]{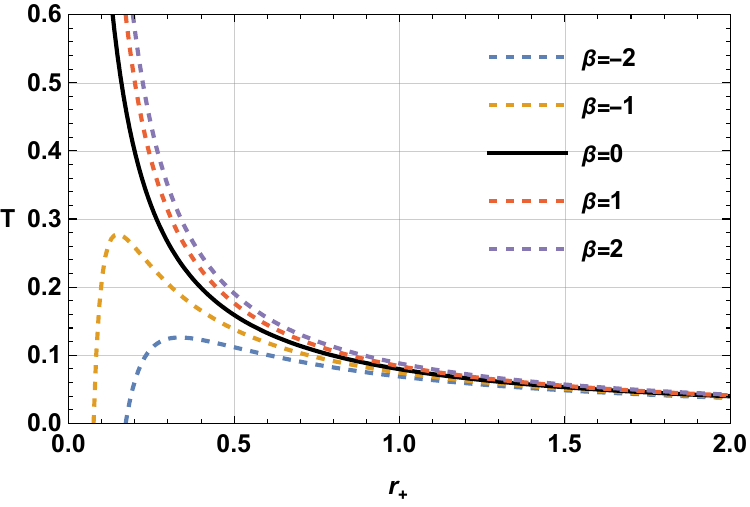}
			\caption{$\omega=0$}
			\label{2a}
		\end{subfigure}
		\begin{subfigure}{0.30\textwidth}
			\includegraphics[width=\linewidth]{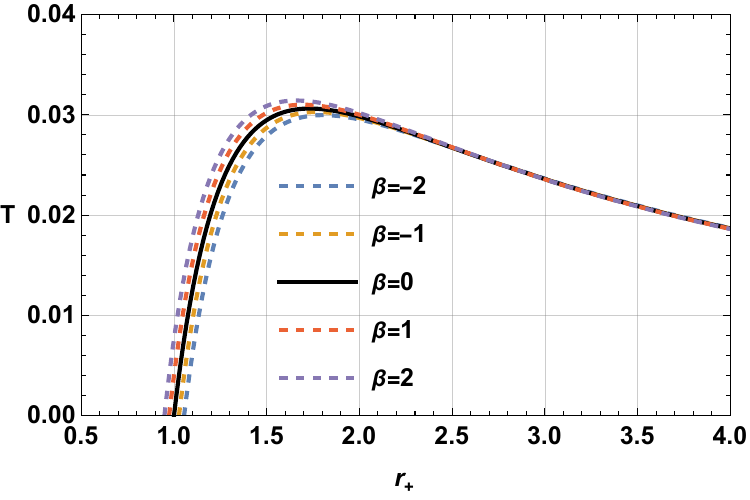}
			\caption{$\omega=\frac{1}{3}$}
			\label{2b}
		\end{subfigure}
		\begin{subfigure}{0.30\textwidth}
			\includegraphics[width=\linewidth]{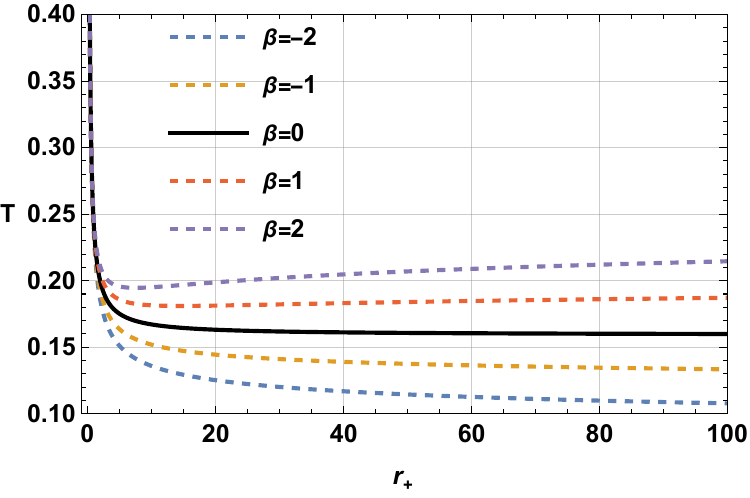}
			\caption{$\omega=-\frac{2}{3}$}
			\label{2c}
		\end{subfigure}
		\begin{subfigure}{0.30\textwidth}
			\includegraphics[width=\linewidth]{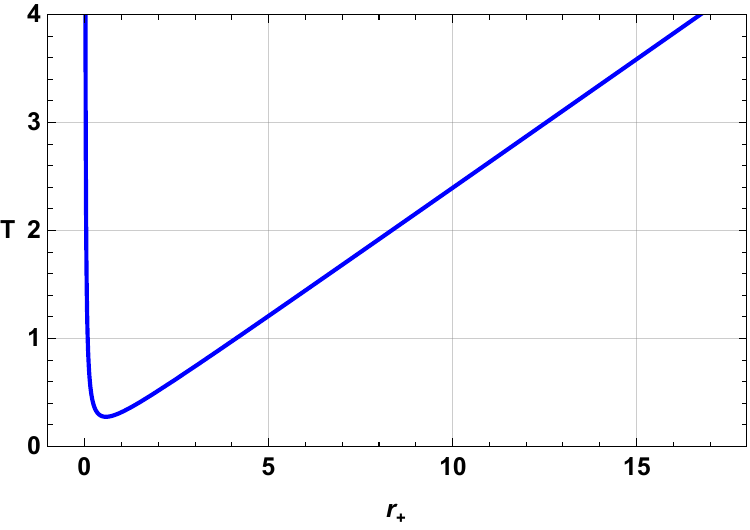}
			\caption{$\omega=-1$}
			\label{2d}
			\end{subfigure}
		\begin{subfigure}{0.30\textwidth}
		\includegraphics[width=\linewidth]{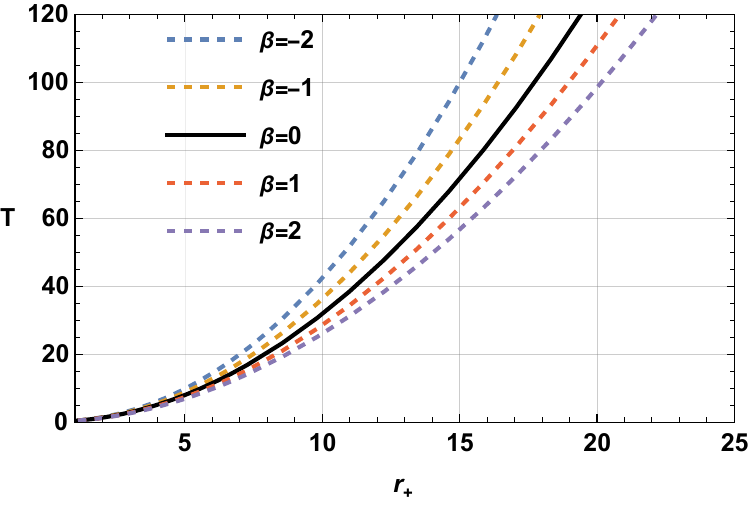}
		\caption{$\omega=-\frac{4}{3}$}
			\label{2e}
		\end{subfigure}
		\caption{ $T$ vs $r_+$ plot for different values of $\omega$. The impact of model  parameter $\beta$ on $T$ vs $r_+$ plots are shown for a specific value of $\omega$}
		\label{2}
	\end{figure}

\begin{figure}[h!]	
		\centering
		\begin{subfigure}{0.30\textwidth}
			\includegraphics[width=\linewidth]{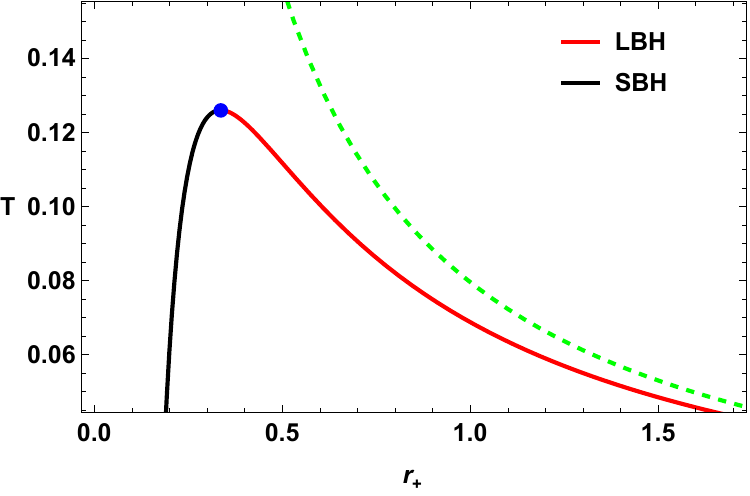}
			\caption{$\omega=0$}
			\label{3a}
		\end{subfigure}
		\begin{subfigure}{0.30\textwidth}
			\includegraphics[width=\linewidth]{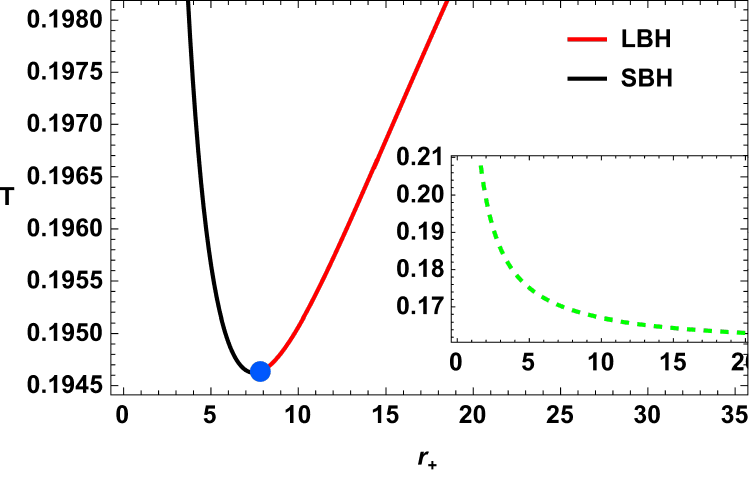}
			\caption{$\omega=\frac{1}{3}$}
			\label{3b}
		\end{subfigure}
		\caption{Impact of model parameter $\beta$ on phase transitioning behaviour of the black hole.}
		\label{3}
	\end{figure}

	The mass $M$ of the  black hole can be derived from the equation by setting $N(r=r_+)=0$ in eq.(\ref{lapse1}) ,
	\begin{equation}
		M=\frac{1}{2} r_+ \left(r_+^{-\frac{8 (\beta  \omega +\pi  (6 \omega +2))}{16 \pi -\beta  (\omega -3)}}+1\right)
		\label{mass1}
	\end{equation}
	The expression for  temperature can be evaluated as :
	\begin{equation}
	T=\frac{1}{4\pi}\frac{\partial N}{\partial r}=\frac{-\beta  \omega +3 \beta +(\beta  (3-9 \omega )-48 \pi  \omega ) r_+^{\frac{8 (\beta  \omega +\pi  (6 \omega +2))}{\beta  (\omega -3)-16 \pi }}+16 \pi }{4 \pi  r_+ (16 \pi -\beta  (\omega -3))}
		\label{temp1}
	\end{equation}
	The entropy in this model is evaluated to be :
	\begin{equation}
	S=\int \frac{d M}{T}= \pi r_+^2
	\label{entropy1}
	\end{equation}
	In this paper, we consider five values of $\omega$, which carry specific physical meaning to each $\omega$ value. in general relativity.  Table \ref{table1} represents the significance of each $\omega$ value.
	\begin{table}[h]
\centering
\begin{tabular}{|c|c|}
\hline
\textbf{$\omega$ values } & \textbf{Black hole surrounded by} \\ \hline
$\omega=0$            & Dust field           \\ 
$\omega=\frac{1}{3}$            & Radiation field           \\ 
$\omega=-\frac{2}{3}$           & Quintessence field          \\ 
$\omega=-1$          & Cosmological field          \\ 
$\omega=-\frac{4}{3}$    & Phantom field         \\ \hline
\end{tabular}
\caption{Physical meaning of different $\omega$ values}
\label{table1}
\end{table}
	The effect of the model parameter $\beta$  on the black hole mass for a specific value of $\omega$ is shown in FIG.\ref{fig1}. For $\omega=-1$, the mass becomes independent of the model parameter. 
	Similarly, temperature $T$ is plotted against horizon radius $r_+$ in FIG.\ref{2} to study the impact of the model parameter. For $\omega=0$(dust field) and $\omega=-2/3$(quintessence field) case, we observed significant changes in the phase transitioning behaviour of the black hole.The black-coloured solid line in FIGs \ref{2a},\ref{2c} shows the $T$ vs $r_+$ plot in GR case where only one black hole phase is observed. These two cases are explicitly represented in FIG.\ref{3}.
	In FIG.\ref{3a} we have considered $\omega=0,K=1$ and $\beta=-2$  for which two black hole branches are observed.  We see a small black hole branch(SBH) for $r_+<0.33563$(blue dot) and a large black hole branch(LBH) for $r_+>0.33563$ represented by black and red solid lines respectively.  The blue dot represents the exact point at which phase transition occurs. The green dashed line shows the $T$ vs $r_+$ plot for this particular class of black holes in GR where model parameter $\beta$ is set to be zero. We can clearly see the difference created by the negative values of the model parameter. Again in FIG.\ref{3b} we consider $\omega=-2/3,K=1$ and $\beta=2$ where a small black hole branch(SBH) is observed for the range $	r_+< 7.47415$(blue dot) represented by a black solid line and a large black hole branch(LBH) is found for the range $r_+>7.47415$. Here also the green dashed line represents the scenario for $\beta=0$. The introduction of $f(R,T)$ gravity indeed alters the phase transitions and critical behaviours of black holes compared to General Relativity (GR).\\

	This behaviour can be more prominantly studied when we analysed the Gibbs free energy(F) of these black holes as it provides valuable insight into the criticality of phase transitions in black holes. The expression of  free energy is calculated by using 
	\begin{equation}
		F=M-TS
	\end{equation}
	Using equation (\ref{mass1}), (\ref{temp1},) and (\ref{entropy1}), F is calculated as :
	\begin{equation}
		F=\frac{r_+^{\frac{3 (\beta  (3 \omega -1)+16 \pi  \omega )}{\beta  (\omega -3)-16 \pi }} \left(K (\beta  (7 \omega +3)+16 \pi  (3 \omega +2))+(16 \pi -\beta  (\omega -3)) r_+^{\frac{8 (\beta  \omega +\pi  (6 \omega +2))}{16 \pi -\beta  (\omega -3)}}\right)}{64 \pi -4 \beta  (\omega -3)}
	\end{equation}
	\begin{figure}[h]	
		\centering
		\begin{subfigure}{0.30\textwidth}
			\includegraphics[width=\linewidth]{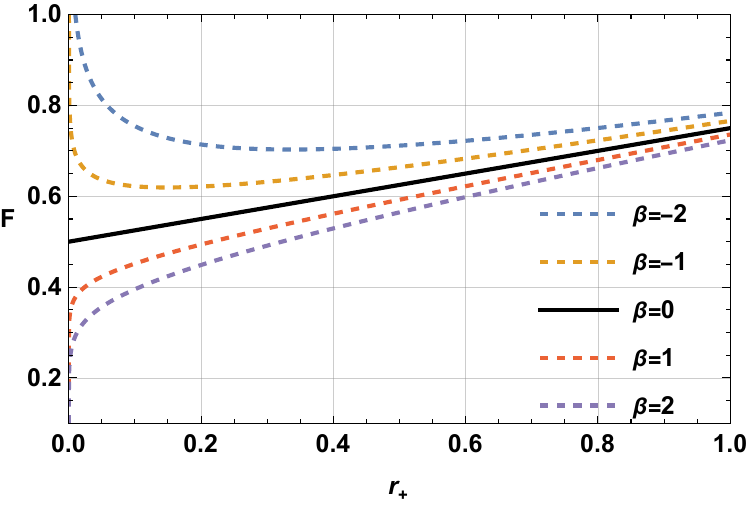}
			\caption{$\omega=0$}
			\label{4a}
		\end{subfigure}
		\begin{subfigure}{0.30\textwidth}
			\includegraphics[width=\linewidth]{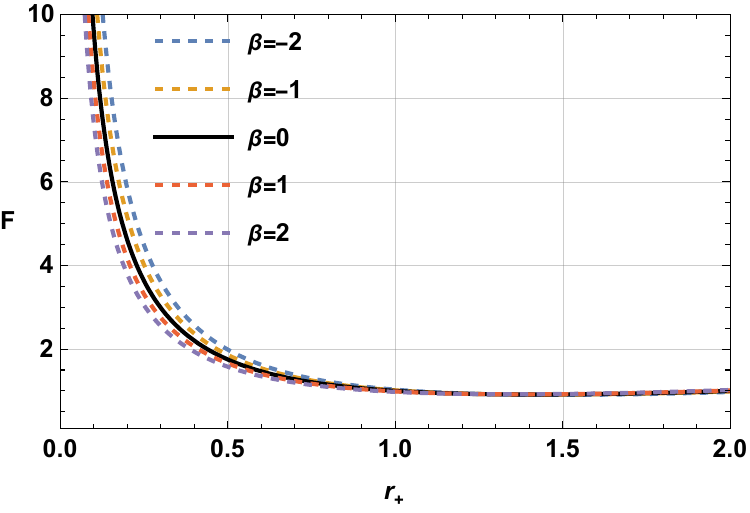}
			\caption{$\omega=\frac{1}{3}$}
			\label{4b}
		\end{subfigure}
		\begin{subfigure}{0.30\textwidth}
			\includegraphics[width=\linewidth]{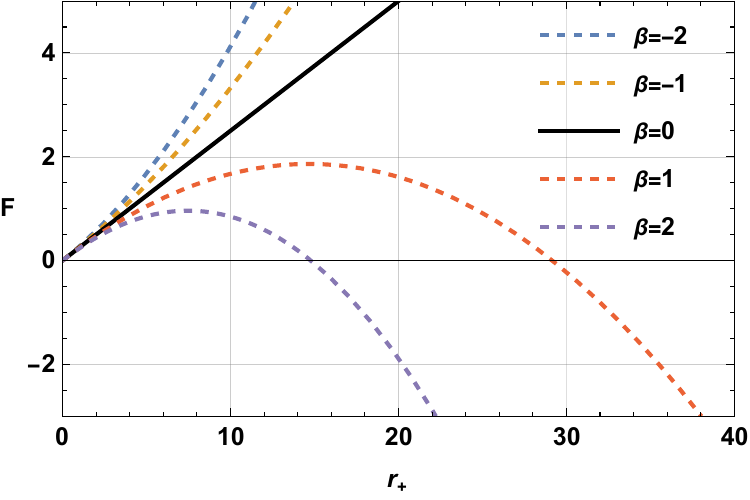}
			\caption{$\omega=-\frac{2}{3}$}
			\label{4c}
		\end{subfigure}
		\begin{subfigure}{0.30\textwidth}
			\includegraphics[width=\linewidth]{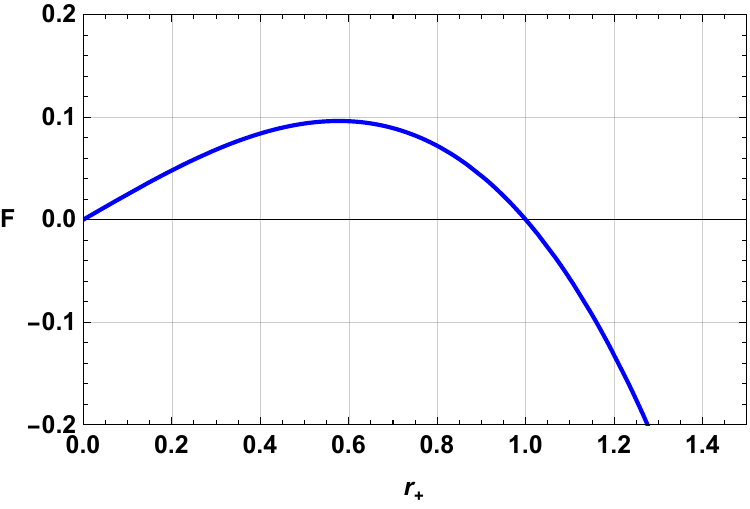}
			\caption{$\omega=-1$}
			\label{4d}
			\end{subfigure}
		\begin{subfigure}{0.30\textwidth}
		\includegraphics[width=\linewidth]{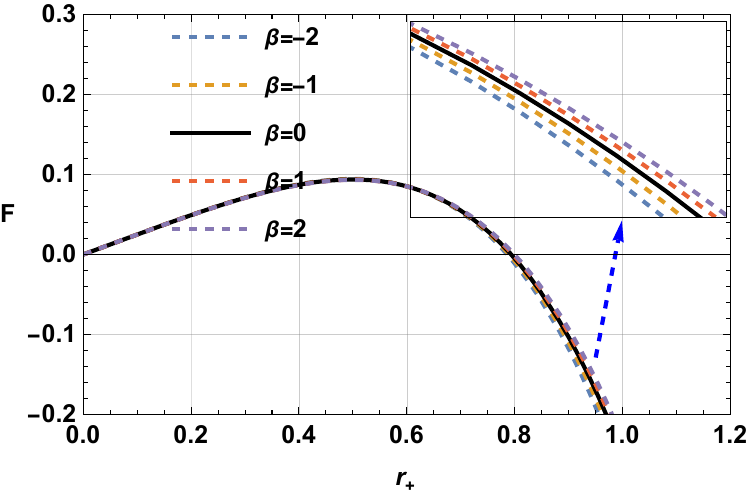}
		\caption{$\omega=-\frac{4}{3}$}
			\label{4e}
		\end{subfigure}
		\caption{ $F$ vs $r_+$ plot for different values of $\omega$. The impact of model  parameter $\beta$ on $F$ vs $r_+$ plots are shown for a specific value of $\omega$}
		\label{4}
	\end{figure}
	FIG.\ref{4} represents the free energy vs horizon radius plots. Here again, we see the difference in F vs $r_+$ plots for $\omega=0,-2/3$ cases.In FIG.\ref{4a} no Hawking Page point is observed but in FIG.\ref{4c}, we observe Hawking Page points for positive values of model parameter $beta$. To obtain information about  the thermal stability of these black holes, we calculate the specific heat $(C)$ of the black hole using the formula :
	\begin{equation}
		C=\frac{d M}{d T}=\frac{\mathcal{A}}{\mathcal{B}}
	\end{equation}
	The expression for $\mathcal{A}$ comes out to be :
	\begin{equation}
	\mathcal{A}=-2 \pi  r_+^2 (16 \pi -\beta  (\omega -3)) \left(\beta  K (9 \omega -3)-16 \pi  \left(r_+^{\frac{8 (\beta  \omega +\pi  (6 \omega +2))}{16 \pi -\beta  (\omega -3)}}-3 K \omega \right)+\beta  (\omega -3) r_+^{\frac{8 (\beta  \omega +\pi  (6 \omega +2))}{16 \pi -\beta  (\omega -3)}}\right)
	\end{equation}
	and 
	\begin{multline}
		\mathcal{B}=-\left(\beta ^2 \left(K \left(-63 \omega ^2-6 \omega +9\right)+(\omega -3)^2 r_+^{\frac{8 (\beta  \omega +\pi  (6 \omega +2))}{16 \pi -\beta  (\omega -3)}}\right)\right)\\+32 \pi  \beta  \left(3 K \left(8 \omega ^2+3 \omega -1\right)+(\omega -3) r_+^{\frac{8 (\beta  \omega +\pi  (6 \omega +2))}{16 \pi -\beta  (\omega -3)}}\right)-256 \pi ^2 \left(r_+^{\frac{8 (\beta  \omega +\pi  (6 \omega +2))}{16 \pi -\beta  (\omega -3)}}-3 K \omega  (3 \omega +2)\right)
	\end{multline}
	In FIG.\ref{3}, the specific heat is plotted against the horizon radius $r_+$. As illustrated in FIG.\ref{5}, the critical point at which the specific heat diverges shifts with the introduction of the model parameter. The comparison between the $C$ versus $r_+$ plot for $\beta = 0$ and for non-zero values of $\beta$ is shown for a specific value of $\omega$. The black solid line in the $C$ versus $r$ plot for each black hole class represents the Davies point in GR theory. For black holes surrounded by a dust field and a quintessence field, we observe no phase transition (Davies point) in GR theory. However, when considering negative and positive values of $\beta$ respectively in both case, we see small-to-large black hole phase transitions. For example, in the case of the dust field ($\omega = 0$), there is no Davies point for $\beta = 0$. But for $\beta = -2$, a Davies point appears at $r_+ = 0.33563$, as indicated by the blue dashed line in FIG.\ref{5a}.
	\begin{figure}[h]	
		\centering
		\begin{subfigure}{0.30\textwidth}
			\includegraphics[width=\linewidth]{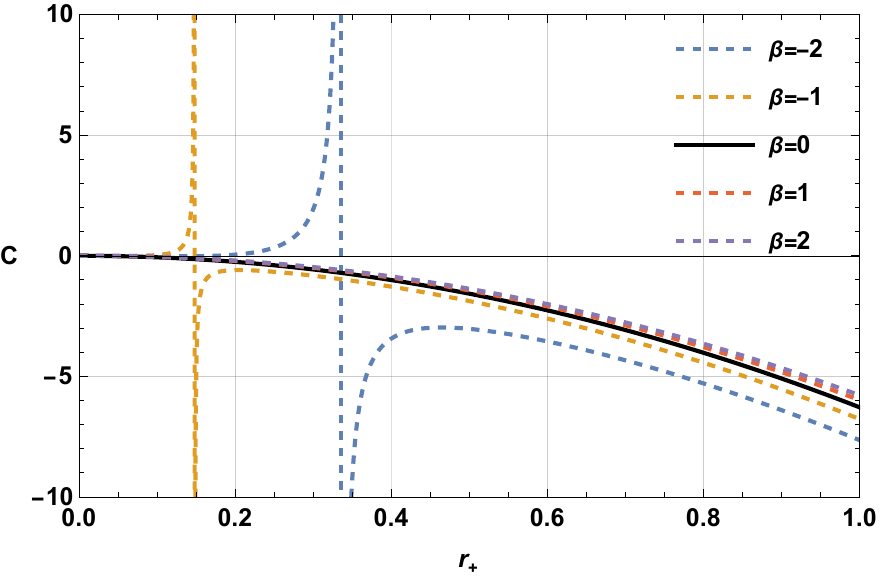}
			\caption{$\omega=0$}
			\label{5a}
		\end{subfigure}
		\begin{subfigure}{0.30\textwidth}
			\includegraphics[width=\linewidth]{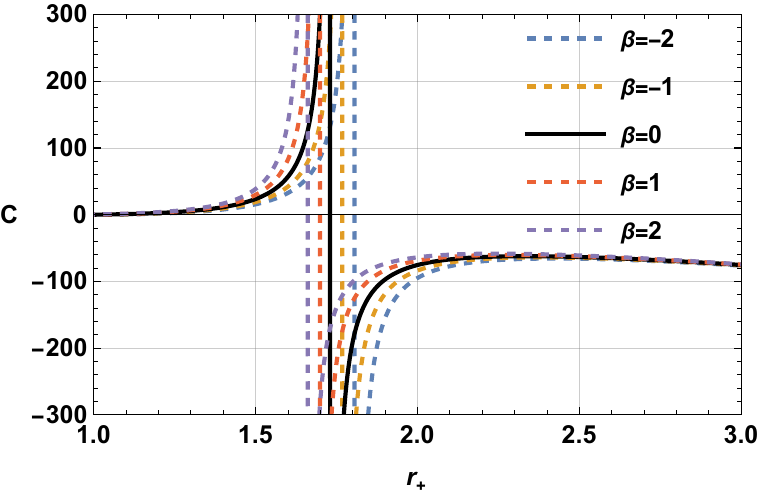}
			\caption{$\omega=\frac{1}{3}$}
			\label{5b}
		\end{subfigure}
		\begin{subfigure}{0.30\textwidth}
			\includegraphics[width=\linewidth]{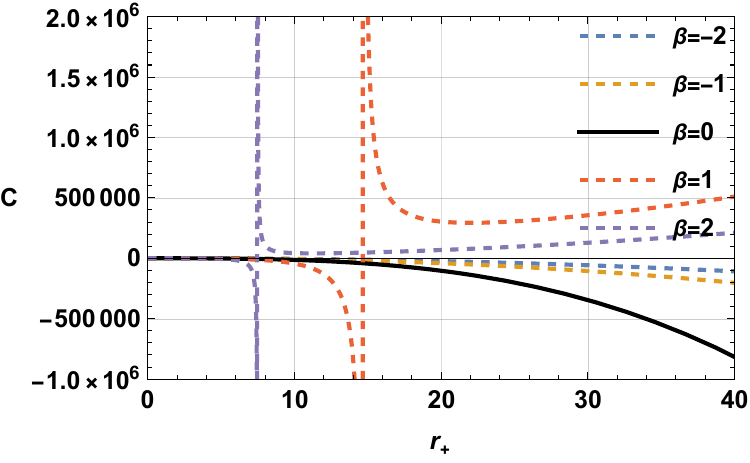}
			\caption{$\omega=-\frac{2}{3}$}
			\label{5c}
		\end{subfigure}
		\begin{subfigure}{0.30\textwidth}
			\includegraphics[width=\linewidth]{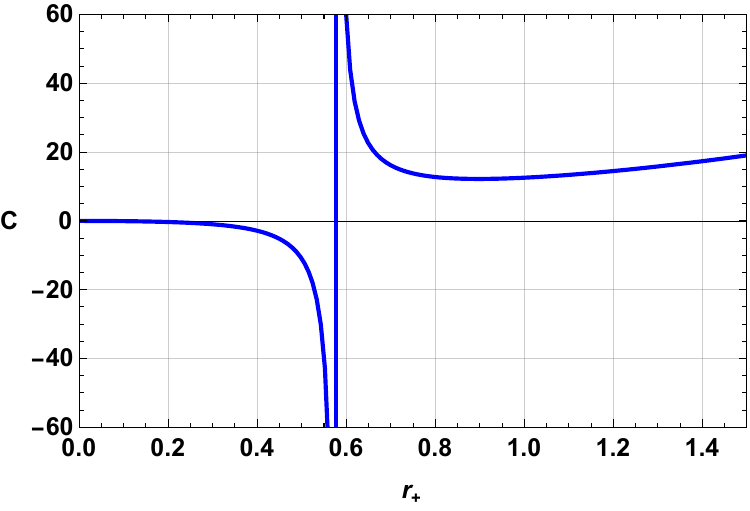}
			\caption{$\omega=-1$}
			\label{5d}
			\end{subfigure}
		\begin{subfigure}{0.30\textwidth}
		\includegraphics[width=\linewidth]{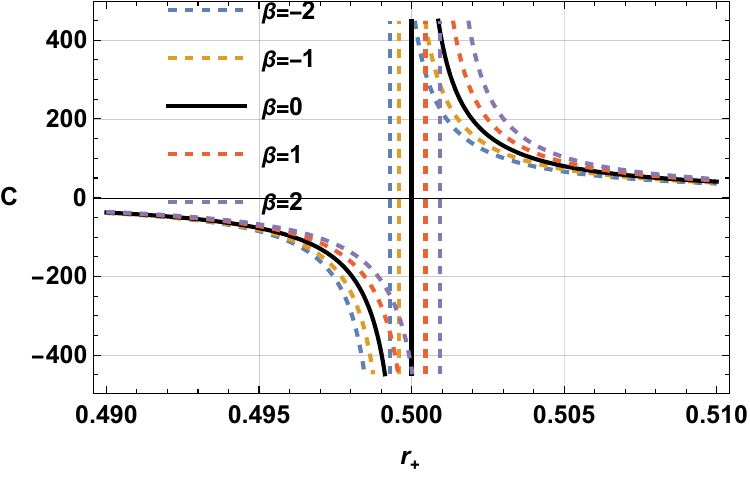}
		\caption{$\omega=-\frac{4}{3}$}
			\label{5e}
		\end{subfigure}
		\caption{ $C$ vs $r_+$ plots for different values of $\omega$. The impact of model  parameter $\beta$  on critical points are shown for a specific value of $\omega$}
		\label{5}
	\end{figure}
Negative values of specific heat indicates an unstable black hole branch and positive value of specific heat indicates the opposite.In FIG \ref{5a}, we found an unstable black hole branch for $\beta=0$ anda  positive value of $\beta$ while keeping $\omega=0,K=1$ fixed.On the other hand we found a stable small black hole(SBH)  branch and an unstable large black hole branch(LBH) fora  negative value of $\beta$. In FIG.\ref{5b}, both a stable small black hole (SBH) branch and an unstable large black hole (LBH) branch are observed for all values of $\beta$. It is important to note that the SBH branch is found to be unstable within a certain range of $r_+$ values; however, we verified that within this range, the temperature is also negative. Consequently, we omit this range and obtain a completely stable SBH branch. The Davies point shifts with variations in the model parameter. In FIG.\ref{5c}, where $\omega = -2/3$ and $K = 1$, we find that positive values of the model parameter lead to a stable LBH branch and an unstable SBH branch. For $\beta = 0$ and $\beta < 0$, we observe a single unstable black hole branch.For black holes surrounded by cosmological constant field($\omega=-1$), we observed that specific heat is independent of the model parameter. For $K=1$, value we found an unstable SBH branch and a stable LBH branch with the 	Davies point located at $r_+=0.5773$ as represented in FIG.\ref{5d}.Similarly in FIG.\ref{5c}, an unstable SBH branch and a stable LBH branch for all values of $\beta$ while keeping $\omega=-4/3,K=1$ constant.Here also the Davies point shifts with the change in the values of $\beta.$
\subsection{Thermodynamic Topology}
The generalized off-shell free energy, first introduced in \cite{29} inspired by the work of \cite{ york}:
\begin{equation}
\mathcal{F} = E - \frac{S}{\tau},
\end{equation}
where $E$ denotes the black hole’s energy (or mass $M$), and $S$ represents its entropy. The parameter $\tau$ acts as a varying time scale, interpreted as the inverse of the black hole's equilibrium temperature within the surrounding shell. From this free energy, a vector field $\phi$ can be constructed as \cite{29}:
\begin{equation}
\phi = \left(\frac{\partial \mathcal{F}}{\partial S}, -\cot \Theta \, \csc \Theta \right),
\end{equation}
where $\Theta$ is a topological angle. The critical points of the system correspond to the zero points of this vector field, occurring at:
\begin{equation}
(\tau, \Theta) = \left(\frac{1}{T}, \frac{\pi}{2}\right),
\end{equation}
where $T$ is the equilibrium temperature of the black hole.

The topological charge $W$ is determined using Duan’s $\phi$-mapping method, where the vector field’s unit vector $n^a$ must satisfy \cite{d1,d2}:
\begin{equation}
n^a n^a = 1.
\end{equation}
From this, the conserved topological current $j^\mu$ is defined in the space of coordinates $x^\nu = \{t, S, \Theta\}$ as \cite{d1,d2} :
\begin{equation}
j^\mu = \frac{1}{2\pi} \epsilon^{\mu \nu \rho} \epsilon_{ab} \partial_\nu n^a \partial_\rho n^b,
\end{equation}
with $\epsilon^{\mu \nu \rho}$ being the Levi-Civita symbol. Conservation of the current is expressed as \cite{d1,d2} :
\begin{equation}
\partial_\mu j^\mu = 0.
\end{equation}
The topological charge $W$ is calculated by integrating the zeroth component of this current \cite{d1,d2,29} :
\begin{equation}
W = \int_\Sigma j^0 \, d^2x = \sum_i w_i,
\end{equation}
where $w_i$ is the winding number for each zero point of the vector field $\phi$, and $\Sigma$ is the region over which the integration is performed.

The winding number $w$ is connected to the deflection angle $\Omega$ as:
\begin{equation}
w = \frac{\Omega}{2\pi},
\end{equation}
where the deflection angle $\Omega$ is computed via \cite {28,29}:
\begin{equation}
\Omega = \int_0^{2\pi} \epsilon_{12} n^1 \partial_\nu n^2 \, d\nu.
\label{deflection}
\end{equation}
Where $n^1$ and $n^2$ are the normalized unit vector and $\nu \in (0,2\pi )$ is the parametrization variable which is used to construct the contours around which we will calculate the winding number. Contours with suitable dimensions are designed to outline the parameter region, as described below.
\begin{equation}
\begin{cases}
r_{+}=r_{1}\cos \nu +r_{0}, \\ 
\\ 
\theta =r_{2}\sin \nu +\frac{\pi }{2},%
\end{cases}%
\end{equation}%
$r_{1}$ and $r_{2}$ are parameters that is used to
control the dimensions of the contour  and $r_{0}$ represents the centre point(defect point in this case) around which the contour is plotted. \\
The sum of all winding numbers provides the total topological charge $W$, which characterizes the structure of the black hole system in thermodynamic topology.
This total charge is nonzero only at the zero points of the vector field $\phi$, indicating the presence of critical points. If no such points are found, the topological charge remains zero, implying the absence of significant thermodynamic transitions.\\

Utilizing the expression for mass in equation (\ref{mass1}), the off-shell free energy is calculated as
\begin{equation}
\mathcal{F}=M-\frac{S}{\tau }=\frac{r_+ \left(K \tau  r_+^{\frac{8 (\beta  \omega +\pi  (6 \omega +2))}{\beta  (\omega -3)-16 \pi }}-2 \pi  r_++\tau \right)}{2 \tau }
\end{equation}

Next, a vector field $\phi$ is constructed as
\begin{equation}
\phi ^{r}=\frac{\tau  (16 \pi -\beta  (\omega -3))-3 K \tau  (\beta  (3 \omega -1)+16 \pi  \omega ) r_+^{\frac{8 (\beta  \omega +\pi  (6 \omega +2))}{\beta  (\omega -3)-16 \pi }}-4 \pi  r_+ (16 \pi -\beta  (\omega -3))}{32 \pi  \tau -2 \beta  \tau  (\omega -3)}
\end{equation}
The zero points $\tau$ of the  $\phi ^{r}$,  can be obtained as 
\begin{equation}
\tau =\frac{4 \pi  (16 \pi -\beta  (\omega -3)) r_+^{\frac{7 \beta  \omega +3 \beta +16 \pi }{\beta  (-\omega )+3 \beta +16 \pi }}}{(16 \pi -\beta  (\omega -3)) r_+^{\frac{8 \beta  \omega }{16 \pi -\beta  (\omega -3)}}-3 K (\beta  (3 \omega -1)+16 \pi  \omega ) r_+^{-\frac{16 (3 \pi  \omega +\pi )}{16 \pi -\beta  (\omega -3)}}}
 \label{tau}
\end{equation}%
%%%%%%%%%%%%%%%%%%%%%%%%%%%%%%%%%%%%%%%%%%%%%%%%%%%%%%%%

\begin{figure}[h!]
\centering
\begin{subfigure}{0.3\textwidth}
		\includegraphics[width=\linewidth]{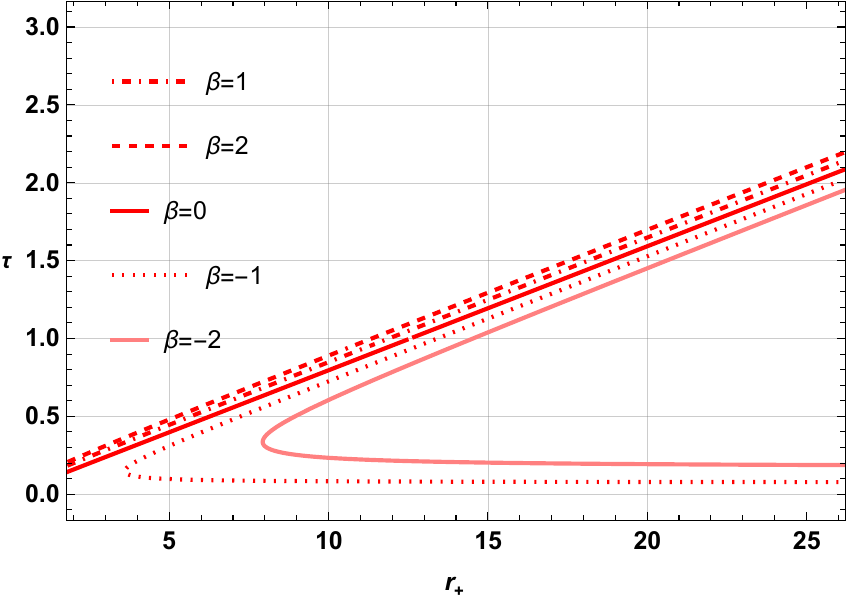}
		\caption{$\omega=0$.}
		\label{t1a}
	\end{subfigure}\hspace{0.3cm} 
\begin{subfigure}{0.3\textwidth}
		\includegraphics[width=\linewidth]{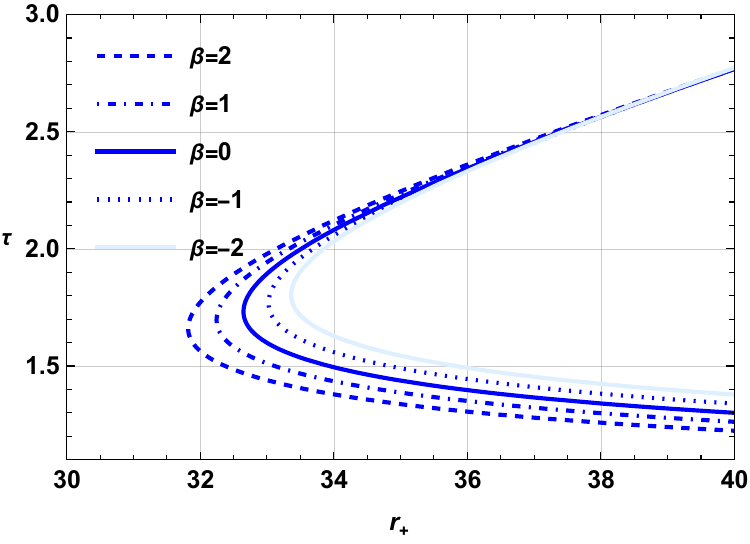}
		\caption{$\omega=\frac{1}{3}$}
		\label{t1b}
	\end{subfigure} \hspace{0.3cm} 
\begin{subfigure}{0.3\textwidth}
		\includegraphics[width=\linewidth]{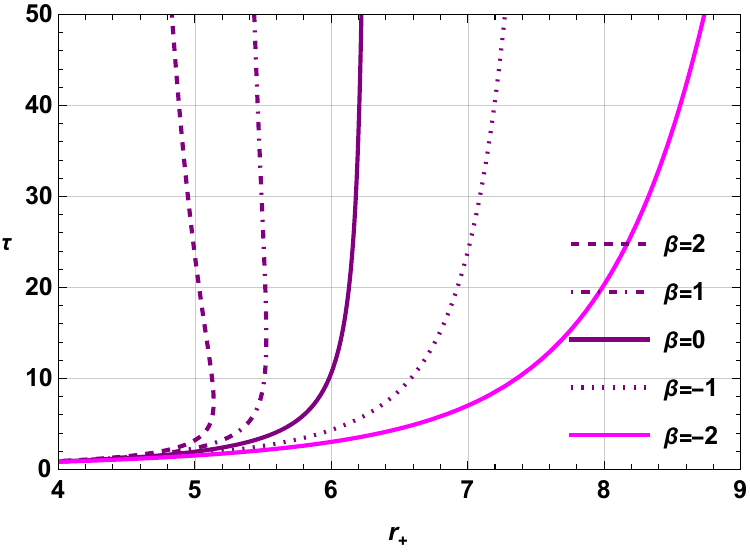}
		\caption{$\omega=-\frac{2}{3}$}
		\label{t1c}
	\end{subfigure}
	\begin{subfigure}{0.3\textwidth}\hspace{0.3cm} 
		\includegraphics[width=\linewidth]{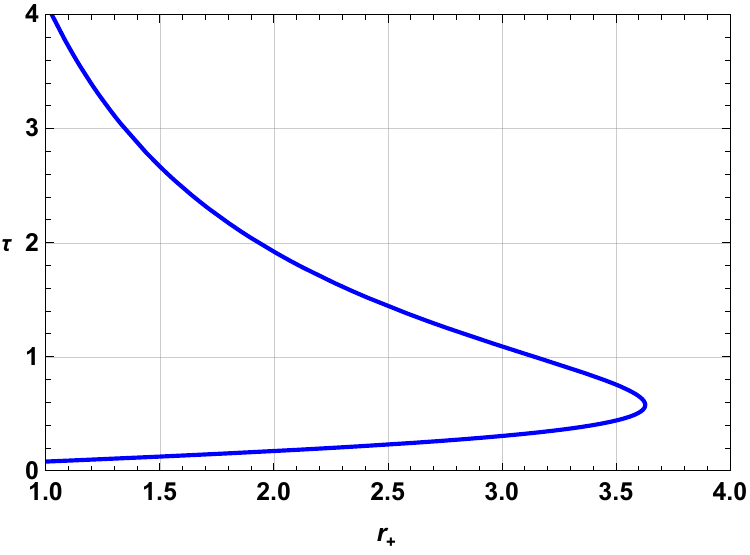}
		\caption{$\omega=-1$}
		\label{t1d}
	\end{subfigure}
	\begin{subfigure}{0.3\textwidth}
		\includegraphics[width=\linewidth]{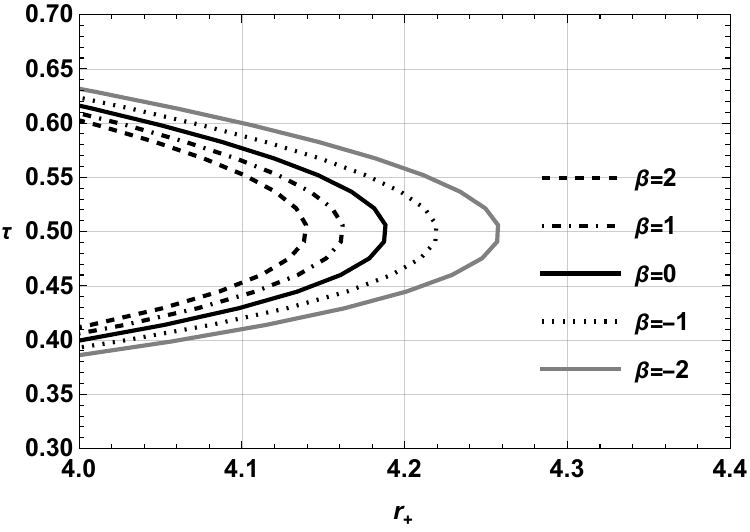}
		\caption{$\omega=-\frac{4}{3}$}
		\label{t1e}
	\end{subfigure}
\caption{ Variation of $\protect\tau $ vs $r_{+}$ plots with model parameter for black holes in $f(R,T)$ gravity }
\label{topo}
\end{figure}

\begin{figure}[h!]
\centering
\begin{subfigure}{0.3\textwidth}
		\includegraphics[width=\linewidth]{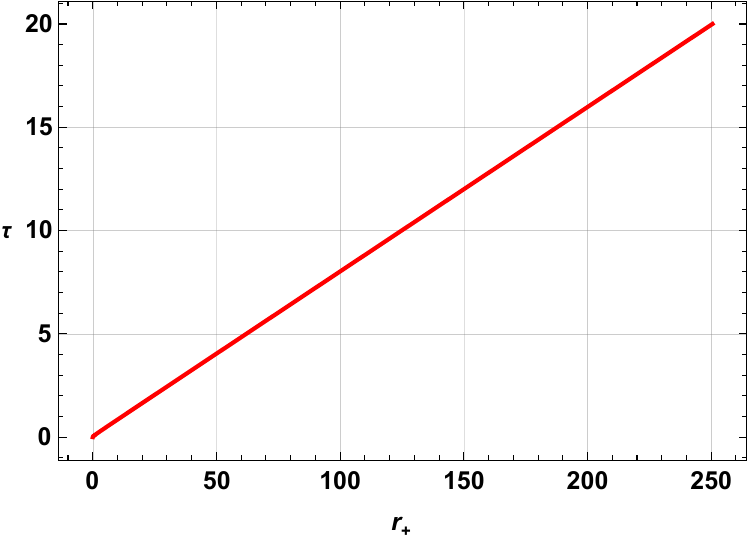}
		\caption{Defect curve}
		\label{t2a}
	\end{subfigure}\hspace{0.3cm} 
\begin{subfigure}{0.3\textwidth}
		\includegraphics[width=\linewidth]{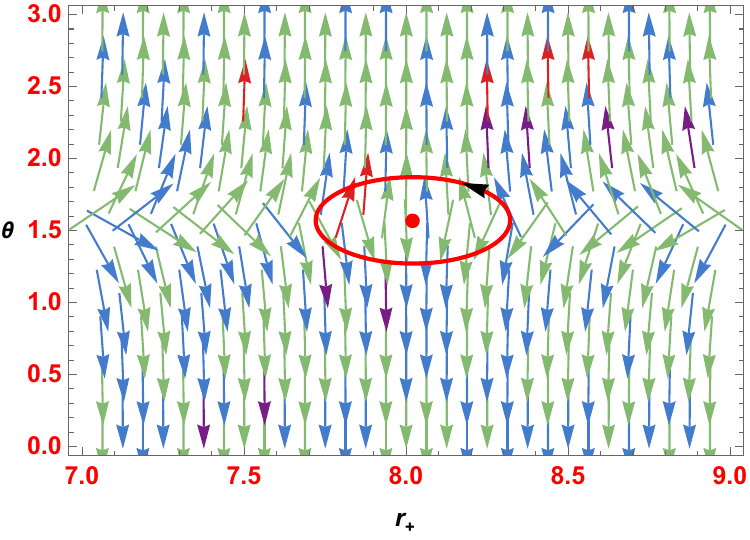}
		\caption{Normalized vectorplot}
		\label{t2b}
	\end{subfigure} \hspace{0.3cm} 
\begin{subfigure}{0.3\textwidth}
		\includegraphics[width=\linewidth]{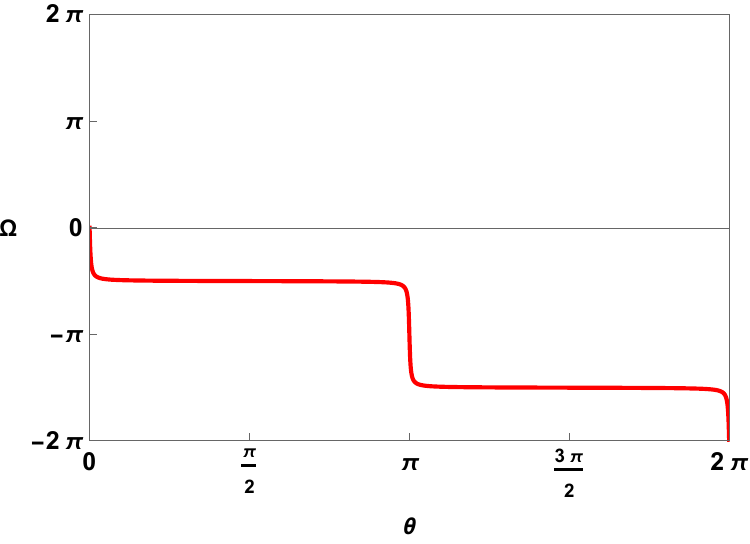}
		\caption{Contourplot}
		\label{t2c}
	\end{subfigure}
	\begin{subfigure}{0.3\textwidth}\hspace{0.3cm} 
		\includegraphics[width=\linewidth]{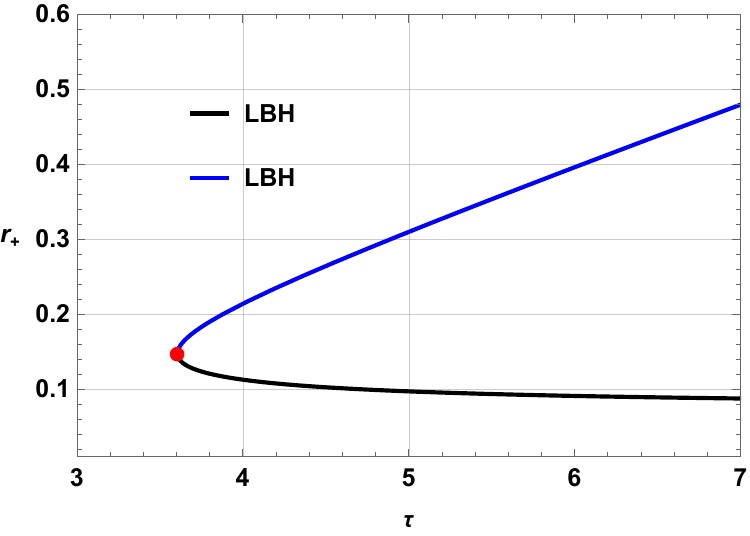}\hspace{0.3cm} 
		\caption{Defect curve}
		\label{t2d}
	\end{subfigure}
	\begin{subfigure}{0.3\textwidth}\hspace{0.3cm} 
		\includegraphics[width=\linewidth]{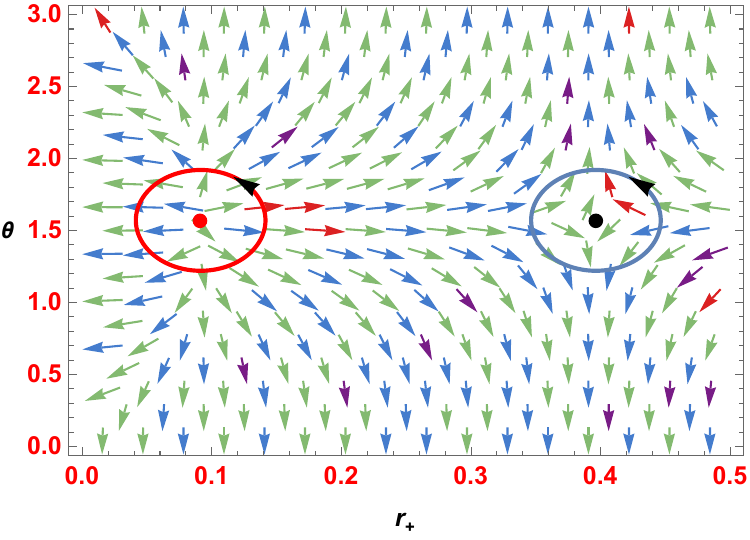}
		\caption{Normalized vectorplot}
		\label{t2e}
	\end{subfigure}
	\begin{subfigure}{0.3\textwidth}
		\includegraphics[width=\linewidth]{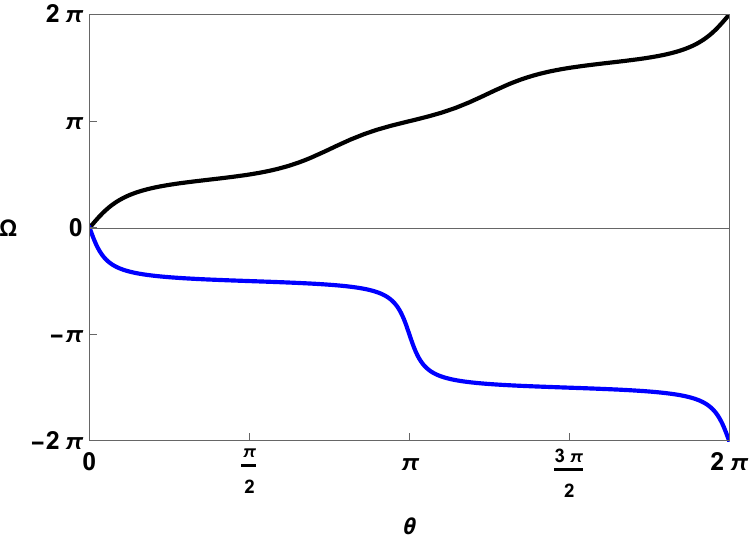}
		\caption{Contourplot}
		\label{t2f}
	\end{subfigure}
\caption{ Topological charge calculation for $w=0$ black holes.}
\label{t2}
\end{figure}

Next, $\tau $ vs $r_{+}$ is plotted in FIG. \ref{topo} for  different values of model parameter $\beta$ for a fixed value of $\omega.$ The effect of model parameters on thermodynamic topology can be studied using these plots. The number of black hole branches eventually determines the topological charge.  As the plot  FIG. \ref{t1a}  reveals there are either two black hole branch or one black hole branch depending on the value of $\beta$ for $\omega=0$. The topological charge calculation for this case is explained in FIG. \ref{t2}.\\

%%%%%%%%%%%%%%%%%%%%%%%%%%%%%%%%%%%%%%%%%%%%%%%
%%%%%%%%%%%%%%%%%%%%%%%%%%%%%%%%%%%%%%%%%%%%%%%%%%%%
In FIG. \ref{t2a}, we plot $\tau$ vs. $r_+$ for the model parameter $\beta = 1$, where we observe a single black hole branch. FIG. \ref{t2b} presents a vector plot of the normalized vector field $n$, with $\tau = 100$. To identify the zero points of the vector field $\phi$ a vector plot is used, where the vector magnitudes become zero at these points, appearing as vanishing or absent arrows. By analyzing the behaviour of the normalized vectors near these points, one can determine the nature of the zero points.Here the zero point is identified to be at $(r_+,\theta)=(8.0205, \pi/2)$ as at that point, all the vector diverges.In FIG. \ref{t2c}, the calculation of the topological charge is shown. To calculate the topological charge, we first calculate the winding number of individual black hole branches. The winding number is related to the deflection of a vector field around its zero points which is calculated using eq.\ref{deflection}.  We conduct a contour integration around the red contour in FIG.\ref{t2b} where we parametrized the contour around the zero point  $(r_+,\theta)=8.0205, \pi/2$.The deflection $\Omega$ is plotted against the parametrized variable $\nu$ in FIG. \ref{t2c} where the $r_+$ and $\theta$ is parametrized using $\nu$ as :
\begin{equation}
\begin{cases}
r_{+}=0.3 \cos \nu +8.0205, \\ 
\\ 
\theta =0.3\sin \nu +\frac{\pi }{2},%
\end{cases}%
\end{equation}%
The contour plot in FIG. \ref{t2c} reveals that the winding number is  $-1$.Since there is only one branch, the topological charge is equivalent to winding number.In FIG. \ref{t2d}, we again plot $\tau$ vs. $r_+$, but this time for the model parameter $\beta = -1$. Here, two black hole branches are identified: a small black hole (SBH) branch for $r_+ < 0.147864$, and a large black hole (LBH) branch for $r_+ > 0.147864$. For $\tau = 6$, the zero points are shown in the vector plot in FIG. \ref{t2e}. To calculate the topological charge, we parameterized the two contours shown in FIG. \ref{t2e}. The red contour is constructed around the zero point located in the SBH branch, while the blue contour is constructed around the zero point in the LBH branch. The winding number calculated around each of these zero points represents the winding number for the entire branch in which they are individually situated. Next, the winding number is calculated by contour integration: for the SBH branch, it is $+1$, represented by the black solid line, and for the LBH branch, it is $-1$, represented by the blue solid line. Adding the winding numbers gives a topological charge $W$ of $1 - 1 = 0$.
A positive winding number corresponds to a stable SBH branch, while a negative winding number indicates an unstable LBH branch. The critical point $(\tau_c, r_c) = (3.60535, 0.14786)$, marked by the red dot in FIG. \ref{t2f}, represents an annihilation point where the stable SBH branch ends and the unstable LBH branch begins.
Thus, the topological charge for a black hole surrounded by a dust field ($\omega = 0$) is $-1$ for a positive value of the model parameter and $0$ for a negative value of the model parameter.

\begin{figure}[h!]
\centering
\begin{subfigure}{0.3\textwidth}
		\includegraphics[width=\linewidth]{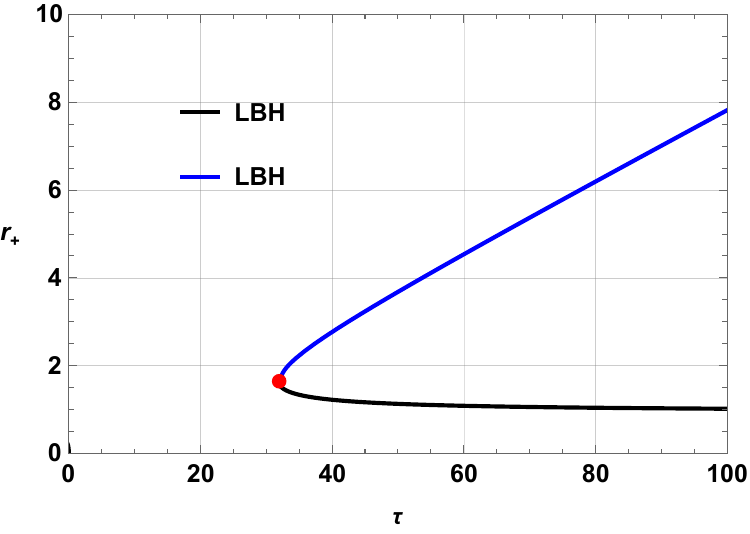}
		\caption{}
		\label{t3a}
	\end{subfigure}\hspace{0.3cm} 
\begin{subfigure}{0.3\textwidth}
		\includegraphics[width=\linewidth]{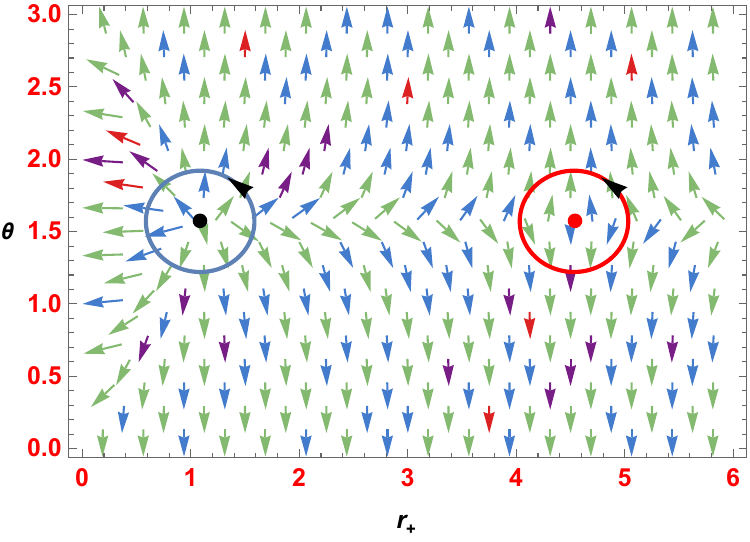}
		\caption{}
		\label{t3b}
	\end{subfigure} \hspace{0.3cm} 
\begin{subfigure}{0.3\textwidth}
		\includegraphics[width=\linewidth]{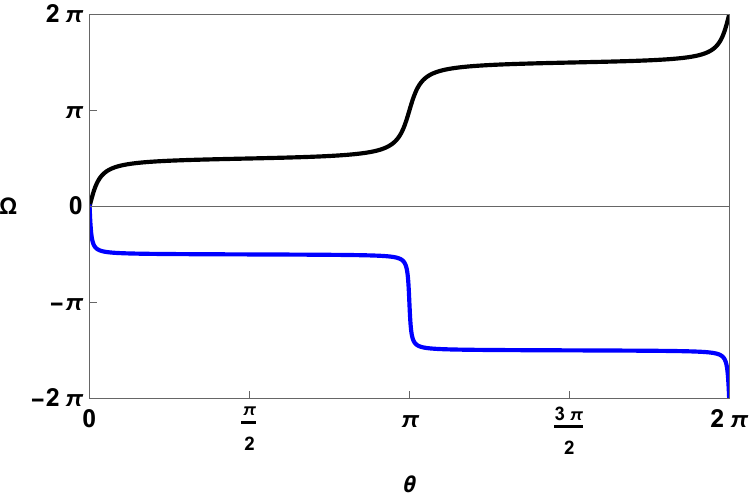}
		\caption{}
		\label{t3c}
	\end{subfigure}
	\caption{ Topological charge calculation for $w=1/3$ black holes.}
\label{t3}
	\end{figure}

FIG. \ref{t1b} shows two black hole branches for all values of $\beta$ when $\omega = \frac{1}{3}$. The topological charge calculation for this case is explained in FIG. \ref{t3}.
In FIG. \ref{t3a}, we plot $\tau$ vs. $r_+$ for the model parameter $\beta = 2$, where we find two black hole branches: a small black hole (SBH) branch for $r_+ < 1.6602$, and a large black hole (LBH) branch for $r_+ > 1.6602$. For $\tau = 60$, zero points are found at $r_+ = 1.0860$ and $r_+ = 4.5316$, as shown in the vector plot in FIG. \ref{t3b}. 
FIG. \ref{t3c} explains the winding number calculation. The winding number for the SBH is calculated to be $+1$, represented by the black solid line, while that for the LBH is $-1$, represented by the blue solid line. By adding the winding numbers, the topological charge $W$ is obtained as $W = 1 - 1 = 0$. 
Here as well, the critical point $(\tau_c, r_c) = (31.8199, 1.6602)$, represented by the red dot in FIG. \ref{t3a}, is an annihilation point.

\begin{figure}[h!]
\centering
\begin{subfigure}{0.3\textwidth}\hspace{0.3cm} 
		\includegraphics[width=\linewidth]{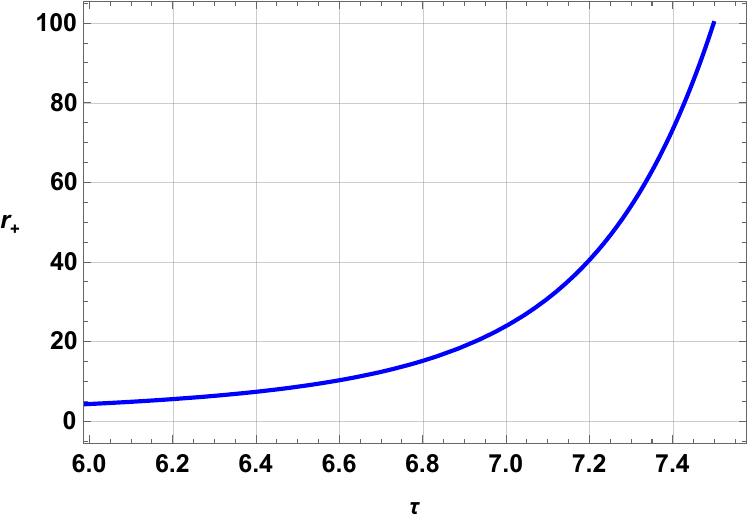}
		\caption{}
		\label{t4a}
	\end{subfigure}\hspace{0.3cm} 
\begin{subfigure}{0.3\textwidth}
		\includegraphics[width=\linewidth]{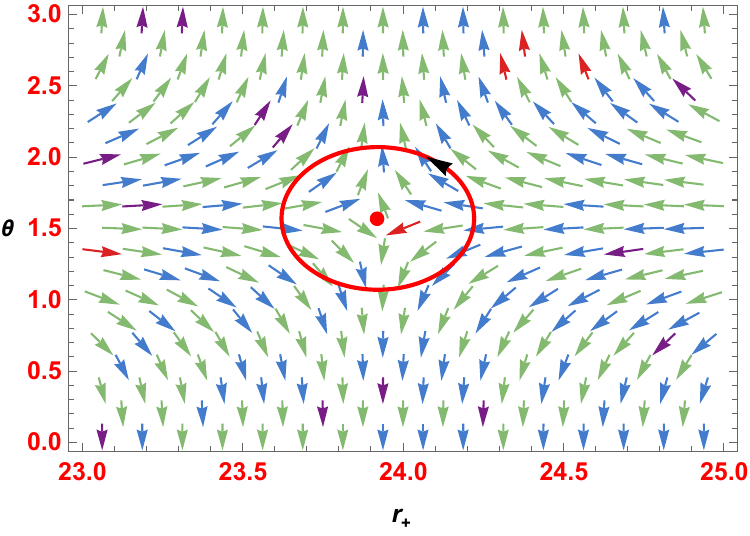}
		\caption{}
		\label{t4b}
	\end{subfigure}
\begin{subfigure}{0.3\textwidth}
		\includegraphics[width=\linewidth]{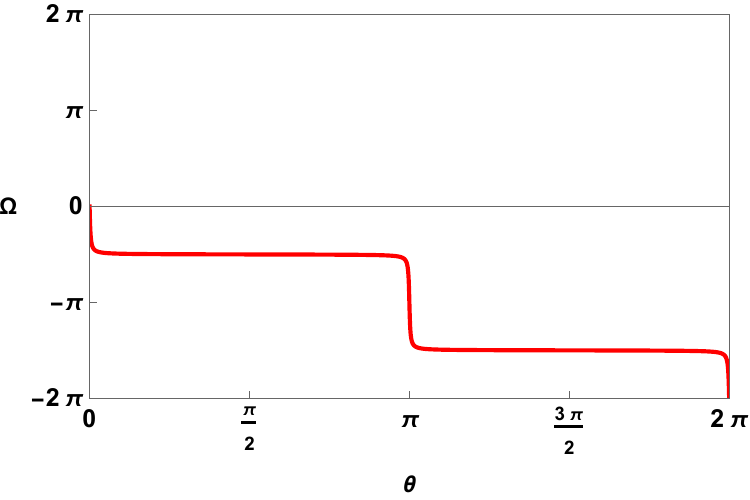}
		\caption{}
		\label{t4c}
	\end{subfigure}
	\begin{subfigure}{0.3\textwidth}\hspace{0.3cm} 
		\includegraphics[width=\linewidth]{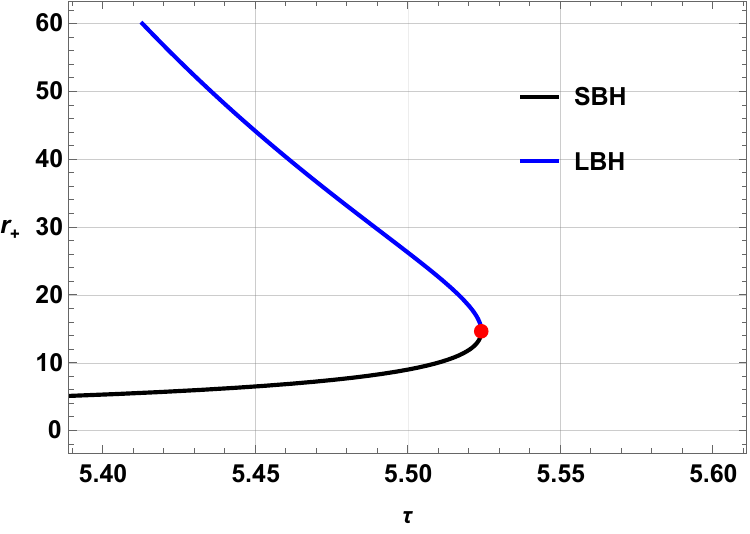}
		\caption{}
		\label{t4d}
	\end{subfigure}\hspace{0.3cm} 
	\begin{subfigure}{0.3\textwidth}\hspace{0.3cm} 
		\includegraphics[width=\linewidth]{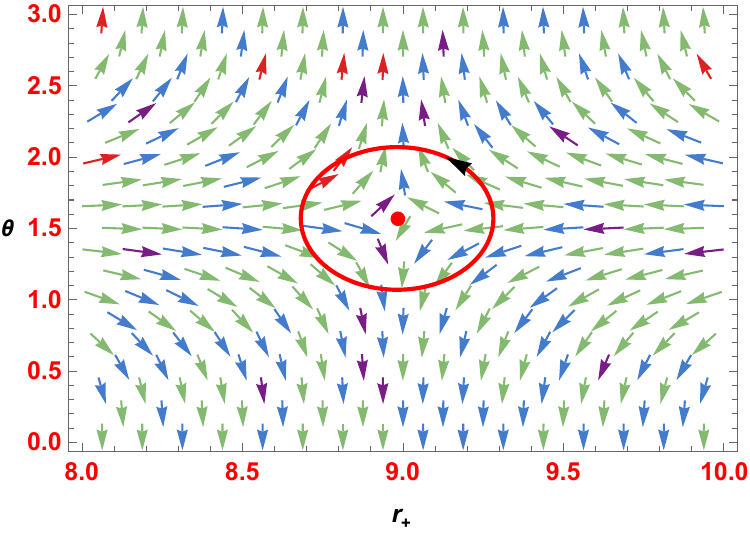}
		\caption{}
		\label{t4e}
	\end{subfigure}
	\begin{subfigure}{0.3\textwidth}\hspace{0.3cm} 
		\includegraphics[width=\linewidth]{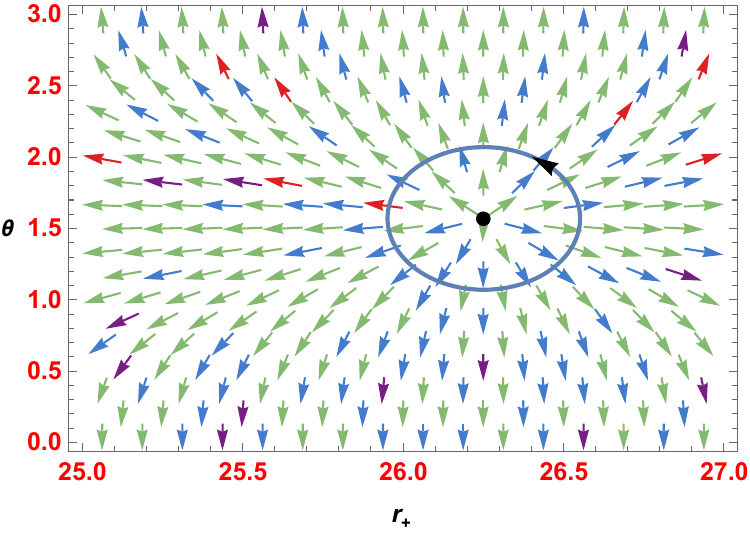}
		\caption{}
		\label{t4f}
	\end{subfigure}
	\begin{subfigure}{0.3\textwidth}
		\includegraphics[width=\linewidth]{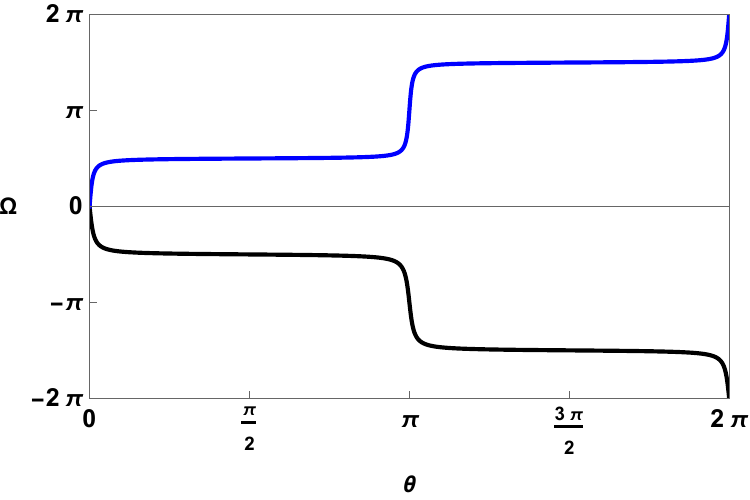}
		\caption{}
		\label{t4g}
	\end{subfigure}
\caption{ Topological charge calculation for $w=-2/3$ black holes.}
\label{t4}
\end{figure}

In FIG. \ref{t1c},  $\tau$ vs. $r_+$  is plotted for  $\omega=-\frac{2}{3}$ and the impact of  model parameter on it is observed. This particular scenario is being discussed in FIG.\ref{t4}. In FIG.\ref{t4a}, we take $\beta=-1$,  where we observe a single black hole branch. FIG. \ref{t4b} presents a vector plot of the normalized vector field $n$, with $\tau = 7$ where the zero point is located at $r_+=23.9214$. In FIG. \ref{t4c}, the calculation of the topological charge is shown around the zero point. The contour plot in FIG. \ref{t2c} reveals that the topological charge is found to be $-1$.
In FIG. \ref{t4d}, we again plot $\tau$ vs. $r_+$, but this time for the model parameter $\beta = 1$. Here, two black hole branches are identified: a small black hole (SBH) branch for $r_+ < 8.9181$, and a large black hole (LBH) branch for $r_+ > 26.2517$. For $\tau = 5.5$, the zero points are shown in the vector plot in FIG. \ref{t4e}. In FIG. \ref{t4f}, the winding number is calculated: for the SBH branch, it is $-1$, represented by the black solid line, and for the LBH branch, it is $1$, represented by the blue solid line. Adding the winding numbers gives a topological charge $W$ of $1 - 1 = 0$.A positive winding number suggests a stable LBH branch, while a negative winding number indicates an unstable SBH branch. The critical point $(\tau_c, r_c) = (3.6275, 0.5773)$, marked by the red dot in FIG. \ref{t4d}, represents an generation point.
Interestingly we observe completely opposite local topology from the $\omega=0$ case where the negative value of the model parameter was considered. Although in both scenario, global topology is the same as the topological charge is found to be zero but the local topology is totally opposite. In a black hole surrounded by quintessence ($\omega=-2/3)$,SBH branch is unstable and LBH branch is stable while in the case of a black hole surrounded by dust field($\omega=0)$,SBH branch is stable and LBH branch is unstable.  In $\omega=0$ we detect an annihilation point but In $\omega=-2/3$, we found a generation point which is also an important distinguishable factor between the topology of both the class of black hole solutions.

%%%%%%%%%%%%%%%%%%%%%%%%%%%%%%%%%%%%%%%%%%%%%%%%%%%%%%%% 

\begin{figure}[h!]
\centering
\begin{subfigure}{0.3\textwidth}
		\includegraphics[width=\linewidth]{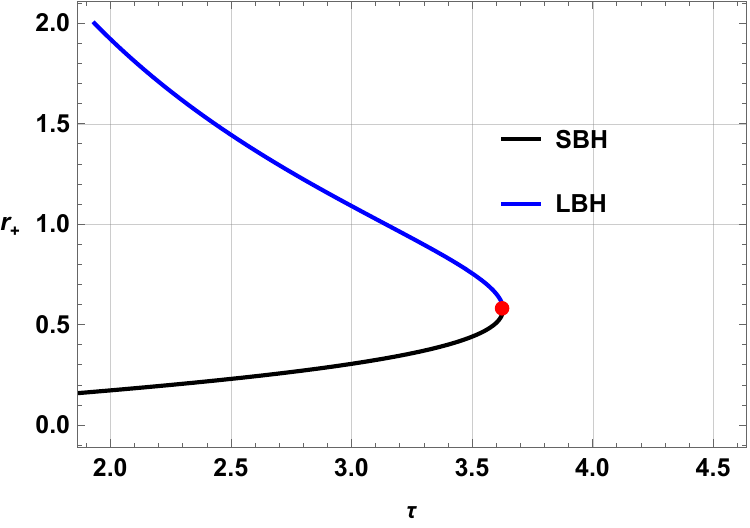}
		\caption{}
		\label{t5a}
	\end{subfigure}\hspace{0.3cm} 
\begin{subfigure}{0.3\textwidth}
		\includegraphics[width=\linewidth]{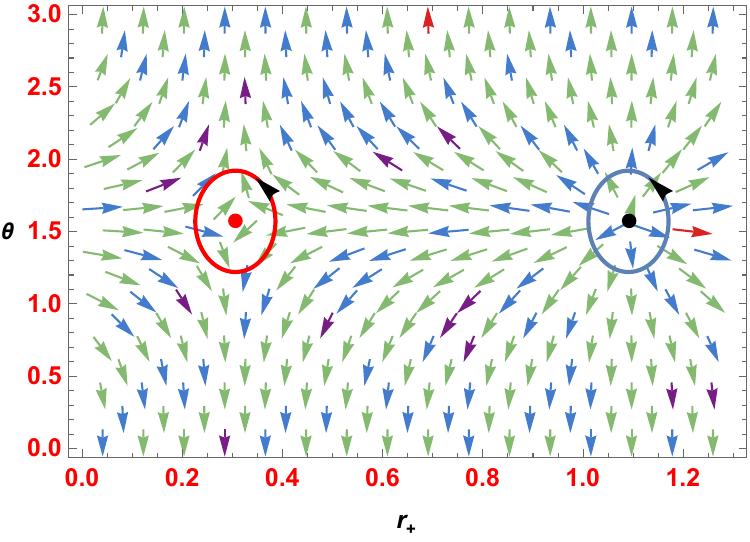}
		\caption{}
		\label{t5b}
	\end{subfigure} \hspace{0.3cm} 
\begin{subfigure}{0.3\textwidth}
		\includegraphics[width=\linewidth]{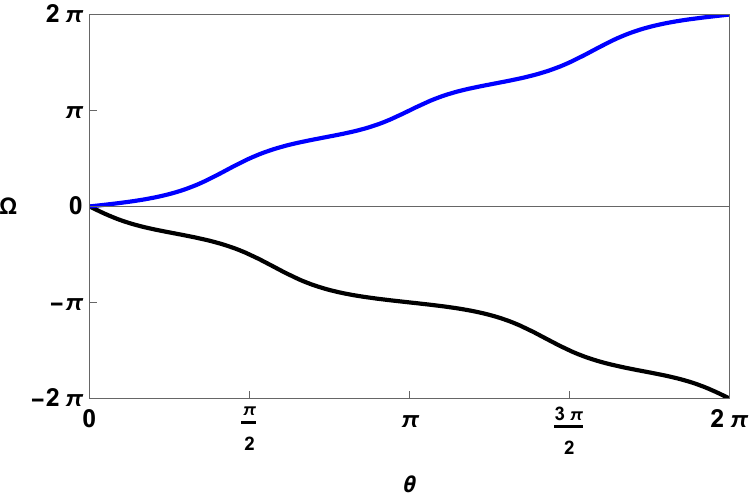}
		\caption{}
		\label{t5c}
	\end{subfigure}
	\caption{ Topological charge calculation for $w=-1$ black holes.}
\label{t5}
	\end{figure}

\begin{figure}[h!]
\centering
\begin{subfigure}{0.3\textwidth}
		\includegraphics[width=\linewidth]{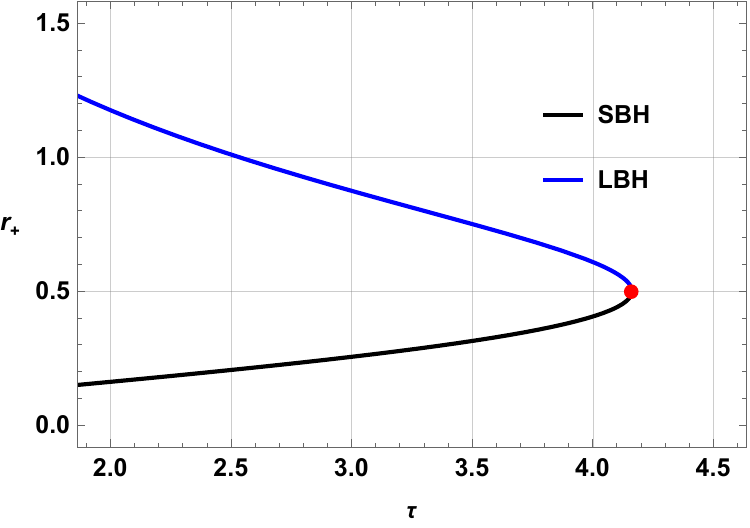}
		\caption{}
		\label{t6a}
	\end{subfigure}\hspace{0.3cm} 
\begin{subfigure}{0.3\textwidth}
		\includegraphics[width=\linewidth]{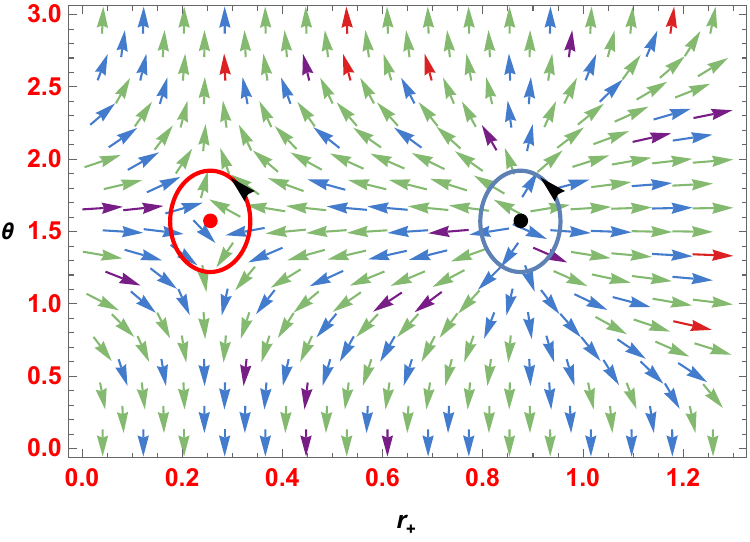}
		\caption{}
		\label{t6b}
	\end{subfigure} \hspace{0.3cm} 
\begin{subfigure}{0.3\textwidth}
		\includegraphics[width=\linewidth]{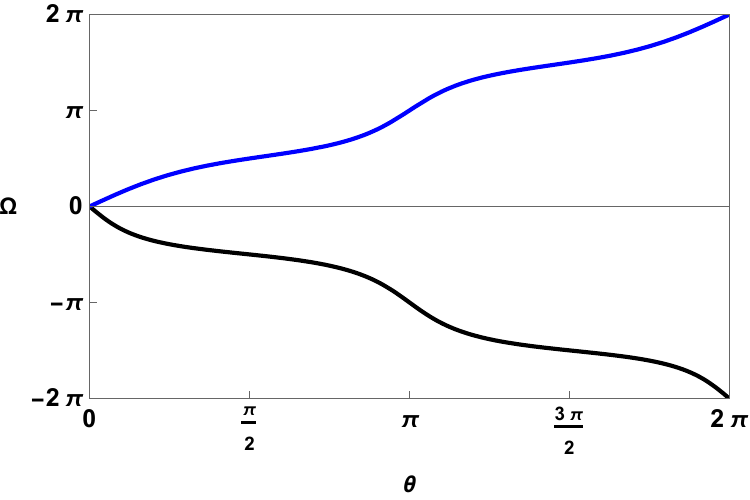}
		\caption{}
		\label{t6c}
	\end{subfigure}
	\caption{ Topological charge calculation for $w=-4/3$ black holes.}
\label{t6}
	\end{figure}

We repeat the same analysis for $\omega=-1$ and $\omega=-\frac{4}{3}$ case. FIG.\ref{t1d} shows that for $\omega=-1$, the expression for $\tau$ become independent of the model parameter. Hence the topological charge is always $0$. The calculations are shown in  FIG.\ref{t5} where we have done the calculation for $\tau=3$ and the zero points are found to be at $r_+=0.3056$ and $r_+=1.0906$. The winding number for SBH and LBH branch is found to be $-1$ and $+1 $ respectively.In  FIG.\ref{t1d}we have considered  $\omega=-\frac{4}{3}$. Here also  the topological charge is found to be always $0$. The calculations are shown in  FIG.\ref{t6} where we have taken $\tau=3,\beta=1$ and the zero points are found to be at $r_+=0.2554$ and $r_+=0.8747$. The winding number for SBH and LBH branch is found to be $-1$ and $+1 $ respectively. In both the case, the critical point is a generation point.

\subsection{Thermodynamic Geometry}
\begin{figure}[h!]
\centering
\begin{subfigure}{0.45\textwidth}
		\includegraphics[height=7cm,width=8cm]{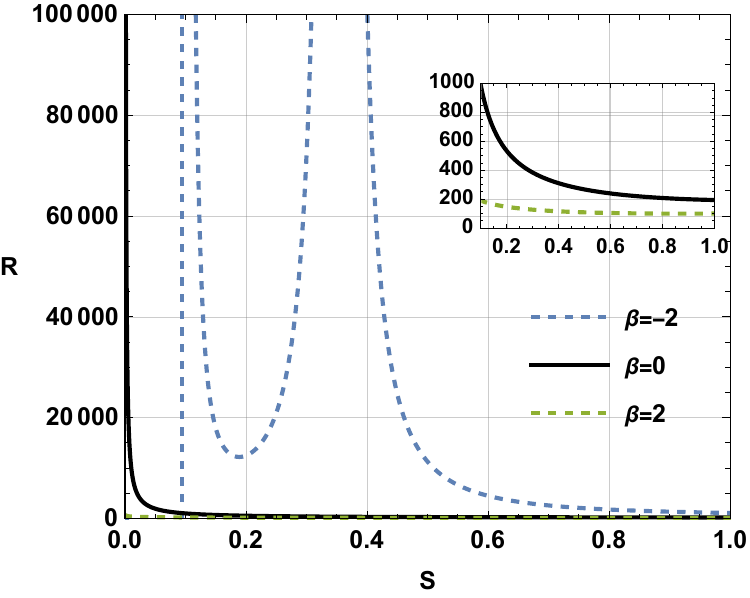}
		\caption{}
		\label{g1a}
	\end{subfigure}\hspace{0.9cm} 
\begin{subfigure}{0.45\textwidth}
		\includegraphics[height=7.3cm,width=8cm]{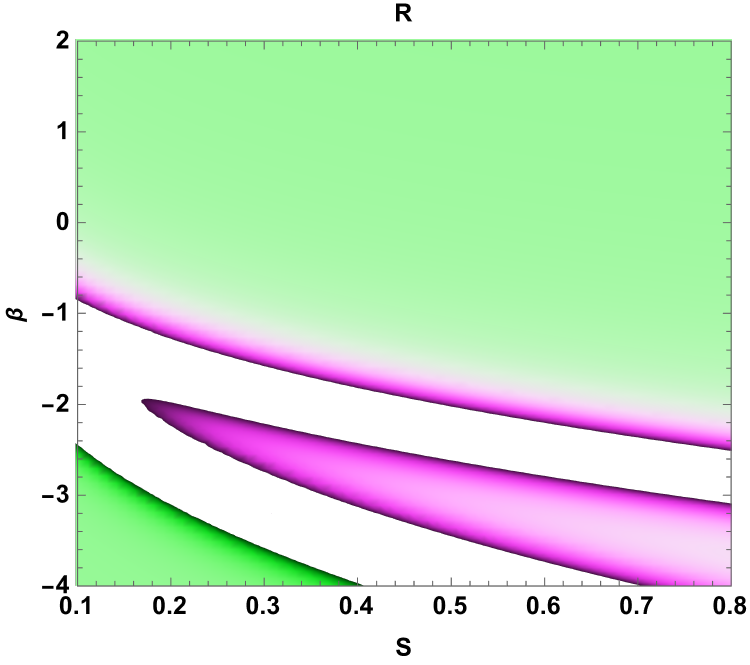}
		\caption{}
		\label{g1b}
	\end{subfigure} 
	\begin{subfigure}{0.45\textwidth}
		\includegraphics[height=7cm,width=9cm]{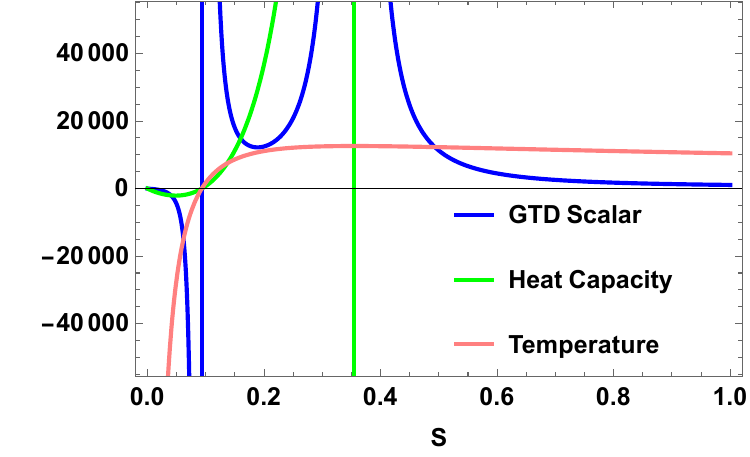}
		\caption{}
		\label{g1c}
	\end{subfigure}
	\caption{Behaviour of GTD scalar  for black hole surrounded with dust field($\omega=0$)}
	\label{g1}
\end{figure}

We explore the thermodynamic geometry of these black holes using the Geometrodynamic Thermodynamics (GTD) formalism, which operates in a multi-dimensional phase space incorporating both extensive and intensive variables of the system. This makes the GTD formalism ideally suited for analyzing all thermodynamic ensembles from a geometric perspective. To describe black holes in the $f(R,T)$ gravity model using the GTD formalism, we first consider a four-dimensional phase space $\mathcal{T}$ with coordinates $M$, $S$, $T$, and $\beta$, representing the mass, entropy, temperature, and the model parameter, respectively. We then express the thermodynamic quantities in terms of entropy $S$ as follows:
\begin{equation}
M(S,\beta)= \frac{\sqrt{S} \left(K \pi ^{\frac{4 (\beta  \omega +\pi  (6 \omega +2))}{16 \pi -\beta  (\omega -3)}} S^{-\frac{4 (\beta  \omega +\pi  (6 \omega +2))}{16 \pi -\beta  (\omega -3)}}+1\right)}{2 \sqrt{\pi }}
\end{equation}
\begin{equation}
T(S,\beta)=\frac{K \pi ^{\frac{4 (\beta  \omega +\pi  (6 \omega +2))}{16 \pi -\beta  (\omega -3)}} S^{-\frac{4 (\beta  \omega +\pi  (6 \omega +2))}{16 \pi -\beta  (\omega -3)}}+1}{4 \sqrt{\pi } \sqrt{S}}-\frac{2 K \pi ^{\frac{\beta  (9 \omega -3)+48 \pi  \omega }{32 \pi -2 \beta  (\omega -3)}} (\beta  \omega +\pi  (6 \omega +2)) S^{-\frac{4 (\beta  \omega +\pi  (6 \omega +2))}{16 \pi -\beta  (\omega -3)}-\frac{1}{2}}}{16 \pi -\beta  (\omega -3)}
\end{equation}
Next, the GTD metric  can be written from the general metric given in  (\ref{gtd}) as :
$$g  =S \left(\frac{\partial M}{\partial S}\right)\left(- \frac{\partial^2 M}{\partial S^2} dS^2 + \frac{\partial^2 M}{\partial \beta^2} d\beta^2 \right) $$  
while writing the elements of the metrix we substitute the value of $K=1$ and $\omega=0,\frac{1}{3},-\frac{2}{3}$ and $-\frac{4}{3}$.\\
 The GTD scalar for $\omega = 0$ is plotted against entropy $S$ in FIG.~\ref{g1a}. We observe that, for negative values of $\beta$, the GTD scalar exhibits a curvature singularity, as illustrated by the blue dashed curve. In contrast, for $\beta = 0$ and positive values of $\beta$, the curve remains regular everywhere without any curvature singularities, as shown by the black solid and green dashed curves, respectively. The location of the singularity identified from the GTD scalar curve corresponds exactly with the divergence point (Davies point) observed in the corresponding heat capacity curve for the same set of values, as depicted in FIG.~\ref{g1c}. Although FIG.~\ref{g1a} shows two singularities for $\beta = -2$, we consider only the singularity in the region where the temperature is positive.
FIG.~\ref{g1b} presents a density plot of the scalar curvature $R$ as a function of entropy $S$ and the model parameter $\beta$. The white patches in the figure indicate regions where $R$ is not defined. The figure reveals that for positive values of the model parameter, $R$ is continuous, suggesting that no Davies point is observed in that region. For negative values of $\beta$, there are two white patches; however, only the upper patch, where the temperature is positive, should be considered. Points within this region mark the critical points where $R$ and specific heat diverge.

\begin{figure}[h!]
\centering
\begin{subfigure}{0.45\textwidth}
		\includegraphics[height=7cm,width=8cm]{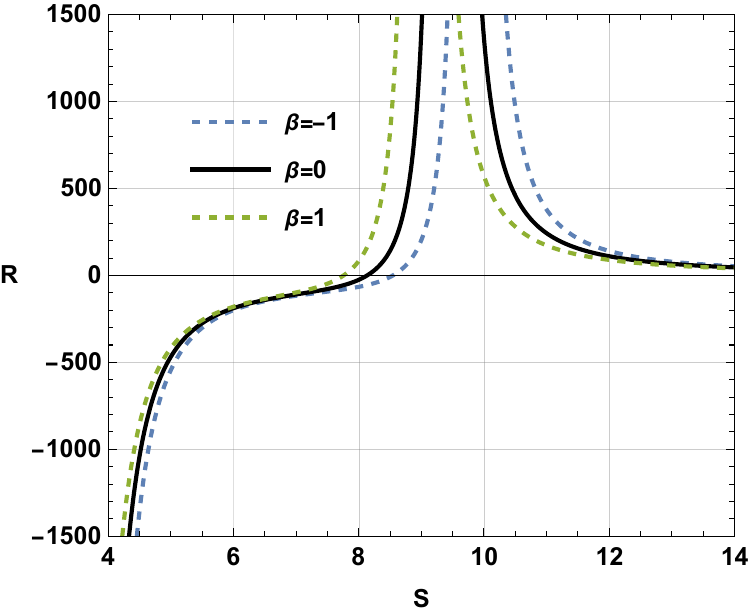}
		\caption{}
		\label{g2a}
	\end{subfigure}\hspace{0.9cm} 
\begin{subfigure}{0.45\textwidth}
		\includegraphics[height=7.3cm,width=8cm]{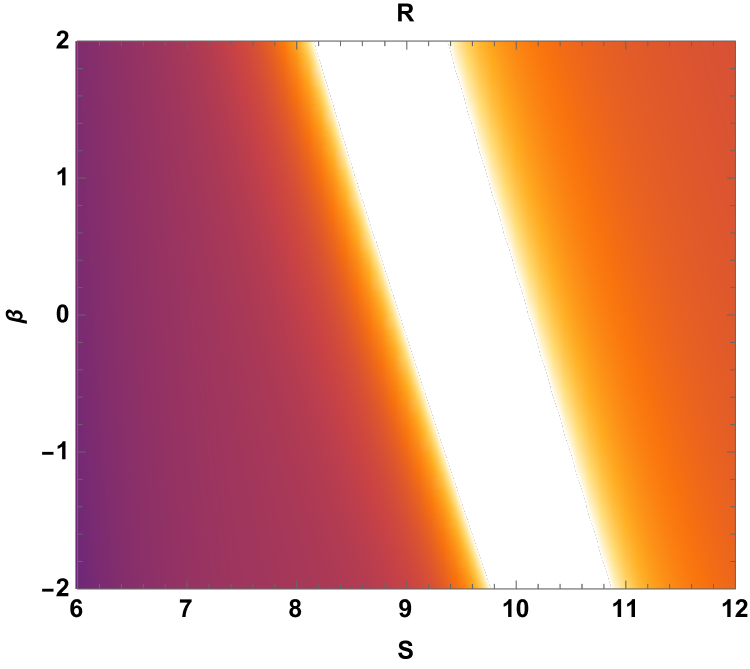}
		\caption{}
		\label{g2b}
	\end{subfigure} 
	\caption{Behaviour of GTD scalar  for black hole surrounded with radiation field($\omega=1/3$)}
	\label{g2}
\end{figure}

The GTD scalar for $\omega = \frac{1}{3}$ is plotted as a function of entropy $S$ in FIG.~\ref{g2a}. We observe that, for every value of $\beta$, the GTD scalar exhibits a curvature singularity. The location of this singularity varies with different values of the model parameter $\beta$.
FIG.~\ref{g2b} presents a density plot of the scalar curvature $R$ as a function of entropy $S$ and the model parameter $\beta$. The white patches in the figure indicate regions where $R$ is not defined. This plot reveals that, for all values of the model parameter, $R$ has a singular point.

\begin{figure}[h!]
\centering
\begin{subfigure}{0.45\textwidth}
		\includegraphics[height=7cm,width=8cm]{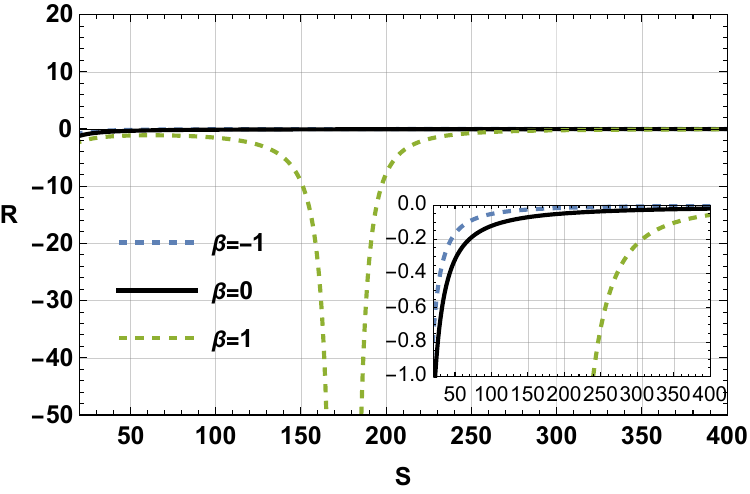}
		\caption{}
		\label{g3a}
	\end{subfigure}\hspace{0.9cm} 
\begin{subfigure}{0.45\textwidth}
		\includegraphics[height=7.3cm,width=8cm]{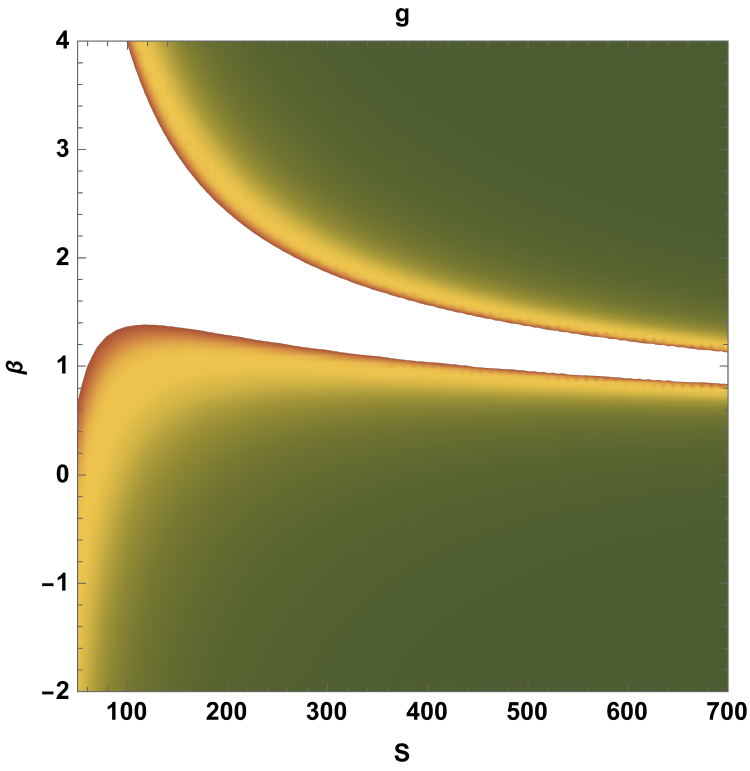}
		\caption{}
		\label{g3b}
	\end{subfigure} 
	\caption{Behaviour of GTD scalar  for black hole surrounded with quintessence field($\omega=-2/3$)}
	\label{g3}
\end{figure}

The GTD scalar for $\omega = -\frac{2}{3}$ is plotted against entropy $S$ in FIG.~\ref{g3a}. For positive values of $\beta$, the GTD scalar displays  curvature singularity, as indicated by the green  dashed curve. Conversely, for $\beta = 0$ and negative values of $\beta$, the curve remains smooth and free from curvature singularities, as shown by the black solid and blue dashed curves, respectively. 
FIG.~\ref{g3b} provides a density plot of the scalar curvature $R$ . The white regions in this plot indicate  that for negative values of the model parameter, $R$ is continuous, suggesting that no Davies point is observed in that region. For positive values of $\beta$, the white region highlights the critical points where $R$ and specific heat diverge.

\begin{figure}[h!]
\centering
\begin{subfigure}{0.45\textwidth}
		\includegraphics[height=7cm,width=8cm]{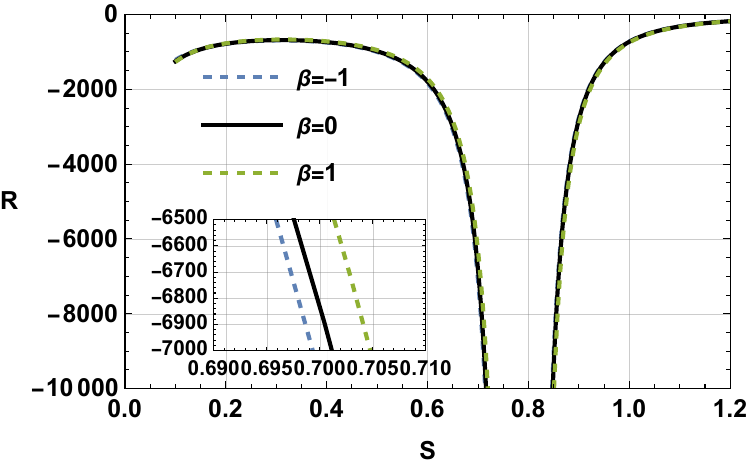}
		\caption{}
		\label{g4a}
	\end{subfigure}\hspace{0.9cm} 
\begin{subfigure}{0.45\textwidth}
		\includegraphics[height=7.3cm,width=8cm]{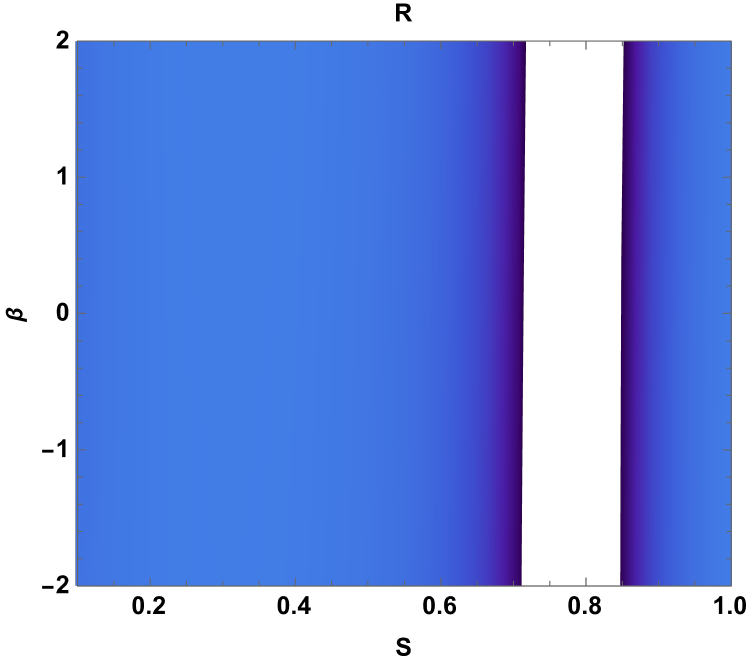}
		\caption{}
		\label{g4b}
	\end{subfigure} 
	\caption{Behaviour of GTD scalar  for black hole surrounded with phantom field($\omega=-4/3)$}
	\label{g4}
\end{figure}

Finally, the plot of $R$ versus $S$ for $\omega = -\frac{4}{3}$ is shown in FIG.~\ref{g4a}. It is evident that, for all values of $\beta$, the GTD scalar consistently exhibits a curvature singularity.
FIG.~\ref{g4b} provides a density plot of the scalar curvature $R$. The white rectangular patches in this figure highlight regions where $R$ is not defined. Notably, these patches suggest that the location of the critical points remains relatively unchanged with varying model parameters. This observation indicates that the critical points are stable across different values of $\beta$.

\subsection{Black hole shadow}
To constrain the $f(R,T)$ models, we utilize data from black hole shadows. This section aims to evaluate and compare the observed angular radius of the Sgr A* black hole, as recently measured by the Event Horizon Telescope (EHT) collaboration, against the shadow radius predicted by theoretical models. By doing so, we seek to constraint the model parameters. We will compute and analyze both the photon sphere and shadow radius, exploring their dependencies on various model parameters. This involves deriving the theoretical expressions for these quantities and assessing how they change with different parameter values. 
This approach enables us to test the $f(R,T)$ model in the context of black hole physics against recent observational results.The photon sphere radius is determined under the assumption of spherical symmetry, following the relation provided below\cite{s0,s04}. 
\begin{equation}
2-\frac{r N'(r)}{N(r)}=0.
\label{eqc1}
\end{equation}
where $N(r)$ is the metric function of the black hole.Solving the equation for $r$, we get the photon radius $r_{ph}$.
Let us take 
\begin{equation}
N(r)=1-\frac{2 M}{r}+K r^{-l}
\end{equation}
where $K$ is a constant, set  to $1$ for simplicity and $l$ is a function of model parameter $\beta$ given as :
\begin{equation}
l=\frac{8 (\beta  \omega +\pi  (6 \omega +2))}{16 \pi -\beta  (\omega -3)}
\label{l}
\end{equation}

\begin{figure}[h!]
\includegraphics[height=6cm,width=8cm]{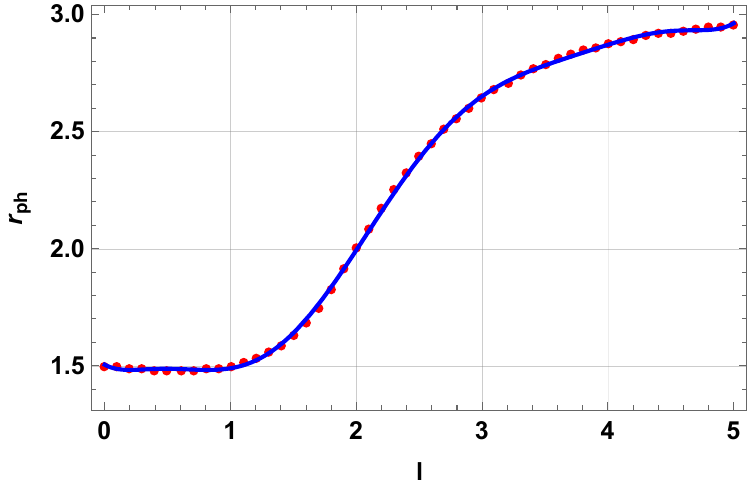}
\caption{$r_{ph}$vs $l$ plot representing numerical data and the fitted curve}.
\label{sh0}
\end{figure}

\begin{figure}[h!]
\includegraphics[scale=0.8]{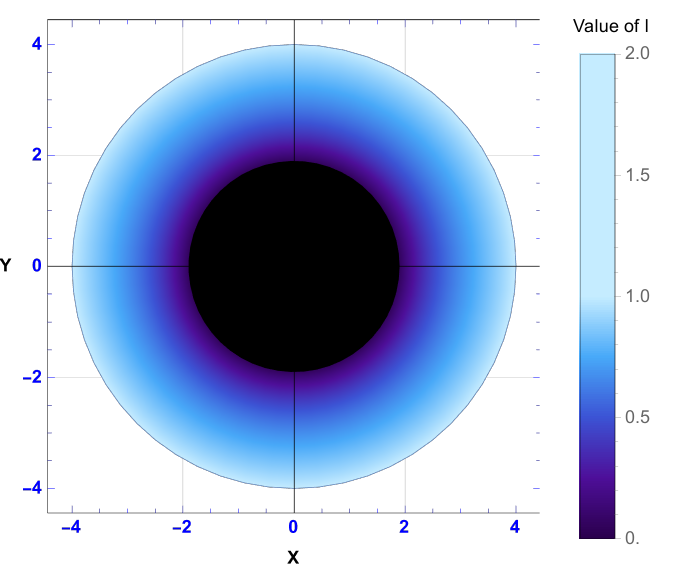}
\caption{Stereoscopic projection of shadow radius in terms of celestial 
coordinates. We have taken} $K=M=1$.
\label{sh1}
\end{figure}

 Solving this equation using traditional algebraic techniques proves to be exceedingly difficult due to its complexity. Consequently, we turn to numerical methods, particularly fitting techniques, to approximate the solution. Numerical fitting techniques involve adjusting a chosen model to the data points obtained from plotting the function, allowing us to find an approximate solution where exact methods fall short. Numerical fitting techniques work by selecting a model and adjusting it to match the data points from the plotted function, helping us find an approximate solution when exact methods aren't feasible. Let us assume $r_{ph}$ can be expressed in terms of $l$ as follows :
$$ r_{ph}=a_0+a_1 \hspace{0.1cm} l +a_2\hspace{0.1cm}  l^2+......a_n  \hspace{0.1cm}  l^n$$
where the coefficients $a_0,a_1.....$ are unknown constants to be determined. The goal of numerical fitting is to obtain the values of these coefficients by adjusting the polynomial to best match the behavior of the exact solution. In FIG.\ref{sh0}, the plot illustrates the relationship between  $r_ph$ and $l$ where the blue line represents the exact solution, and the red dots indicate the numerically computed points based on data.By fitting the curve, we obtain the expression  $r_{ph}$ that accurately approximates the solution in terms of the parameter $l$ upto 8th order.
\begin{equation}
r_{ph}=1.50478 - 0.294205 \hspace{0.1cm}l + 1.35878\hspace{0.1cm} l^2 - 2.67671 \hspace{0.1cm}l^3 + 2.50238 \hspace{0.1cm} l^4 - 1.15041\hspace{0.1cm} l^5 + 0.277115 \hspace{0.1cm} l^6 - 0.0338078 \hspace{0.1cm} l^7 + 0.00165349 \hspace{0.1cm}l^8
\label{eqc2}
\end{equation}
we have taken $K=M=1$.
From the photon radius, we can derive the shadow radius as follows:
\begin{equation}
r_{sh}==\frac{r_{ph}}{\sqrt{N[r_{ph}]}}=\frac{\mathcal{A}}{\mathcal{B}}
\end{equation}
\begin{equation}
\mathcal{A}=\left(-0.0830538 l^6+0.439648 l^5-0.804952 l^4+0.691513 l^3-0.210153 l^2-0.03199 l+1.499\right)
\end{equation}
\begin{multline}
\mathcal{B}^2= \left(-0.0830538 l^6+0.439648 l^5-0.804952 l^4+0.691513 l^3-0.210153 l^2-0.03199 l+1.499^{-l}\right)\\-\left(\frac{24.0808}{-1. l^6+5.29354 l^5-9.69193 l^4+8.32608 l^3-2.53032 l^2-0.385172 l+18.0485}+1 \right)
\end{multline}
Now, for the 2-D 
stereoscopic projection of shadow radius, we define celestial coordinates $X$ 
and $Y$ as given by \cite{s0,s04}
\begin{align}
 X & =\lim_{r_{0}\rightarrow\infty}\left(-\,r_{0}^{2}\sin\theta_{0}\left.\frac{d\phi}{dr}\right|_{r_{0}}\right),\\[5pt]
Y & =\lim_{r_{0}\rightarrow\infty}\left(r_{0}^{2}\left.\frac{d\theta}{dr}\right|_{(r_{0},\theta_{0})}\right).
\end{align}
In this context, $\theta_0$ denotes the angular position of the observer relative to the plane of the black hole. The terms $\frac{d\phi}{dr}$ and $\frac{d\theta}{dr}$ are derived by solving the geodesic equations, which, though straightforward, involve lengthy calculations \cite{s14,s16,s17}. To maintain focus on the primary goal of constraining the model parameter, using the shadow  radius expression we omit these intermediate steps.  Figure \ref{sh1} illustrates the variation of the black hole's shadow radius as a function of the model parameter $l$. The plot reveals a clear trend: as $l$ increases, the shadow radius grows accordingly. The dark region in the plot corresponds to areas where the shadow radius is effectively zero. Conversely, lighter shades of blue represent increasing values of $l$, highlighting the gradual enlargement of the shadow radius. This behavior is evident in the visualization provided in Figure \ref{sh1}.

\begin{figure}[h!]
\includegraphics[height=8cm,width=10cm]{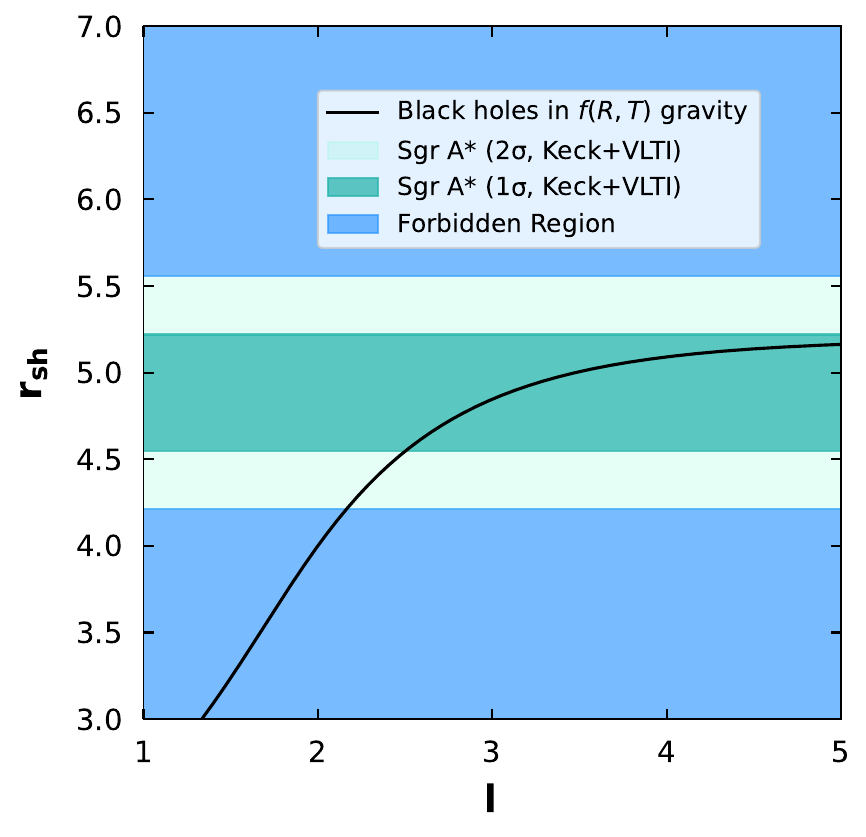}
\vspace{-0.2cm}
\caption{Shadow radius versus parameter $m$ and $c_{2}$ have been plotted in 
the background of Keck and VLTI constrains \cite{s0} from observations of 
Sgr A*. We have chosen $M=1$ and $K=1$ for these plots. The blue portion 
represents the zone forbidden by Keck-VLTI observation.}
\label{sh2}
\end{figure}

\begin{figure}[h!]
\centering
\begin{subfigure}{0.45\textwidth}
		\includegraphics[height=7.3cm,width=9cm]{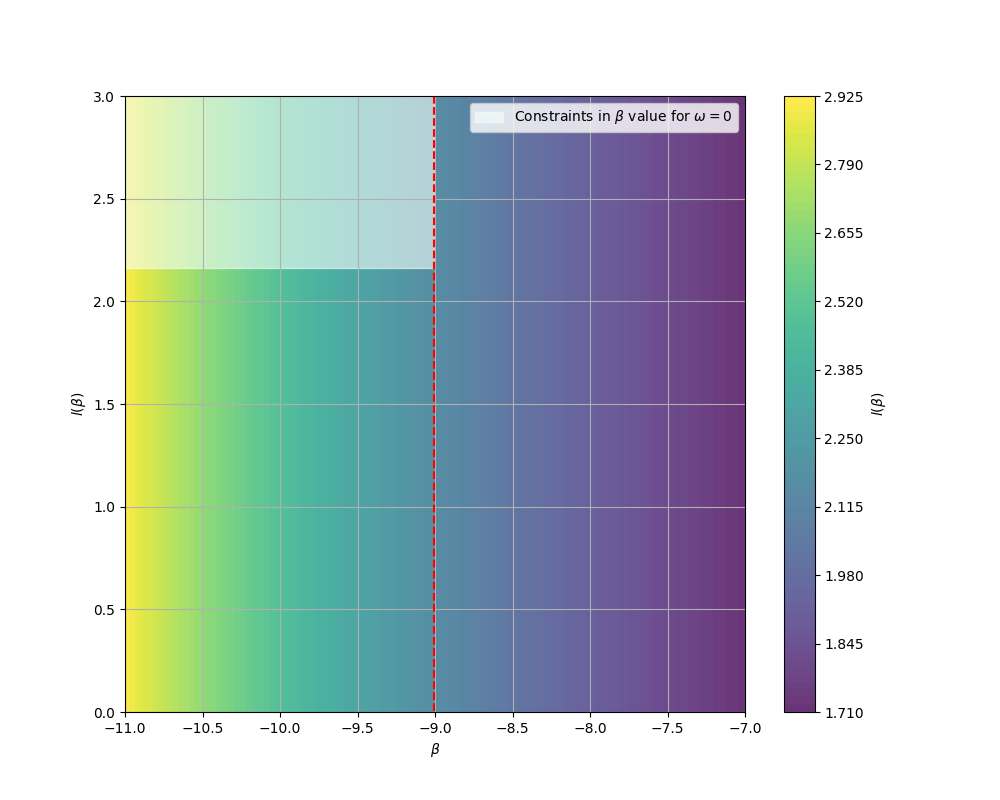}
		\caption{$\omega=0$}
		\label{sh3a}
	\end{subfigure} 
\begin{subfigure}{0.45\textwidth}
		\includegraphics[height=7.3cm,width=9cm]{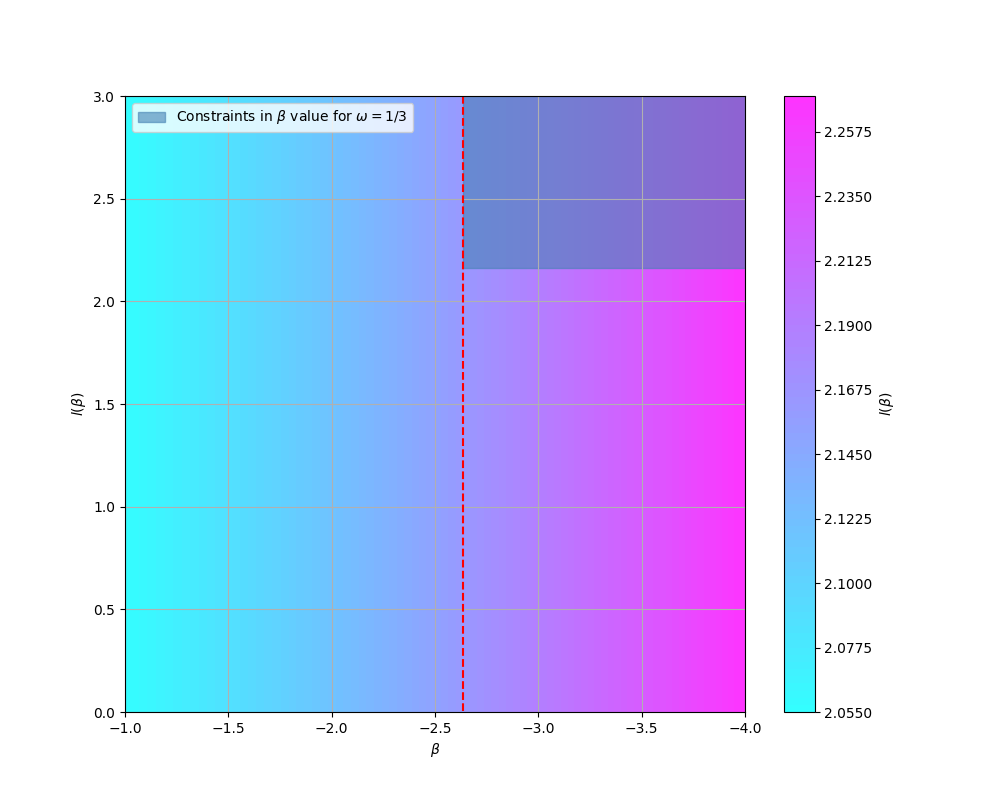}
		\caption{$\omega=1/3$}
		\label{sh3b}
	\end{subfigure} 
\begin{subfigure}{0.45\textwidth}
		\includegraphics[height=7.3cm,width=9cm]{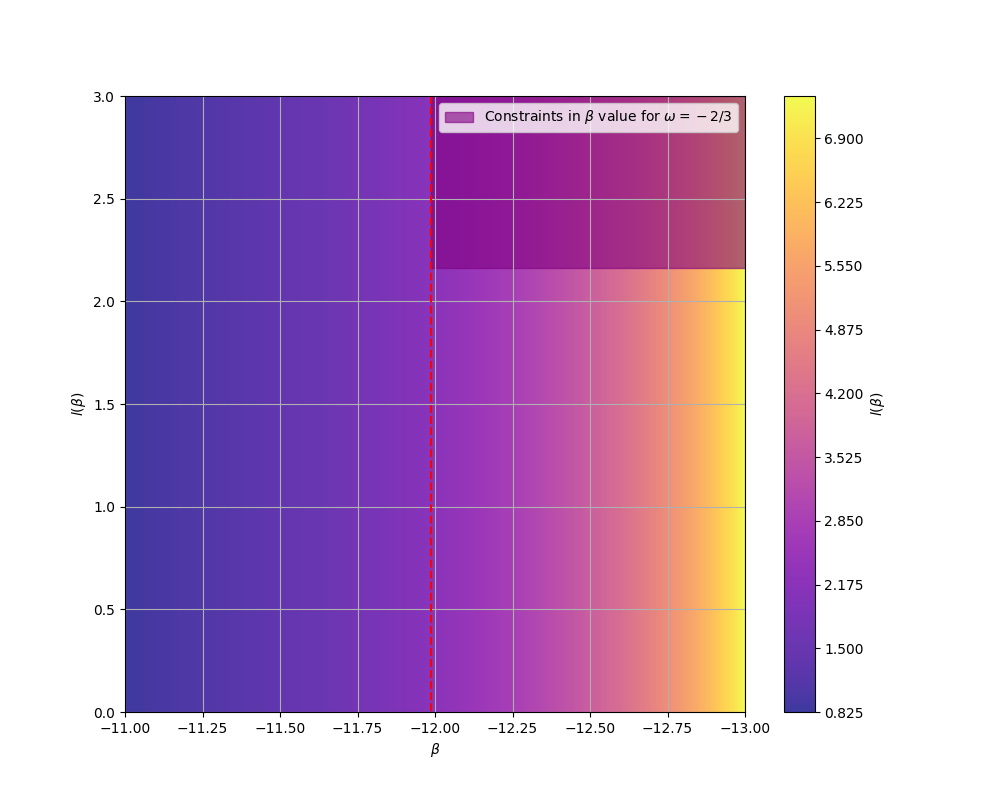}
		\caption{$\omega=-2/3$}
		\label{sh3c}
	\end{subfigure} 
\begin{subfigure}{0.45\textwidth}
		\includegraphics[height=7.3cm,width=9cm]{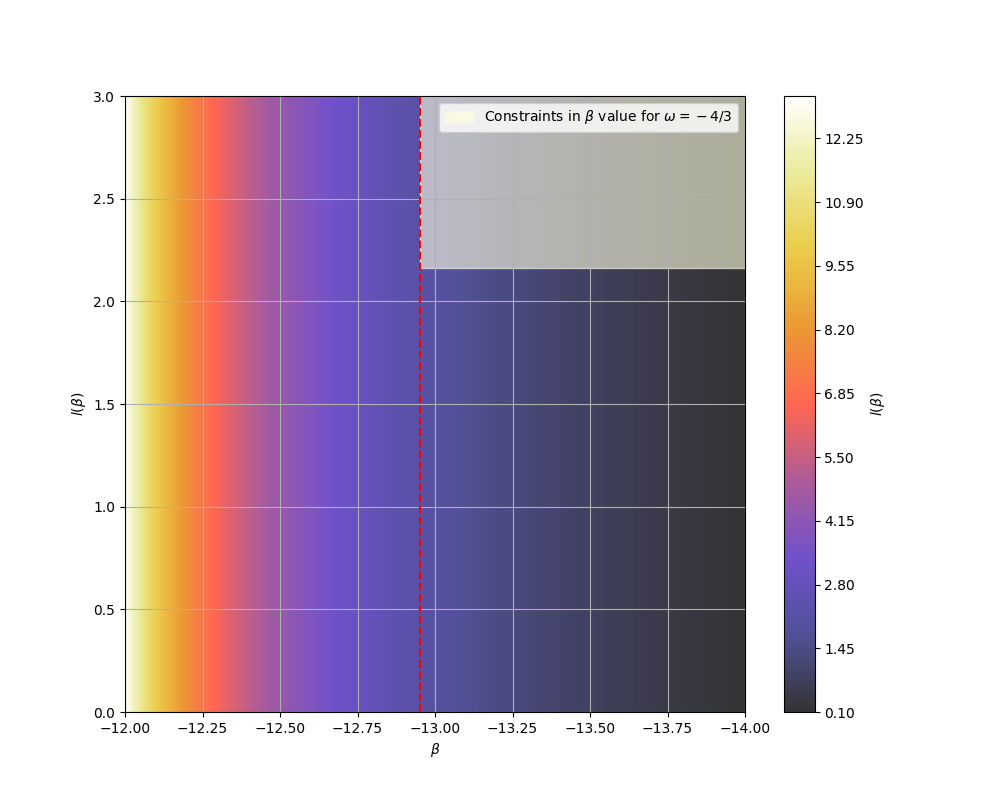}
		\caption{$\omega=-4/3$}
		\label{sh3d}
	\end{subfigure} 
	\caption{Constraints on the model parameter $\beta$.The highlighted regions in each figure indicate the values of $\beta$	 that are consistent with the observational data.} 
	\label{sh3}
\end{figure}

In order to constrain the model parameters, we employ the method outlined in \cite{s0}, briefly summarizing key steps here. This approach requires the mass-to-distance ratio for Sgr A* and a calibration factor correlating the observed shadow radius with the theoretical one. The Event Horizon Telescope (EHT) team introduced a parameter, $\delta$, which quantifies the fractional deviation between the observed shadow radius ($r_s$) and the shadow radius of a Schwarzschild black hole ($r_{sch}$) as follows \cite{s0}:
\begin{equation}
\delta = \frac{r_s}{r_{sch}} - 1 = \frac{r_s}{3\sqrt{3}M} - 1.
\label{eq1}
\end{equation}
From observations by the Keck and VLTI instruments, the estimated values for $\delta$ are given as:
\begin{align*}
\text{Keck}: \delta = -0.04^{+0.09}_{-0.10}, \\
\text{VLTI}: \delta = -0.08^{+0.09}_{-0.09}.
\end{align*}
For simplicity, as done in \cite{s0}, we take the average of these observations:
\begin{equation}
\delta = -0.060 \pm 0.065.
\label{eq2}
\end{equation}
This gives the following confidence intervals for $\delta$:
\begin{align}
-0.125 \lesssim \delta \lesssim 0.005 &\quad (1\sigma),
\label{eq3}\\
-0.190 \lesssim \delta \lesssim 0.070 &\quad (2\sigma).
\label{eq4}
\end{align}
These bounds on $\delta$, when applied to the equation for $\delta$, provide constraints on the shadow radius $r_{sh}$, leading to the following results \cite{s0}:
\begin{align}
4.55 \lesssim \frac{r_{sh}}{M} \lesssim 5.22 &\quad (1\sigma),
\label{eq5}\\
4.21 \lesssim \frac{r_{sh}}{M} \lesssim 5.56 &\quad (2\sigma).
\label{eq6}
\end{align}
The figure in Fig.~\ref{sh2} illustrates how the black hole shadow radius, $r_{sh}$, varies with the parameter $l$, constrained by Keck and VLTI observations of Sgr A*. The solid black curve represents the shadow radius for black holes in $f(R,T)$ gravity as a function of $l$. The blue-shaded region corresponds to the forbidden zone, excluded based on Keck-VLTI observational constraints, while the green and cyan bands denote the $1\sigma$ and $2\sigma$ observational limits for Sgr A*, respectively. In this context, $\sigma$ represents the standard deviation, which quantifies the uncertainty or spread of the measurement. The plot reveals that for lower values of $l$, the shadow radius rapidly approaches or falls within the forbidden region, rendering those values of $l$ inconsistent with observations. In contrast, for higher values of $l$, the shadow radius lies comfortably within the observational limits, satisfying the constraints from Keck and VLTI. From this figure, constraints on $l$ can be extracted, which further allow us to impose restrictions on the model parameter $\beta$.\\
The parameter $l$ is related to the model parameter $\beta$ through Eq.\ \ref{l}. Using the constraints on $l$ discussed earlier, we derive the corresponding constraints on $\beta$ by analyzing the density profile of Eq.\ \ref{l} for four specific cases, each with a standard value of the parameter $\omega$. These density plots provide a clear visualization of the allowed ranges for $\beta$, with the highlighted regions indicating values that are consistent with observational data, as shown in FIG.\ \ref{sh3}. For instance, FIG.\ \ref{sh3a} illustrates that for Model I ($f(R,T) = f_1(R) + f_2(T)$), the constraints on the parameter $\beta$ vary considerably with different values of $\omega$. When $\omega = 0$, $\beta$ is constrained to the range $-13.7042 \leq \beta \leq -9.00669$, reflecting a relatively narrow interval. As $\omega$ increases to $1/3$, the range of $\beta$ shifts and broadens, with $-14.6531 \leq \beta \leq -2.63325$, providing more flexibility in the parameter space. For negative values of $\omega$, such as $\omega = -2/3$, the range tightens again, with $-12.8117 \leq \beta \leq -11.986$, indicating a smaller allowable interval. Similarly, for $\omega = -4/3$, the constraints become even narrower, with $-12.9505 \leq \beta \leq -12.3851$. This analysis highlights the sensitivity of $\beta$ to changes in $\omega$, where tighter bounds are observed for negative values of $\omega$ compared to $\omega = 1/3$. The density plots not only provide a clear picture of these constraints but also help identify the regions of parameter space where $\beta$ aligns with observational data, serving as a valuable tool for narrowing down the possible range of $\beta$ in the model to ensure consistency with observed measurements.
\section{Model II : $f(R,T)=f_1(R)+f_2(R)f_3(T)$}
	We consider $f_1(R)=\alpha R$,  $f_2(R)=\beta R$, and $f_3(T)= \gamma T$ where $\alpha$  $\beta$ and $\gamma$ are the model parameters. The field equations using this model can be calculated as :
	$$ (\alpha+\beta \gamma T)  R_{\mu  \nu }-\frac{1}{2}  g_{\mu  \nu }( \alpha  R+\beta \gamma R T) =8 \pi  T_{\mu  \nu } - \beta \gamma R T _{\mu \nu}- \beta \gamma R \Theta _{\mu \nu}$$
	$$(\alpha+\beta \gamma T)  G^\mu{}_ \nu=8 \pi  T^\mu{}_ \nu - \beta \gamma R T^\mu{}_ \nu+\beta R \gamma (2 T^\mu{}_ \nu+\frac {1}{3}(p_r+2 p_t)$$	
	and finally the first field equation is obtained as
	\begin{equation}
	 G^t{}_t= G^r{}_r=\frac{1}{\alpha+\beta R T}\left(8 \pi  \rho + \beta  R \rho  \gamma (\omega+1) \right)
	 \label{fe3}
	\end{equation}
	\begin{equation}
	 G^{\theta }{}_{\theta }=\frac{1}{\alpha+\beta R T}\left(-4 \pi \rho (3 \omega +1)-\frac{\beta  R \rho  \gamma}{2}(\omega+1)\right)
	 \label{fe4}
	\end{equation}

	Following the same steps as we have done previously and after few simplifications we reached the following equation
	\begin{equation}
  -\frac{r^2 B''(r)+3 r B'(r)+B(r)-1}{r^2 B''(r)+r B'(r)-B(r)+1}=-\frac{12 \pi  \omega }{\beta  \gamma  R (\omega +1)+4 \pi  (3 \omega +2)}
  \label{sol}
\end{equation}
For spherically symmetric ansatz, the expression for Ricci scalar is
	$$R=\frac{2-N(r)-4 r N'(r)-r^2 N''(r)}{r^2}$$
Substituting the value of $R$ results in a highly complicated differential equation that is difficult to solve. To simplify the process, we will treat $R=R_0$ as a constant, making the differential equation easier to solve.Thus solving eq.(\ref{sol}), we get the Lapse function of the black hole solution as :
\begin{multline}
N(r)=1+c_1 r^{\frac{4 \pi  \left(3 \sqrt{\kappa  \omega +\kappa +8 \pi } \sqrt{\kappa  \omega +\kappa +24 \pi  \omega +8 \pi } \sqrt{\frac{\omega ^2}{\kappa ^2 (\omega +1)^2+8 \pi  \kappa  \left(3 \omega ^2+5 \omega +2\right)+64 \pi ^2 (3 \omega +1)}}-3 \omega -2\right)-\kappa  (\omega +1)}{\kappa  \omega +\kappa +8 \pi }}\\+c_2 r^{-\frac{4 \pi  \left(3 \sqrt{\kappa  \omega +\kappa +8 \pi } \sqrt{\kappa  \omega +\kappa +24 \pi  \omega +8 \pi } \sqrt{\frac{\omega ^2}{\kappa ^2 (\omega +1)^2+8 \pi  \kappa  \left(3 \omega ^2+5 \omega +2\right)+64 \pi ^2 (3 \omega +1)}}+3 \omega +2\right)+\kappa  (\omega +1)}{\kappa  \omega +\kappa +8 \pi }}
\end{multline}
where $\kappa=R_0 \beta \gamma$ is the model parameter.If we substitute $\omega=1/3$ and $\kappa=0$, we get 
\begin{equation}
N(r)=1+\frac{c_1}{r}+\frac{c_2}{r^2}
\end{equation}
comparing with the standard RN black hole, we get $c_1=-2M$ and $c_2$ is the effective charge. We have set $c_2=1$.Again by setting $\omega = 0$ and the model parameter $\kappa = 0$, we recover the simple Schwarzschild black hole solution with mass $c_1+c_2$.Moreover when the model parameter $\kappa$ is set to zero, then the solution reduces to Kiselev black hole in GR i.e
$$N(r)=1-\frac{2M}{r}+c_2 r^{-(3 \omega+1)}$$
Now, substituting the value of $N(r)$ in the field equation eq.(\ref{fe3}), we obtain the value of the energy density $\rho$
\begin{equation}
\rho(r)=\frac{24 \pi  \alpha  R \omega }{\kappa ^2 R (\omega +1)^2 r^{\frac{3 (\kappa +8 \pi ) (\omega +1)}{\kappa  \omega +\kappa +8 \pi }}+64 \pi ^2 R r^{\frac{3 (\kappa +8 \pi ) (\omega +1)}{\kappa  \omega +\kappa +8 \pi }}+8 \pi  \kappa  \left(2 R (\omega +1) r^{\frac{3 (\kappa +8 \pi ) (\omega +1)}{\kappa  \omega +\kappa +8 \pi }}+3 \omega  (3 \omega -1)\right)}
\label{rho}
\end{equation}
\begin{figure}[h!]	
		\centering
		\begin{subfigure}{0.45\textwidth}
			\includegraphics[width=\linewidth]{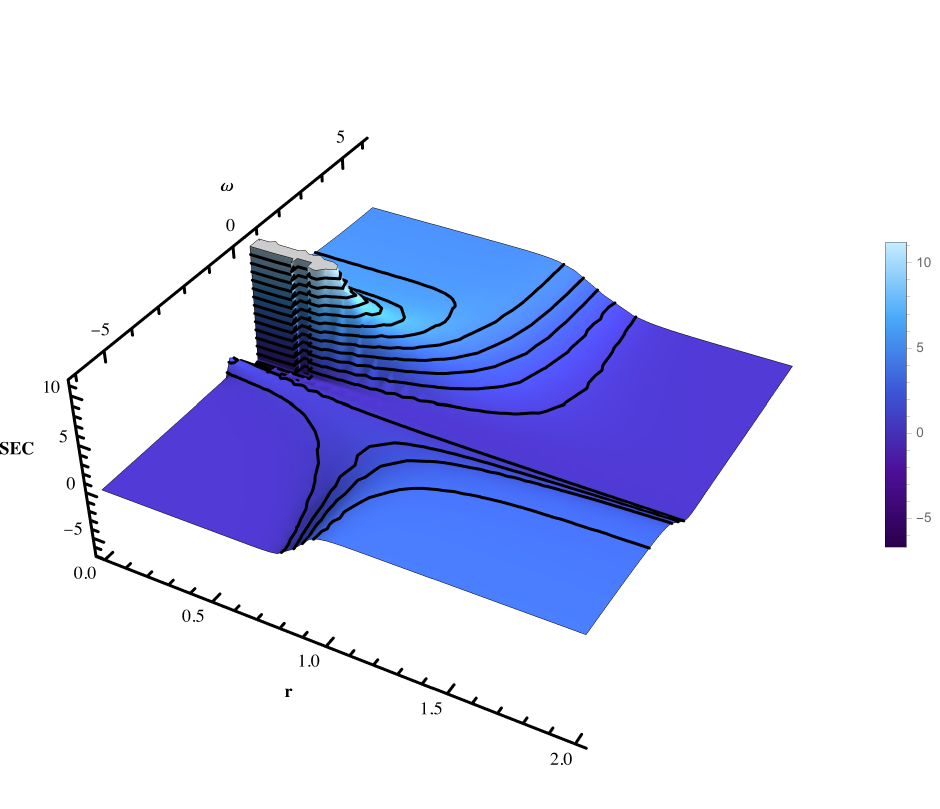}
			\caption{$\alpha<0$}
			\label{seca}
		\end{subfigure}
		\begin{subfigure}{0.45\textwidth}
			\includegraphics[width=\linewidth]{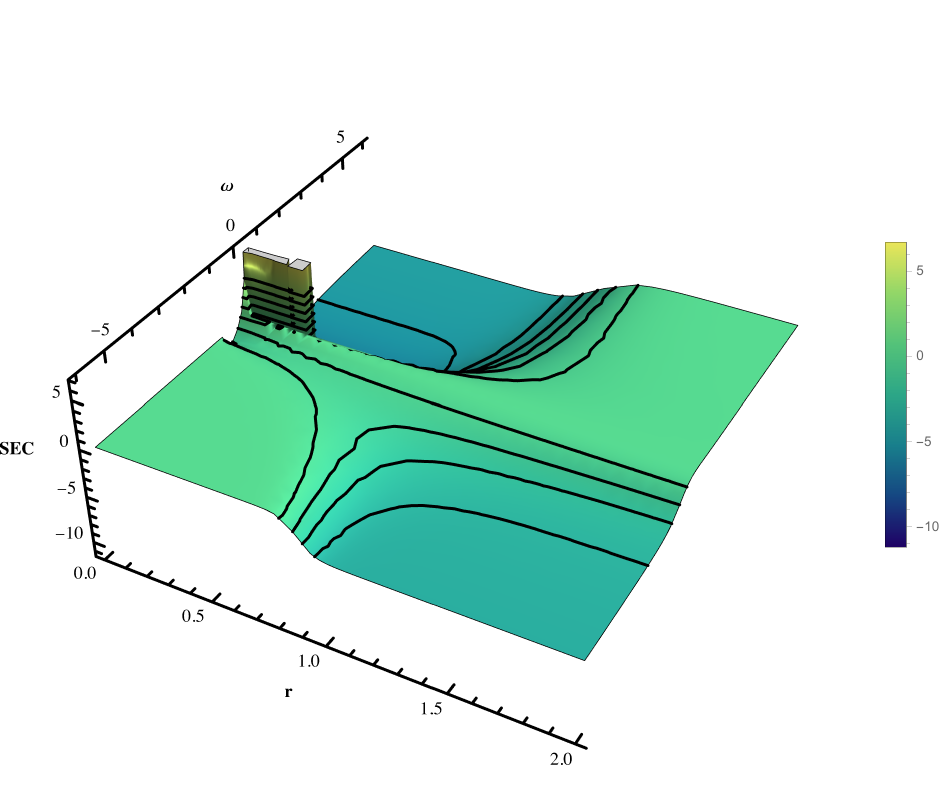}
			\caption{$\alpha>0$}
			\label{secb}
		\end{subfigure}
		\begin{subfigure}{0.45\textwidth}
			\includegraphics[width=\linewidth]{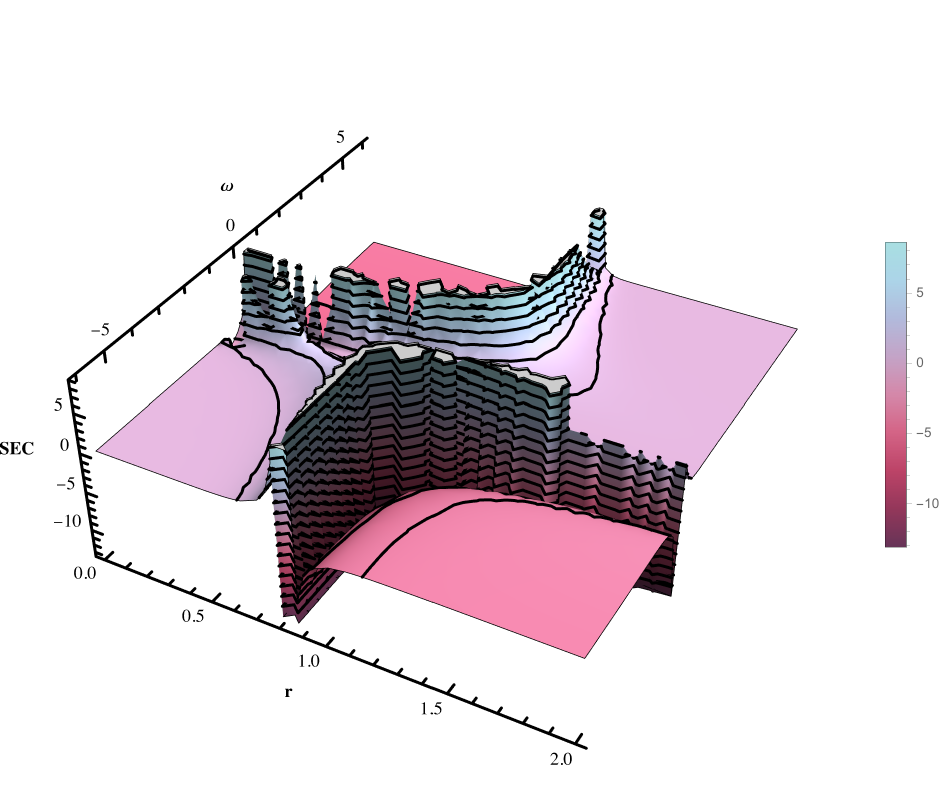}
			\caption{$\kappa<0, R$ and $\alpha >0$}
			\label{secc}
		\end{subfigure}
		\begin{subfigure}{0.45\textwidth}
			\includegraphics[width=\linewidth]{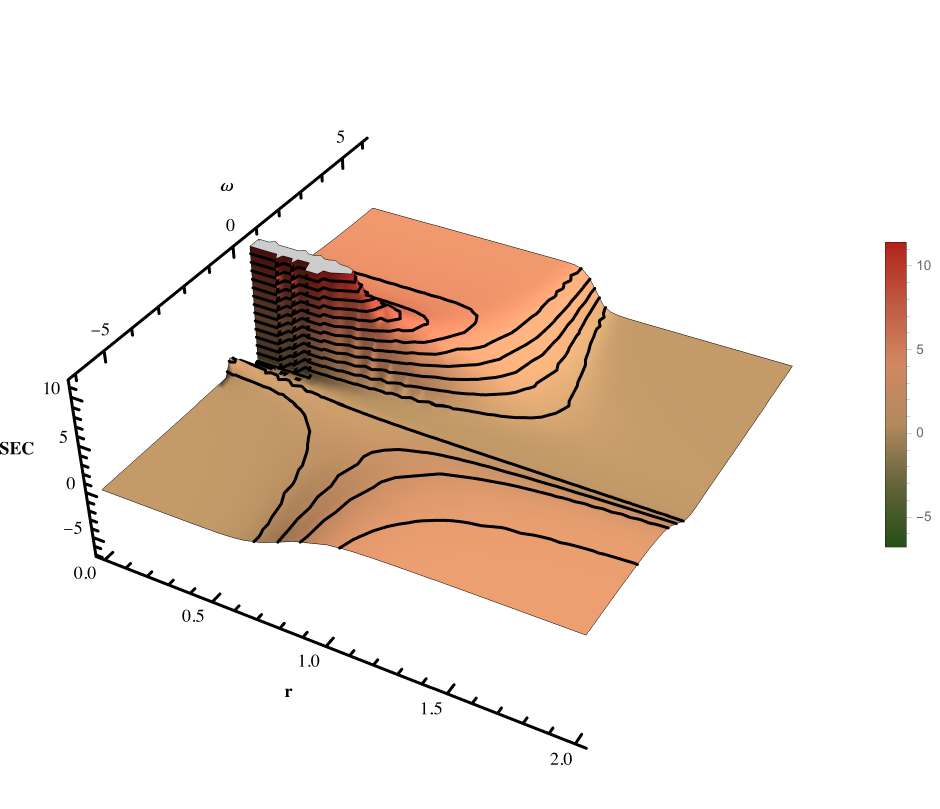}
			\caption{$R, \kappa<0$ and $\alpha >0$}
			\label{secd}
			\end{subfigure}
		\caption{ Visualization of SEC condition eq.(\ref{secfig})}
		\label{SECC}
	\end{figure}
	
For anisotropic fluids, the components of the energy-momentum tensor must fulfill specific criteria to accurately describe a physically realistic matter distribution. It is widely recognized that certain forms of exotic matter do not adhere to particular energy conditions imposed on the energy-momentum tensor. In the context of the strong energy condition (SEC), the relevant conditions for anisotropic fluids are described by a set of equations\cite{SEC}.
\begin{equation}
\text{SEC}:\:\: \rho + p_{n} \geq 0, \:\:\: \rho + \sum_{n} p_{n} \geq 0,
    \label{eq16}
\end{equation}
where $n=1,2,3...$ . In our context, 
\begin{align}
p_r=&-\rho\\
p_t=&\frac{1}{2}(3w+1)\rho
\end{align}
Using these expression we obtain the following SEC conditions
\begin{align}
    \rho+p_r=&0, \label{sec1}\\    
    \rho+p_t=&\frac{36 (\omega+1) \pi  \alpha  R \omega }{\kappa ^2 R (\omega +1)^2 r^{\frac{3 (\kappa +8 \pi ) (\omega +1)}{\kappa  \omega +\kappa +8 \pi }}+64 \pi ^2 R r^{\frac{3 (\kappa +8 \pi ) (\omega +1)}{\kappa  \omega +\kappa +8 \pi }}+8 \pi  \kappa  \left(2 R (\omega +1) r^{\frac{3 (\kappa +8 \pi ) (\omega +1)}{\kappa  \omega +\kappa +8 \pi }}+3 \omega  (3 \omega -1)\right)}\label{sec2}\\
    \rho+p_r+2p_t=&\frac{24 (3\omega+1) \pi  \alpha  R \omega }{\kappa ^2 R (\omega +1)^2 r^{\frac{3 (\kappa +8 \pi ) (\omega +1)}{\kappa  \omega +\kappa +8 \pi }}+64 \pi ^2 R r^{\frac{3 (\kappa +8 \pi ) (\omega +1)}{\kappa  \omega +\kappa +8 \pi }}+8 \pi  \kappa  \left(2 R (\omega +1) r^{\frac{3 (\kappa +8 \pi ) (\omega +1)}{\kappa  \omega +\kappa +8 \pi }}+3 \omega  (3 \omega -1)\right)}
    \label{sec3}
\end{align}
The condition in which the SEC is satisfied for eq.(\ref{sec2}) is
\begin{equation}
\frac{(\omega+1)   \alpha  R \omega }{\kappa ^2 R (\omega +1)^2 r^{\frac{3 (\kappa +8 \pi ) (\omega +1)}{\kappa  \omega +\kappa +8 \pi }}+64 \pi ^2 R r^{\frac{3 (\kappa +8 \pi ) (\omega +1)}{\kappa  \omega +\kappa +8 \pi }}+8 \pi  \kappa  \left(2 R (\omega +1) r^{\frac{3 (\kappa +8 \pi ) (\omega +1)}{\kappa  \omega +\kappa +8 \pi }}+3 \omega  (3 \omega -1)\right)} \geq 0
\label{•}
\end{equation}
    and for eq.(\ref{sec3}) 
    \begin{equation}
    \frac{ (3\omega+1)   \alpha  R \omega }{\kappa ^2 R (\omega +1)^2 r^{\frac{3 (\kappa +8 \pi ) (\omega +1)}{\kappa  \omega +\kappa +8 \pi }}+64 \pi ^2 R r^{\frac{3 (\kappa +8 \pi ) (\omega +1)}{\kappa  \omega +\kappa +8 \pi }}+8 \pi  \kappa  \left(2 R (\omega +1) r^{\frac{3 (\kappa +8 \pi ) (\omega +1)}{\kappa  \omega +\kappa +8 \pi }}+3 \omega  (3 \omega -1)\right)}  \geq 0
    \label{secfig}
    \end{equation}
It is noteworthy that although the strong energy condition (SEC) depends on the model parameter $\alpha$, the black hole solution itself remains independent of this parameter. FIG \ref{SECC} illustrates the relationship between the SEC and the parameters $\omega$, $\kappa$, $R$, and $\alpha$, based on the condition described in Eq. (\ref{secfig}). Regions where the SEC is violated correspond to negative values of the SEC axis.
In FIG \ref{seca}, we plot the left-hand side (LHS) of Eq. (\ref{secfig}) for positive values of $\alpha$, while Figure \ref{secb} shows the same for negative values of $\alpha$. It is important to observe that negative values of $\alpha$ reflect on the SEC axis compared to positive $\alpha$.
FIG \ref{secb} further demonstrates the behavior of the SEC when $\kappa$ is negative, while $R$ and $\alpha$ remain positive. Meanwhile, FIG \ref{secd} shows the scenario where both $R$ and $\alpha$ are negative, with $\kappa$ positive. \\
	\begin{figure}[h!]	
		\centering
		\begin{subfigure}{0.3\textwidth}
			\includegraphics[width=\linewidth]{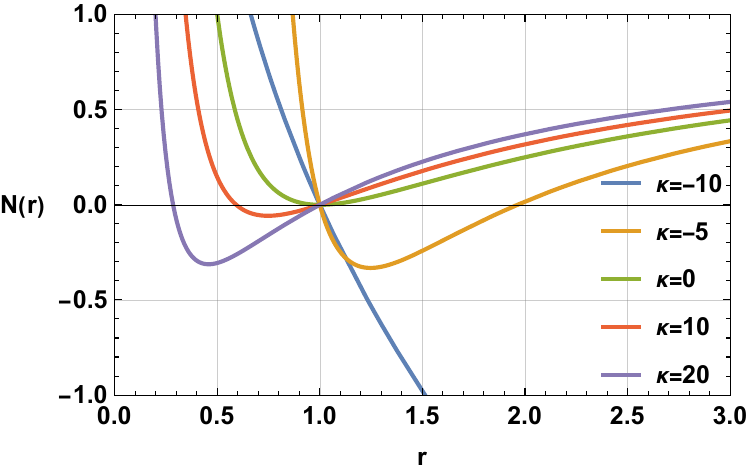}
			\caption{$\omega=\frac{1}{3},M=1$}
			\label{ha}
		\end{subfigure}
		\begin{subfigure}{0.3\textwidth}
			\includegraphics[width=\linewidth]{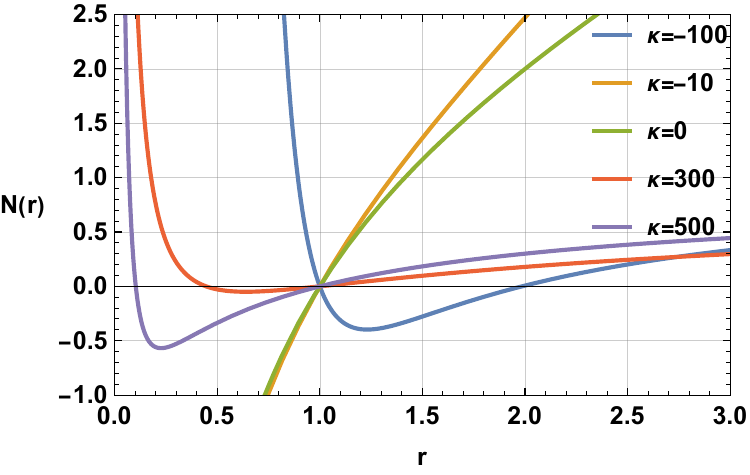}
			\caption{$\omega=-\frac{2}{3},M=1$}
			\label{hb}
		\end{subfigure}
		\begin{subfigure}{0.3\textwidth}
			\includegraphics[width=\linewidth]{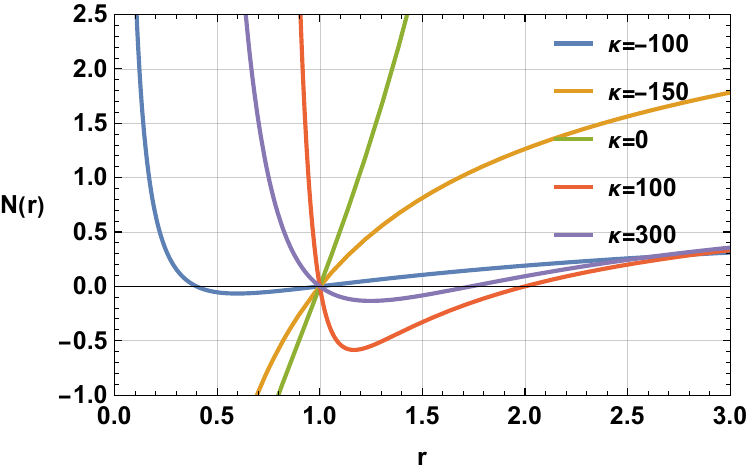}
			\caption{$\omega=-\frac{4}{3},M=1$}
			\label{hc}
		\end{subfigure}
		\caption{ Horizon structure of the black hole solution obtained in model II}
		\label{horizon}
	\end{figure}
	Next, we turn our attention to the horizon structure of the black hole. The horizon locations are determined by solving the equation $N(r) = 0$ for $r$, which yields the values where the metric function vanishes. For the case $\omega = 1/3$, it is observed that a black hole solution always exists for all values of $\kappa$. The number of horizons depends on the value of $\kappa$: for all positive values of $\kappa$, both the Cauchy and event horizons are present, whereas for very small negative values of $\kappa$, only the event horizon is found.
When considering $\omega = -2/3$ and $\omega = -4/3$, the black hole exhibits a more intricate horizon structure. For a specific range of $\kappa$, a degenerate horizon appears, indicating a transition where two horizons coincide. Outside this range, the black hole possesses two distinct horizons. Importantly, in all cases explored, a black hole solution persists for all values of $\kappa$, as an event horizon always exists, ensuring the presence of a black hole.
%%%%%%%%%%%%%%%%%%%%%%%%%%%%%%%%%%%%%%%%%%

\subsection{Thermodynamical properties}
	The mass $M$ of the  black hole in this model is found to be ,
	\begin{equation}
		M=\frac{1}{2} r^{-\frac{24 \pi  \sqrt{\frac{\omega ^2}{(\kappa  \omega +\kappa +8 \pi ) (\kappa  \omega +\kappa +24 \pi  \omega +8 \pi )}} \sqrt{\kappa  \omega +\kappa +24 \pi  \omega +8 \pi }}{\sqrt{\kappa  \omega +\kappa +8 \pi }}} \left(r^{\frac{4 \pi  \left(3 \sqrt{\kappa  \omega +\kappa +8 \pi } \sqrt{\kappa  \omega +\kappa +24 \pi  \omega +8 \pi } \sqrt{\frac{\omega ^2}{(\kappa  \omega +\kappa +8 \pi ) (\kappa  \omega +\kappa +24 \pi  \omega +8 \pi )}}+3 \omega +2\right)+\kappa  \omega +\kappa }{\kappa  \omega +\kappa +8 \pi }}+1\right)
		\label{mass2}
	\end{equation}
	\begin{figure}[t!]	
		\centering
		\begin{subfigure}{0.30\textwidth}
			\includegraphics[width=\linewidth]{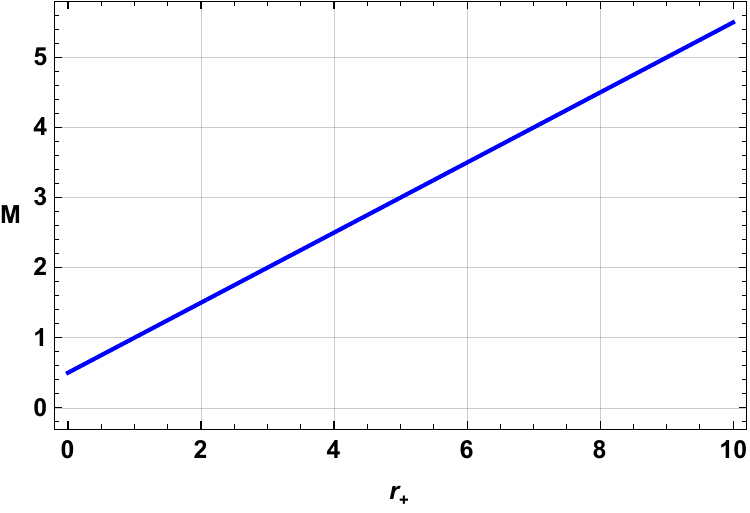}
			\caption{$\omega=0$}
			\label{21a}
		\end{subfigure}
		\begin{subfigure}{0.30\textwidth}
			\includegraphics[width=\linewidth]{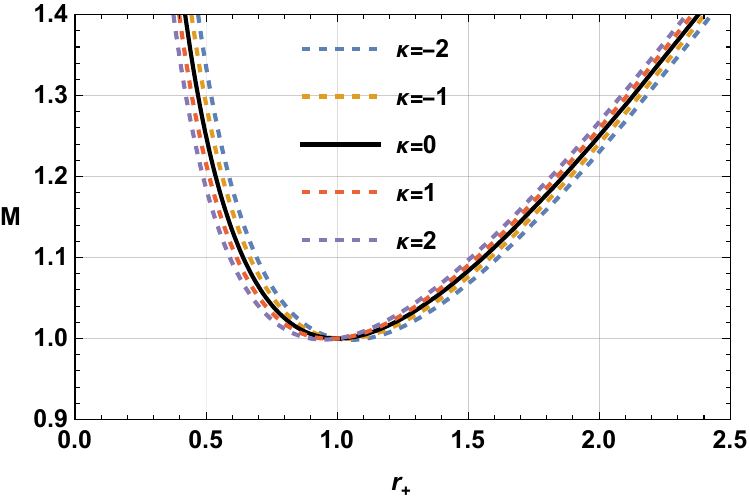}
			\caption{$\omega=\frac{1}{3}$}
			\label{21b}
		\end{subfigure}
		\begin{subfigure}{0.30\textwidth}
			\includegraphics[width=\linewidth]{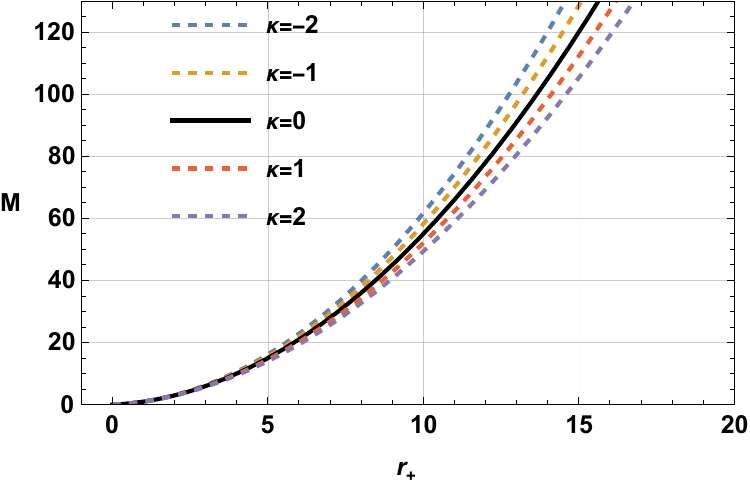}
			\caption{$\omega=-\frac{2}{3}$}
			\label{21c}
		\end{subfigure}
		\begin{subfigure}{0.30\textwidth}
			\includegraphics[width=\linewidth]{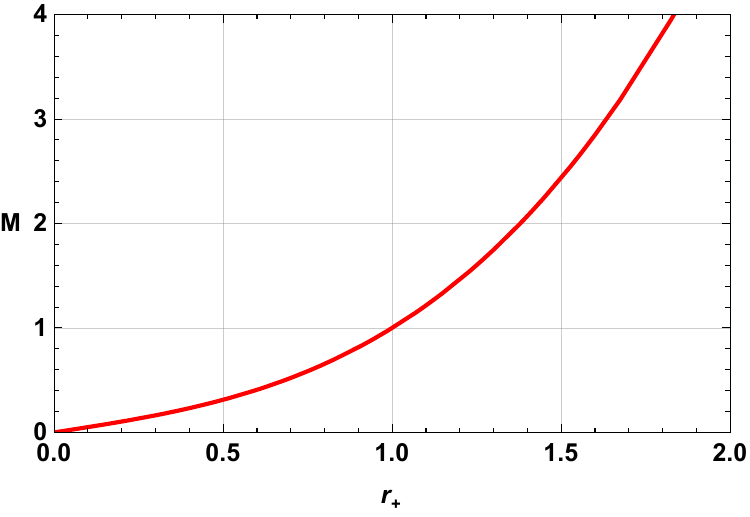}
			\caption{$\omega=-1$}
			\label{21d}
			\end{subfigure}
		\begin{subfigure}{0.30\textwidth}
		\includegraphics[width=\linewidth]{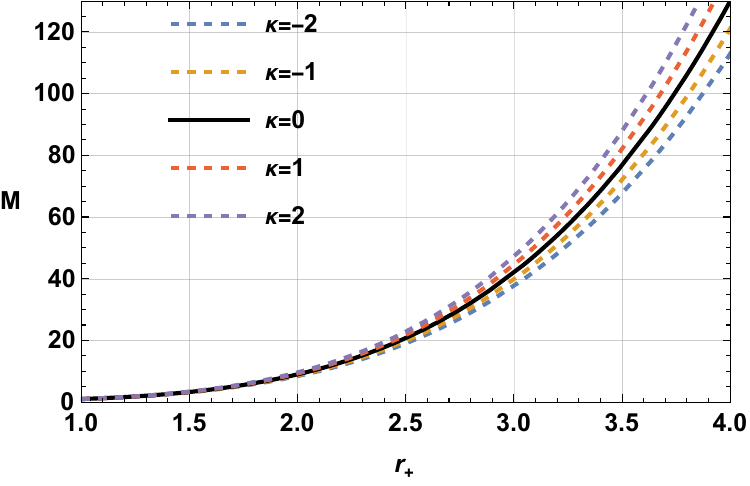}
		\caption{$\omega=-\frac{4}{3}$}
			\label{21f}
		\end{subfigure}
		\caption{ $M$ vs $r_+$ plot for different values of $\omega$. The impact of model  parameter $\kappa$ is shown for a specific value of $\omega$}
		\label{21}
	\end{figure}
	here we have substitute $R_0  \beta \gamma=\kappa$, and we will treat this quantity as the model parameter.
	The expression for  temperature can be evaluated as :
	\begin{equation}
	T=\frac{\alpha}{\beta}
		\label{temp2}
	\end{equation}
	where 
	\begin{multline}
	\alpha=r^{-a} \Bigg( 4\pi \Bigg( r^b \Bigg( -3 \sqrt{\kappa \omega + \kappa + 8\pi} 
\sqrt{ \frac{\omega^2}{ (\kappa \omega + \kappa + 8\pi)( \kappa \omega + \kappa + 24\pi \omega + 8\pi)}} 
\sqrt{\kappa \omega + \kappa + 24\pi \omega + 8\pi} + 3\omega + 2 \Bigg) \\
- 6 \sqrt{\kappa \omega + \kappa + 8\pi} 
\sqrt{ \frac{\omega^2}{ (\kappa \omega + \kappa + 8\pi)( \kappa \omega + \kappa + 24\pi \omega + 8\pi)}} 
\sqrt{\kappa \omega + \kappa + 24\pi \omega + 8\pi} \Bigg) 
+ \kappa (\omega + 1) r^b \Bigg)
	\end{multline}
	and 
	\begin{equation}
	\beta=4 \pi  (\kappa  \omega +\kappa +8 \pi )
	\end{equation}
	here
	\begin{equation}
	a=\frac{2 \left(\pi  \left(6 \sqrt{\kappa  \omega +\kappa +8 \pi } \sqrt{\kappa  \omega +\kappa +24 \pi  \omega +8 \pi } \sqrt{\frac{\omega ^2}{(\kappa  \omega +\kappa +8 \pi ) (\kappa  \omega +\kappa +24 \pi  \omega +8 \pi )}}+6 \omega +8\right)+\kappa  \omega +\kappa \right)}{\kappa  \omega +\kappa +8 \pi }
	\end{equation}
	and 
	\begin{equation}
	b=\frac{4 \pi  \left(3 \sqrt{\kappa  \omega +\kappa +8 \pi } \sqrt{\kappa  \omega +\kappa +24 \pi  \omega +8 \pi } \sqrt{\frac{\omega ^2}{(\kappa  \omega +\kappa +8 \pi ) (\kappa  \omega +\kappa +24 \pi  \omega +8 \pi )}}+3 \omega +2\right)+\kappa  \omega +\kappa }{\kappa  \omega +\kappa +8 \pi }
	\end{equation}
	from the expression of mass and temperature we can calculate entropy expression using :
	\begin{equation}
	S=\int \frac{d M}{T}
	\end{equation}
	where we will observe dependency of entropy on model parameter and $\omega.$
	The effect of the model  parameter $\kappa$  on the black hole mass for a specific value of $\omega$ is shown in FIG.\ref{21}.For $\omega=-1$ and $\omega=0$, the mass becomes independent of the model parameter. 
Similarly, the temperature $T$ is plotted against the horizon radius $r_+$ in Fig. \ref{22}. Except for the $\omega = 0$ case, we observe behavior similar to that seen in Model I. In this case, the temperature becomes independent of the model parameter $\kappa$, although the model still alters the phase structure. The black hole for the $\omega = 0$ case starts to behave like a Schwarzschild black hole. While two phases were observed in Model I, here we find only one black hole phase. The remaining cases follow the same trend as studied in Model I.
		\begin{figure}[h]	
		\centering
		\begin{subfigure}{0.30\textwidth}
			\includegraphics[width=\linewidth]{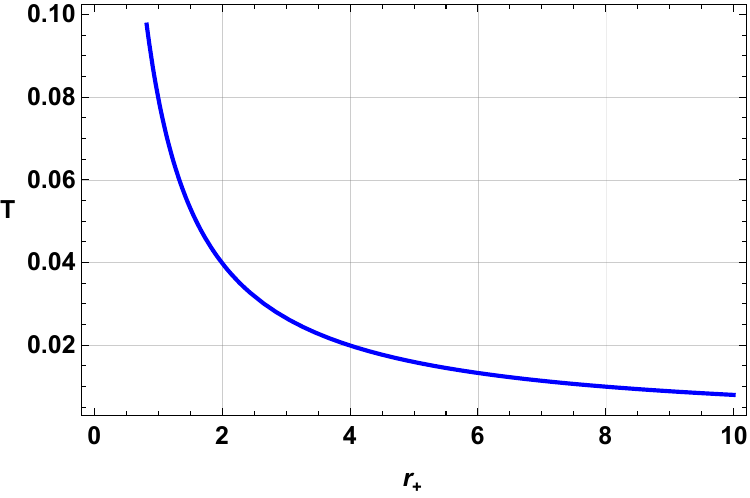}
			\caption{$\omega=0$}
			\label{22a}
		\end{subfigure}
		\begin{subfigure}{0.30\textwidth}
			\includegraphics[width=\linewidth]{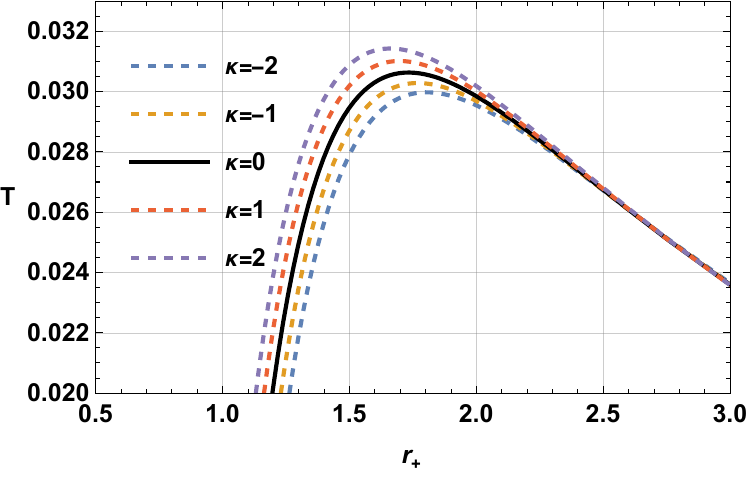}
			\caption{$\omega=\frac{1}{3}$}
			\label{22b}
		\end{subfigure}
		\begin{subfigure}{0.30\textwidth}
			\includegraphics[width=\linewidth]{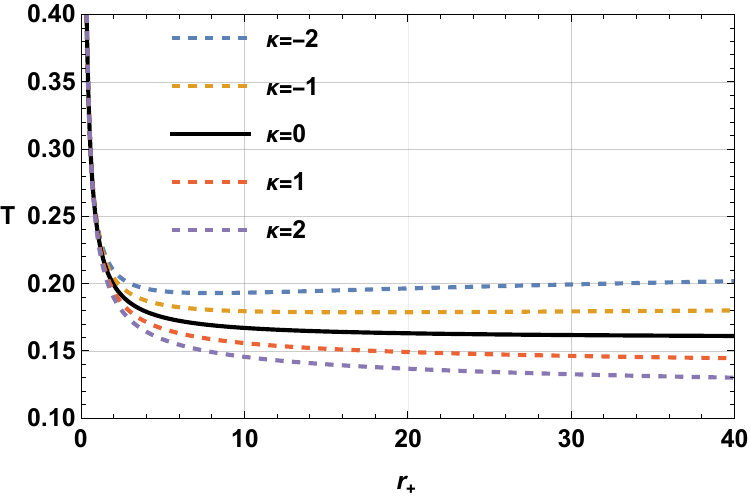}
			\caption{$\omega=-\frac{2}{3}$}
			\label{22c}
		\end{subfigure}
		\begin{subfigure}{0.30\textwidth}
			\includegraphics[width=\linewidth]{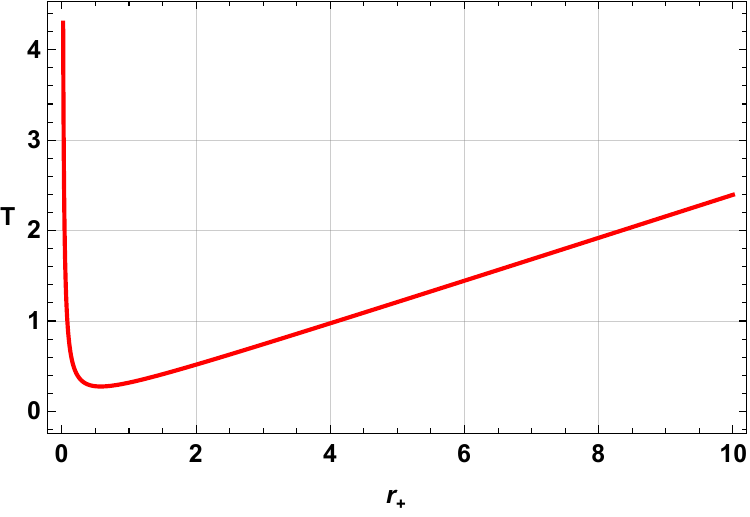}
			\caption{$\omega=-1$}
			\label{22d}
			\end{subfigure}
		\begin{subfigure}{0.30\textwidth}
		\includegraphics[width=\linewidth]{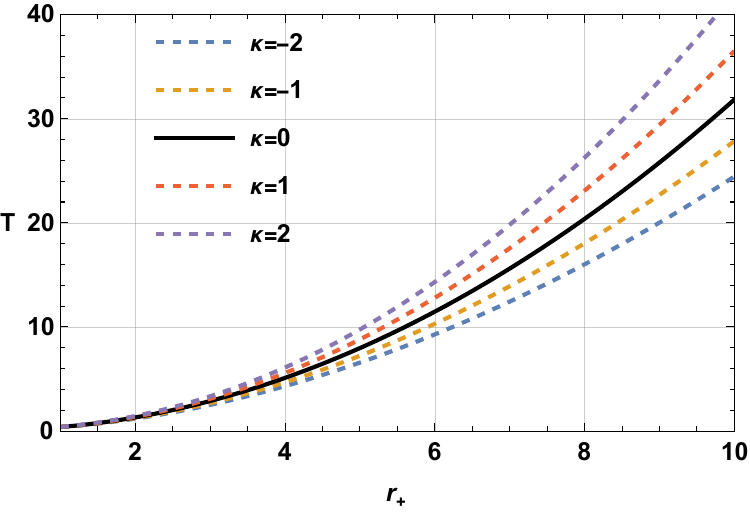}
		\caption{$\omega=-\frac{4}{3}$}
			\label{22e}
		\end{subfigure}
		\caption{ $T$ vs $r_+$ plot for different values of $\omega$. The impact of model  parameter $\kappa$ on $T$ vs $r_+$ plots are shown for a specific value of $\omega$}
		\label{22}
	\end{figure}
	The next step is to calculate the specific heat $C$ of the black holes in this model. As shown in FIG.\ref{25}, the critical point where the specific heat diverges shifts when the model parameter is introduced. This behavior mirrors what we observed in the case of model I, but with notable differences.
For black holes with $\omega = 0$, the behavior differs significantly from that seen in model I. In model I, we observe two distinct black hole branches: a Large Black Hole (LBH) branch, which is unstable, and a Small Black Hole (SBH) branch, which is stable. However, in model II, we find only one unstable black hole branch when $\omega = 0$.
Despite this difference, the overall behavior in the remaining cases is consistent with the trends observed in model I. The critical points and stability patterns across the other four scenarios follow a similar pattern, with shifts in the critical point of specific heat divergence corresponding to changes in the model parameter. \\

	\begin{figure}[h]	
		\centering
		\begin{subfigure}{0.30\textwidth}
			\includegraphics[width=\linewidth]{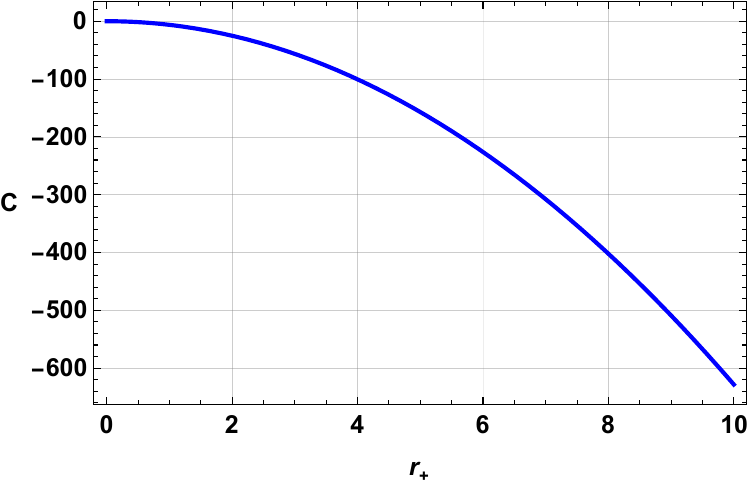}
			\caption{$\omega=0$}
			\label{25a}
		\end{subfigure}
		\begin{subfigure}{0.30\textwidth}
			\includegraphics[width=\linewidth]{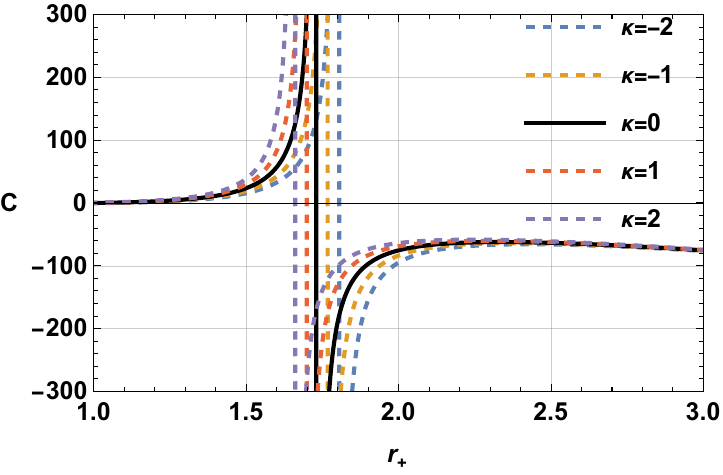}
			\caption{$\omega=\frac{1}{3}$}
			\label{25b}
		\end{subfigure}
		\begin{subfigure}{0.30\textwidth}
			\includegraphics[width=\linewidth]{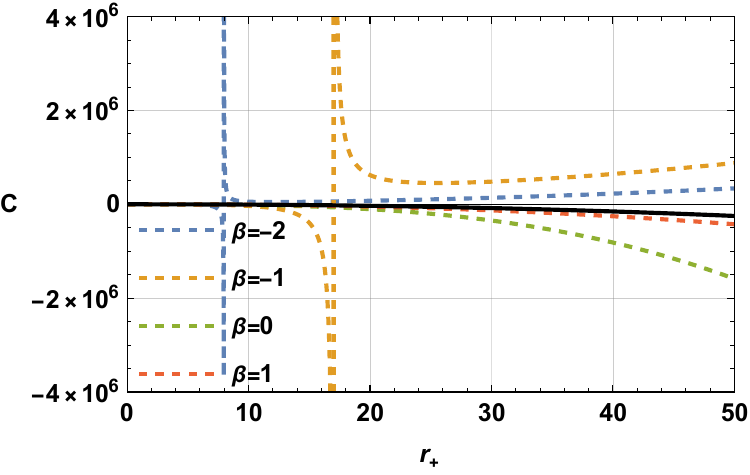}
			\caption{$\omega=-\frac{2}{3}$}
			\label{25c}
		\end{subfigure}
		\begin{subfigure}{0.30\textwidth}
			\includegraphics[width=\linewidth]{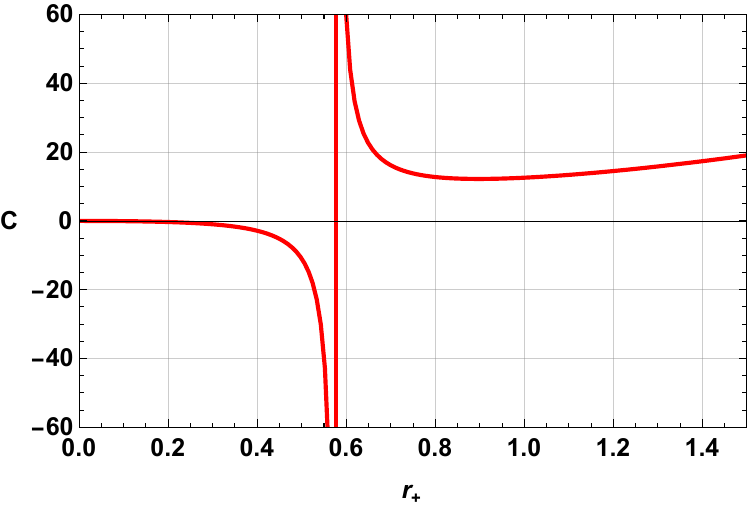}
			\caption{$\omega=-1$}
			\label{25d}
			\end{subfigure}
		\begin{subfigure}{0.30\textwidth}
		\includegraphics[width=\linewidth]{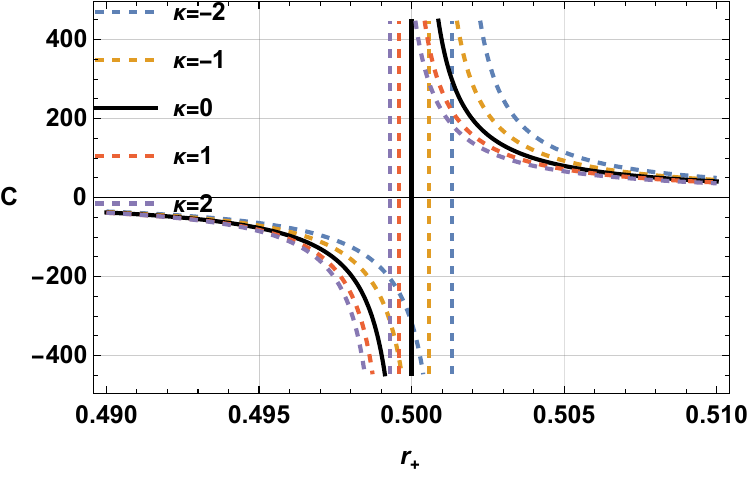}
		\caption{$\omega=-\frac{4}{3}$}
			\label{25e}
		\end{subfigure}
		\caption{ $C$ vs $r_+$ plots for different values of $\omega$. The impact of model  parameter $\beta$  on critical points are shown for a specific value of $\omega$}
		\label{25}
	\end{figure}

Next, we investigate the thermodynamic topology of these black holes and identify two distinct topological classes, characterized by topological charges $W = -1$ and $W = 0$. For the case $\omega = 0$, we obtain the simple Schwarzschild black hole with a topological charge of $W = -1$, which has no creation or annihilation points.
For $\omega = 1/3$ in the $f(R,T)$ framework, we find that the topological charge is $W = 0$ for all values of the model parameter. When examining $\omega = -2/3$, we encounter the $W = 0$ topological class for negative values of the model parameter $\kappa$. This class includes a small black hole (SBH) with a winding number of $w_1 = -1$ and a large black hole (LBH) with $w_2 = 1$. The transition between the winding numbers $-1$ and $1$ signifies a generation point. For positive values of $\kappa$, the topological class $W = -1$ emerges. In the general relativity (GR) framework, the only existing topological class in this case is also $W = -1$.
For $\omega = -1$, the topology becomes independent of the model parameter $\kappa$, yielding a universal class with $W = 0$, consisting of an SBH with winding number $w_1 = -1$ and an LBH with $w_2 = 1$. The phase transition point in this scenario is similarly identified as a generation point. Lastly, for $\omega = -4/3$, we find the topological class $W = 0$ for all values of $\kappa$, maintaining the same SBH and LBH configuration in both the $f(R,T)$ gravity and GR frameworks. A generation point is also observed in this case.
It is noteworthy that the local topology of the black hole solutions changes depending on the chosen model, as we see deviations in thermodynamic topology compared to model I. While the model parameter $\kappa$ influences the local topology of these black holes, it does not alter their global topology.\\

In this model also, we study the thermodynamic geometry of the black hole using the Geometrothermodynamics (GTD) formalism. We examine the Ruppeiner metric for the system, but we find that the singular points of the Ruppeiner curvature do not coincide with the critical points of the system. 
In our analysis, we demonstrate that the singular point at which the GTD scalar curvature diverges corresponds to the point where the heat capacity changes sign. Additionally, the critical point at which the scalar curvature $R$ diverges depends on the values of the model parameters in this particular model also.\\
%%%%%%%%%%%%%%%%%%%%%%%%%%%%%%%%%%%%%%%%%%%%%%%%%%%%%%%% 

	\subsection{Black hole shadow}
To constrain this  $f(R,T)$ model, we utilize data from black hole shadows.  We first plot the shadow for black hole solutions in this model for $\omega$ value $1/3,-2/3$ and $-4/3$ to observe the effect of the model parameter on the shadow radius of the black hole. The following equations are needed to be solved to obtained the photon radius.\\

\begin{figure}[h]	
		\centering
		\begin{subfigure}{0.30\textwidth}
			\includegraphics[width=\linewidth]{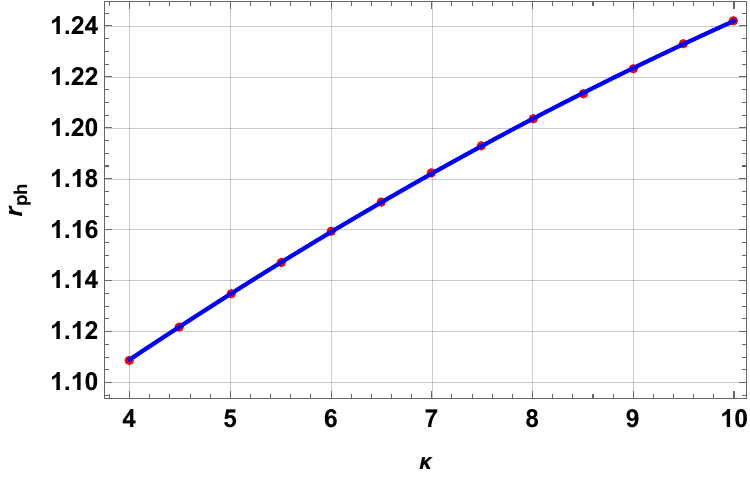}
			\caption{$\omega=\frac{1}{3}$}
			\label{sh20a}
		\end{subfigure}
		\begin{subfigure}{0.30\textwidth}
			\includegraphics[width=\linewidth]{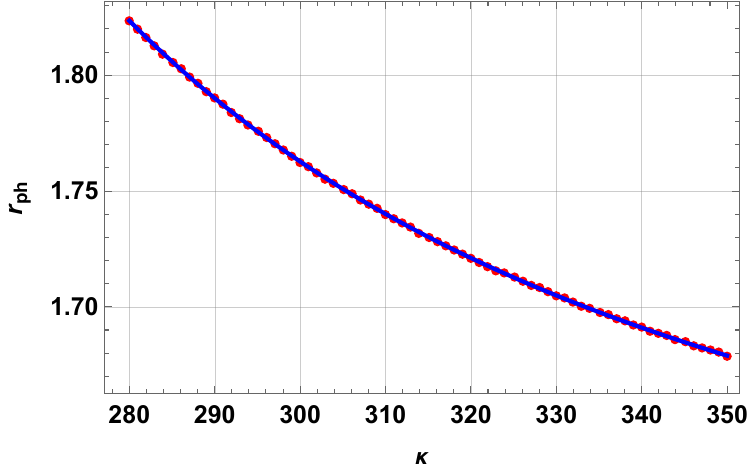}
			\caption{$\omega=-\frac{2}{3}$}
			\label{sh20b}
		\end{subfigure}
		\begin{subfigure}{0.30\textwidth}
			\includegraphics[width=\linewidth]{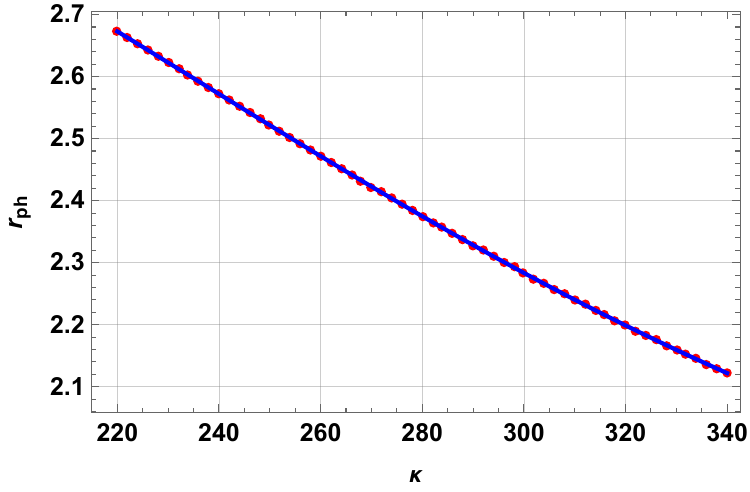}
			\caption{$\omega=-\frac{4}{3}$}
			\label{sh20c}
		\end{subfigure}
		\caption{ $r_{ph}$vs $\kappa$ plot representing numerical data and the fitted curve}
		\label{sh20}
	\end{figure}

For $\omega=\frac{1}{3}$,
\begin{multline}
-2 (\kappa + 6 \pi) \bigg( r^{\frac{3 \pi \sqrt{\kappa + 6 \pi} \sqrt{\kappa + 12 \pi} \sqrt{\frac{1}{\kappa^2 + 18 \pi \kappa + 72 \pi^2}} + \kappa + 9 \pi}{\kappa + 6 \pi}} 
- 2 r^{\frac{6 \pi \sqrt{\frac{1}{\kappa^2 + 18 \pi \kappa + 72 \pi^2}}}{\sqrt{\frac{\kappa + 6 \pi}{\kappa + 12 \pi}}}} + 1 \bigg)\\
- \bigg( \kappa - 2 \kappa r^{\frac{6 \pi \sqrt{\frac{1}{\kappa^2 + 18 \pi \kappa + 72 \pi^2}}}{\sqrt{\frac{\kappa + 6 \pi}{\kappa + 12 \pi}}}} 
- 3 \pi \bigg( \sqrt{\kappa + 6 \pi} \sqrt{\kappa + 12 \pi} \sqrt{\frac{1}{\kappa^2 + 18 \pi \kappa + 72 \pi^2}} 
+ 2 \bigg( \sqrt{\kappa + 6 \pi} \sqrt{\kappa + 12 \pi} \sqrt{\frac{1}{\kappa^2 + 18 \pi \kappa + 72 \pi^2}} - 3 \bigg) \\
r^{\frac{6 \pi \sqrt{\frac{1}{\kappa^2 + 18 \pi \kappa + 72 \pi^2}}}{\sqrt{\frac{\kappa + 6 \pi}{\kappa + 12 \pi}}}} + 3 \bigg) \bigg) = 0
\end{multline}
For $\omega=-\frac{2}{3}$,
\begin{multline}
\left(\kappa -2 \kappa  r^{\frac{48 \pi  \sqrt{\frac{1}{\kappa +24 \pi }}}{\sqrt{\kappa +24 \pi }}}+24 \pi  \left(2 r^{\frac{48 \pi  \sqrt{\frac{1}{\kappa +24 \pi }}}{\sqrt{\kappa +24 \pi }}}+1\right)\right)+2 (\kappa +24 \pi ) \left(-2 r^{48 \pi  \sqrt{\frac{\kappa -24 \pi }{\kappa +24 \pi }} \sqrt{\frac{1}{\kappa ^2-576 \pi ^2}}}+r^{\frac{24 \pi  \sqrt{\kappa -24 \pi } \sqrt{\kappa +24 \pi } \sqrt{\frac{1}{\kappa ^2-576 \pi ^2}}+\kappa }{\kappa +24 \pi }}+1\right) \\r^{\frac{24 \pi  \left(\sqrt{\frac{1}{\kappa +24 \pi }}-\sqrt{\kappa -24 \pi } \sqrt{\frac{1}{\kappa ^2-576 \pi ^2}}\right)}{\sqrt{\kappa +24 \pi }}}=0
\end{multline}
For $\omega=-\frac{4}{3}$,
\begin{multline}
2 (24 \pi - \kappa) \Big( 2 r^{96 \pi \sqrt{\frac{1}{\kappa^2 + 48 \pi \kappa - 1728 \pi^2}} \sqrt{\frac{96 \pi}{\kappa - 24 \pi} + 1}} 
- r^{48 \pi \sqrt{\frac{1}{\kappa^2 + 48 \pi \kappa - 1728 \pi^2}} \sqrt{\frac{96 \pi}{\kappa - 24 \pi} + 1} + \frac{1}{\kappa - 24 \pi} + 1} - 1 \Big) \\
- \Big( \kappa \Big( 2 r^{96 \pi \sqrt{\frac{1}{\kappa^2 + 48 \pi \kappa - 1728 \pi^2}} \sqrt{\frac{96 \pi}{\kappa - 24 \pi} + 1}} - 1 \Big) 
+ 24 \pi \Big( 2 \sqrt{-\kappa - 72 \pi} \sqrt{24 \pi - \kappa} \sqrt{\frac{1}{\kappa^2 + 48 \pi \kappa - 1728 \pi^2}} \\
+ 2 \Big( 2 \sqrt{-\kappa - 72 \pi} \sqrt{24 \pi - \kappa} \sqrt{\frac{1}{\kappa^2 + 48 \pi \kappa - 1728 \pi^2}} + 1 \Big) 
r^{96 \pi \sqrt{\frac{1}{\kappa^2 + 48 \pi \kappa - 1728 \pi^2}} \sqrt{\frac{96 \pi}{\kappa - 24 \pi} + 1}} - 1 \Big) \Big) = 0
\end{multline}
 Solving this equation are not straightforward due to the existence of complex terms in the power of $r$. Consequently, we use numerical fitting techniques, to approximate the solution. as we have done previously.  We assume $r_{ph}$ can be expressed in terms of $\kappa$ as follows :
$$ r_{ph}=a_0+a_1 \hspace{0.1cm} \kappa +a_2\hspace{0.1cm}  \kappa^2+a_3\hspace{0.1cm}  \kappa^3+a_4\hspace{0.1cm}  \kappa^4+a_5\hspace{0.1cm}  \kappa^5+a_6  \hspace{0.1cm}  \kappa^6$$
where the coefficients $a_0,a_1.....$ are unknown constants to be determined.  In FIG.\ref{sh20}, the plot illustrates the relationship between  $r_ph$ and $\kappa$ for all three case where the blue line represents the exact solution, and the red dots indicate the numerically computed points based on data.

\begin{figure}[h]	
		\centering
		\begin{subfigure}{0.30\textwidth}
			\includegraphics[width=\linewidth]{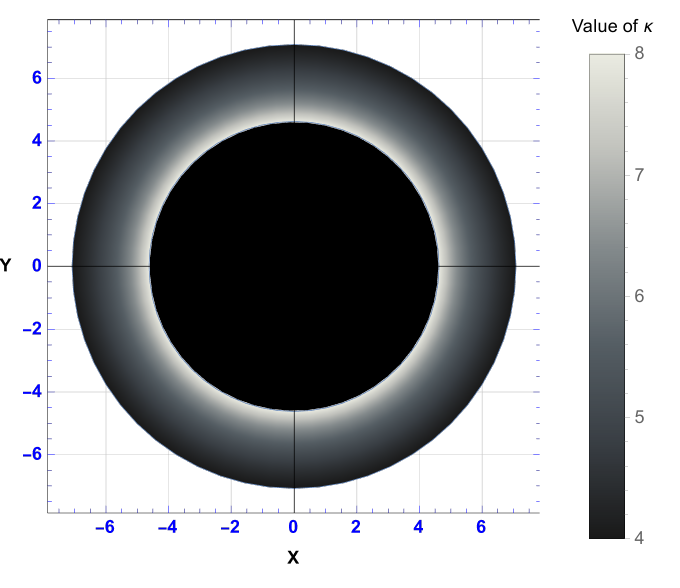}
			\caption{$\omega=\frac{1}{3}$}
			\label{sh21a}
		\end{subfigure}
		\begin{subfigure}{0.30\textwidth}
			\includegraphics[width=\linewidth]{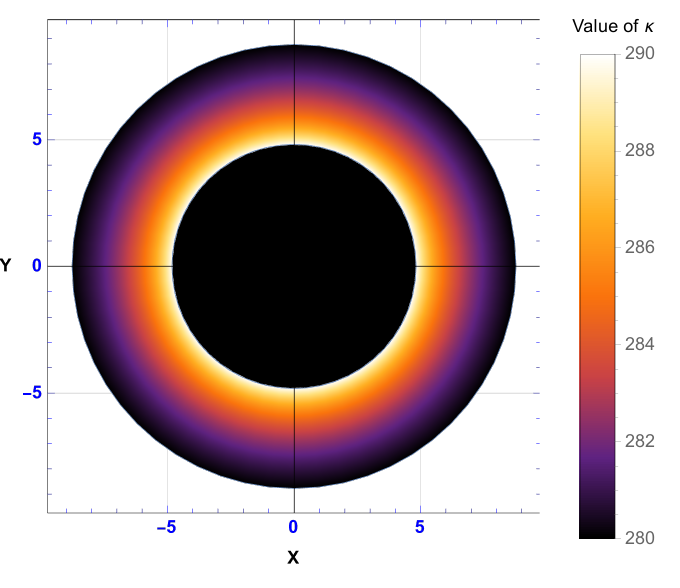}
			\caption{$\omega=-\frac{2}{3}$}
			\label{sh21b}
		\end{subfigure}
		\begin{subfigure}{0.30\textwidth}
			\includegraphics[width=\linewidth]{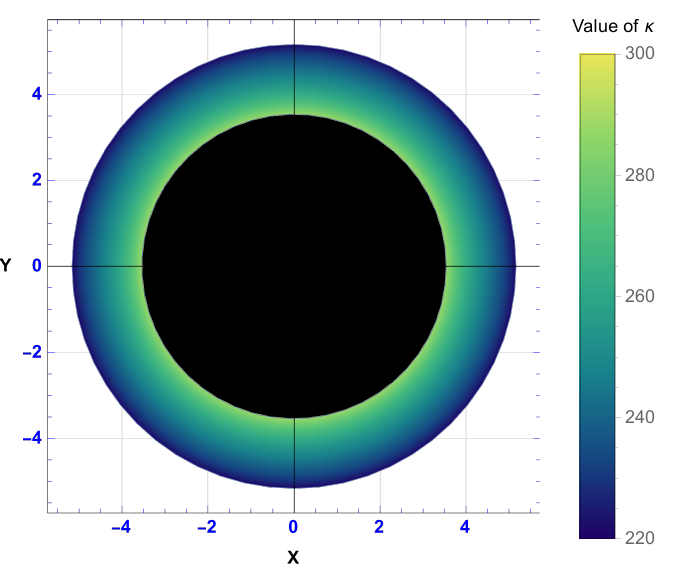}
			\caption{$\omega=-\frac{4}{3}$}
			\label{sh21c}
		\end{subfigure}
		\caption{Stereoscopic projection of shadow radius in terms of celestial 
coordinates. We have taken $K=M=1$. }
		\label{sh21}
	\end{figure}

	The expressions  for photon orbit radius obtained from these figures are given by\\
	
	for $\omega=\frac{1}{3}$
\begin{equation}
r_{ph}=1.00023 + 0.0262167 \hspace{0.1cm}\kappa + 0.00088749 \hspace{0.1cm} \kappa^2 - 0.000210915 \hspace{0.1cm}\kappa^3 + 0.0000144762 \hspace{0.1cm} \kappa^4 - 5.05055 \times 10^{-7}\hspace{0.1cm} \kappa^5 + 7.4656 \times 10^{-9} \hspace{0.1cm} \kappa^6 
\end{equation}

for $\omega=-\frac{2}{3}$
\begin{equation}
r_{ph}=48.0887 -0.555373  \hspace{0.1cm}\kappa + 0.00205398 \hspace{0.1cm} \kappa^2 + 8.06782 \times 10^{-7} \hspace{0.1cm}\kappa^3 + 2.43789 \times 10^{-8} \hspace{0.1cm} \kappa^4 + 6.00831 \times 10^{-11}\hspace{0.1cm} \kappa^5 - 4.78583 \times 10^{-14} \hspace{0.1cm} \kappa^6 
\end{equation}

for $\omega=-\frac{4}{3}$
\begin{equation}
r_{ph}=0.988581 +0.0418523  \hspace{0.1cm}\kappa  -0.000315406 \hspace{0.1cm} \kappa^2 + 1.09742 \times 10^{-6} \hspace{0.1cm}\kappa^3 -2.13863 \times 10^{-9} \hspace{0.1cm} \kappa^4 + 2.31041 \times 10^{-12}\hspace{0.1cm} \kappa^5-1.09189 \times 10^{-15} \hspace{0.1cm} \kappa^6 
\end{equation}
we have taken $K=M=1$.
From the photon radius, we can derive the shadow radius as follows:
\begin{equation}
r_{sh}=\frac{r_{ph}}{\sqrt{B[r_{ph}]}}
\end{equation}
\begin{equation}
\mathcal{A}=\left(-0.0830538 l^6+0.439648 l^5-0.804952 l^4+0.691513 l^3-0.210153 l^2-0.03199 l+1.499\right)
\end{equation}
\begin{multline}
\mathcal{B}^2= \left(-0.0830538 l^6+0.439648 l^5-0.804952 l^4+0.691513 l^3-0.210153 l^2-0.03199 l+1.499^{-l}\right)\\-\left(\frac{24.0808}{-1. l^6+5.29354 l^5-9.69193 l^4+8.32608 l^3-2.53032 l^2-0.385172 l+18.0485}+1 \right)
\end{multline}
Now, for the 2-D 
stereoscopic projection of shadow radius is plotted in celestial coordinates $X$ 
and $Y$.  Figure \ref{sh21} illustrates how the shadow radius changes with respect to the model parameter $\kappa$. The plot reveals that for  $\omega=1/3$,  as the parameter $\kappa$ increases, the shadow radius of the black hole also increases. On the other hand, for  $\omega=-2/3$ and $\omega=-4/3$,  as the parameter $\kappa$ increases, the shadow radius of the black hole also decreases.

\begin{figure}[h]	
		\centering
		\begin{subfigure}{0.30\textwidth}
			\includegraphics[width=\linewidth]{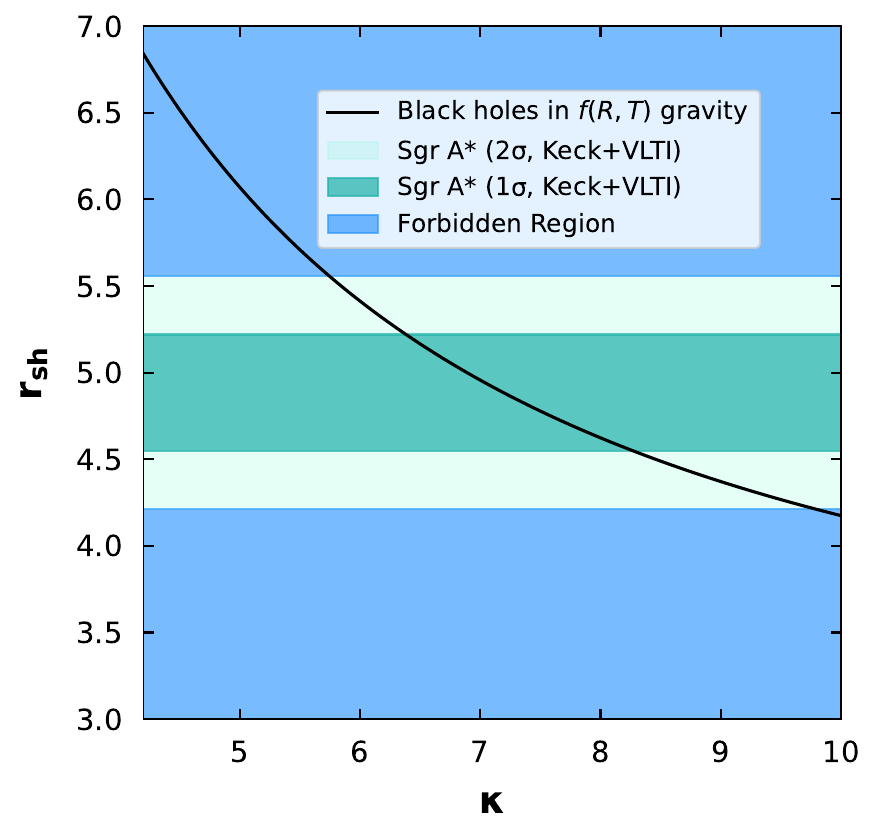}
			\caption{$\omega=\frac{1}{3}$}
			\label{sh22a}
		\end{subfigure}
		\begin{subfigure}{0.30\textwidth}
			\includegraphics[width=\linewidth]{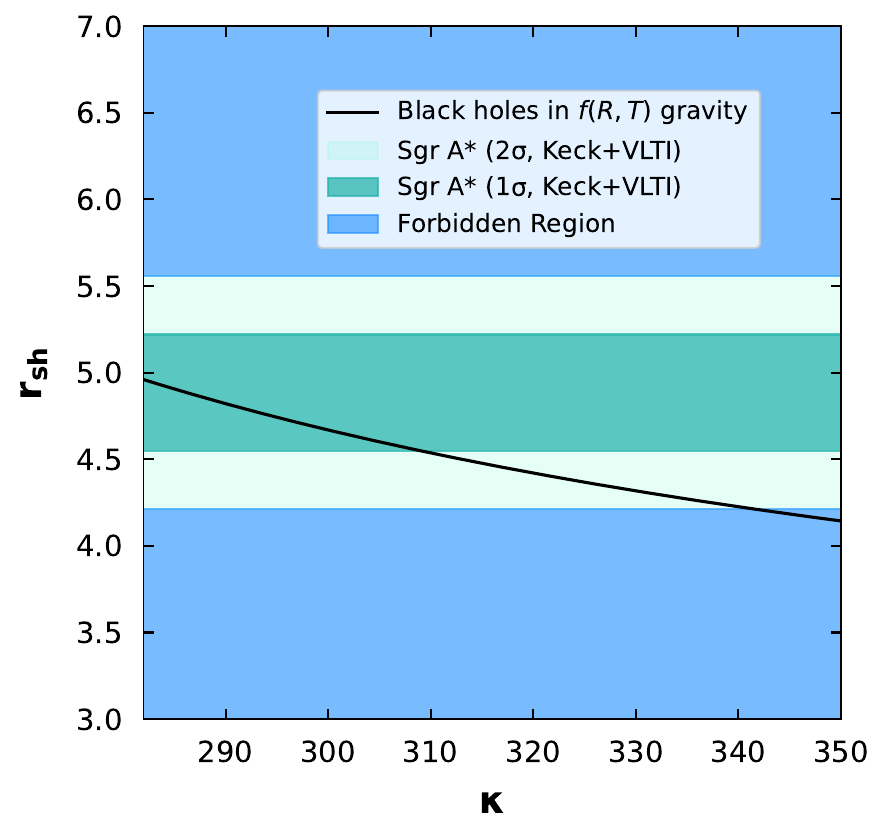}
			\caption{$\omega=-\frac{2}{3}$}
			\label{sh22b}
		\end{subfigure}
		\begin{subfigure}{0.30\textwidth}
			\includegraphics[width=\linewidth]{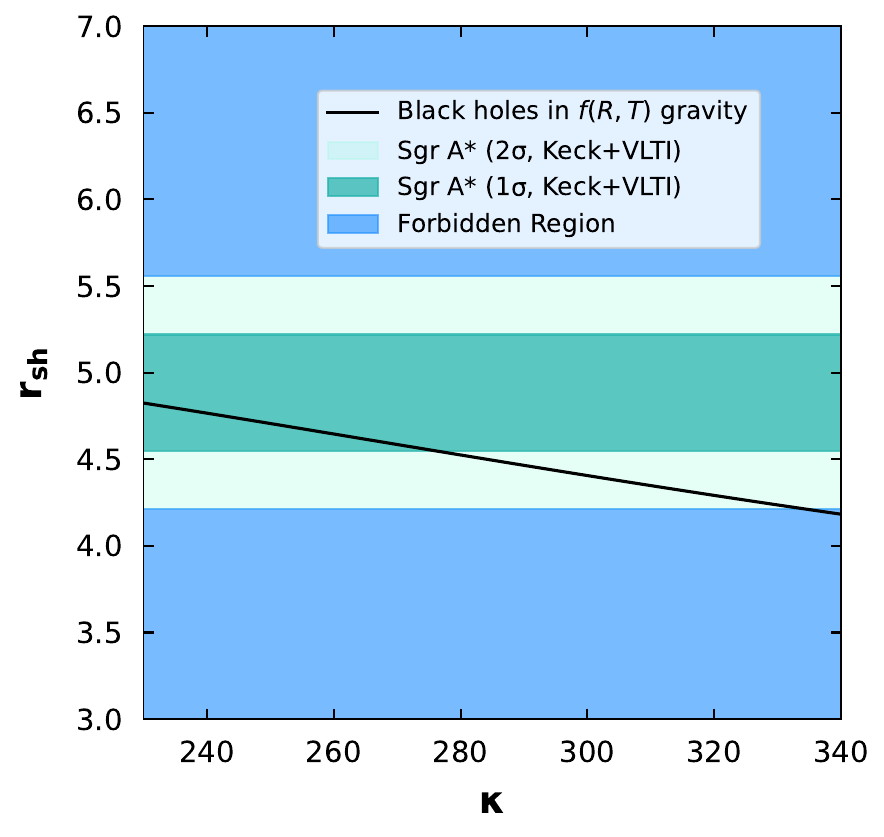}
			\caption{$\omega=-\frac{4}{3}$}
			\label{sh22c}
		\end{subfigure}
		\caption{Shadow radius versus parameter $m$ and $c_{2}$ have been plotted in 
the background of Keck and VLTI constrains \cite{s0} from observations of 
Sgr A*. We have chosen $M=1$ and $K=1$ for these plots. The blue portion 
represents the zone forbidden by Keck-VLTI observation.}
\label{sh22}
\end{figure}

%%%%%
In order to constrain the model parameters,  we put bounds on $\delta$, to provide constraints on the shadow radius $r_{sh}$. We present a plot in Fig.~\ref{sh22}, illustrating the shadow radius constrained by the Keck and VLTI observations for all the three cases.

The constraints on the parameter $\kappa$ are illustrated in FIG.\ref{sh22}. Specifically, FIG.\ref{sh22a} shows that for $\omega = 1/3$, $\kappa$ falls within a relatively narrow range of $5.74018 < \kappa < 9.80015$, consistent with observational data. However, as $\omega$ takes on negative values, the range of $\kappa$ expands significantly. For instance, when $\omega = -2/3$, the allowed range broadens to $256.03 < \kappa < 341.818$, providing a much larger parameter space. Similarly, for $\omega = -4/3$, $\kappa$ is constrained within $75 < \kappa \leq 334.827$, reflecting a wider but still bounded interval.\\

The comparative analysis of the parameter constraints in Model I : $f(R,T) = f_1(R) + f_2(T)$) and Model II : $f(R,T) = f_1(R) + f_2(R)f_3(T)$) highlights distinct behaviors driven by the values of $\omega$. In Model I, the parameter $\beta$ exhibits a strong sensitivity to $\omega$, with the constraints tightening significantly as $\omega$ becomes more negative. This indicates that $\beta$ requires precise fine-tuning to maintain observational consistency, particularly for $\omega = -2/3$ and $\omega = -4/3$. Conversely, in Model II, the parameter $\kappa$ demonstrates greater flexibility, with broader allowable ranges for negative values of $\omega$, especially at $\omega = -2/3$ and $\omega = -4/3$.  Model II does not impose any constraints on $\kappa$ for $\omega = 0$, which contrasts with Model I, where $\beta$ remains constrained across all $\omega$. Overall, Model I enforces tighter restrictions on its parameter space, whereas Model II provides more freedom, especially in scenarios involving negative $\omega$. 

The comparative constraints for $\beta$ and $\kappa$ across different values of $\omega$ are summarized in Table \ref{comparison}.  

\begin{table}[h!]
    \centering
    \renewcommand{\arraystretch}{1.5} % Adjust the row spacing
    \resizebox{\textwidth}{!}{%
    \begin{tabular}{|c|c|c|c|c|c|}
        \hline
        \textbf{Model} & \textbf{Parameter} & $\boldsymbol{\omega = 0}$ & $\boldsymbol{\omega = 1/3}$ & $\boldsymbol{\omega = -2/3}$ & $\boldsymbol{\omega = -4/3}$ \\ \hline
        \textbf{A.} $f(R,T) = f_1(R) + f_2(T)$ & $\beta$ & $-13.7042\leq\beta \leq -9.00669$ & $-14.6531\leq\beta \leq -2.63325$ & $-12.8117\leq\beta \leq-11.986$ & $ -12.9505\leq \beta \leq-12.3851$ \\ \hline
        \textbf{C.} $f(R,T) = f_1(R) + f_2(R)f_3(T)$ & $\kappa$ & -- & $5.74018 < \kappa < 9.80015$ & $256.03 < \kappa < 341.818$ & $75<\kappa \leq 334.827$ \\ \hline
    \end{tabular}%
    }
    \caption{Comparison of parameter constraints for Model I (\textbf{A.}) and Model II (\textbf{C.}) under different values of $\omega$.}
    \label{comparison}
\end{table}

In one interesting work ref.\cite{h7}, charged black hole solutions in $f(R,T)$ gravity coupled to nonlinear electrodynamics were analyzed with the model $R + \beta T$, which is equivalent to Model I in our work ($\alpha R + \beta T$) where our black hole solution is independent of $\alpha$. In their analysis, the parameter $\beta$ was constrained by examining the black hole shadow radius for three different powers $p$: $p = 2, 4$, and $6$. The allowable ranges for $\beta$ were found to be $0 \leq \beta_{p=2} \lesssim 1.6 \times 10^4$, $0 \leq \beta_{p=4} \lesssim 3 \times 10^{10}$, and $0 \leq \beta_{p=6} \lesssim 7.8 \times 10^{16}$, with the results showing that the shadow radius increases with $\beta$.In comparison, our work evaluates $\beta$ for Kiselev-type black holes in $f(R,T)$ gravity for different values of the equation of state parameter $\omega$. We determined that for $\omega = 0$, $\beta$ lies within $-13.7042 \leq \beta \leq -9.00669$. When $\omega$ increases to $1/3$, the range broadens to $-14.6531 \leq \beta \leq -2.63325$. For negative values of $\omega$, such as $\omega = -2/3$ and $\omega = -4/3$, the allowable ranges tighten to $-12.8117 \leq \beta \leq -11.986$ and $-12.9505 \leq \beta \leq -12.3851$, respectively.
The comparison shows that the values of $\beta$ differ significantly between the two studies. The parameter $\beta$ reaches very large values, going up to $10^{16}$ for certain cases (like $p = 6$). On the other hand, our study, gives much smaller ranges. This large difference likely comes from the fact that the earlier study includes a cosmological constant (AdS space), while our work is done without it, leading to tighter constraints on $\beta$.Moreover the nature of the black hole soltion and presence of the nonlinear electrodynamic source also attribute to the difference in both the studies.

\section{conclusion}
In this paper, we explored two $f(R,T)$ gravity models and derived black hole solutions within these frameworks. First, we examined models of the type $f(R,T) = f_1(R) + f_2(T)$. Specifically, we chose $f_1(R) = \alpha R$ and $f_2(T) = \beta T$. We focused on obtaining key thermodynamic parameters, including the black hole’s mass, temperature, free energy, and heat capacity. Five distinct values of $\omega$, each corresponding to unique physical interpretations within general relativity (GR), were considered in our analysis.
Our primary objective was to investigate how the $f(R,T)$ model influenced the thermodynamic characteristics of black holes and to assess the degree of deviation from GR based on the values of the model parameters. For $\omega = 0$, we observed Davies-type phase transitions when the model parameter $\beta$ was negative. However, for $\omega = -2/3$, similar phase transitions occurred with a positive value of $\beta$. In both cases, these phase transitions were absent in the corresponding GR solutions.
For $\omega = 1/3$ and $\omega = -4/3$, we observed similar behavior in both GR and the $f(R,T)$ framework, with Davies-type transitions present in both cases. Notably, the critical points at which these transitions occurred shifted depending on the model parameters. For $\omega = -1$, both GR and the $f(R,T)$ gravity model coincided, yielding a black hole solution independent of the model parameters. In this case, we also detected Davies-type phase transitions, further highlighting the consistency between the two frameworks in this particular scenario.\\

Next, we investigated the thermodynamic topology of these black holes in Model I and identified two distinct topological classes, characterized by topological charges $W = -1$ and $W = 0$. For the case $\omega = 0$ in $f(R,T)$ gravity, we observed that the topological charge was $W = 0$ for negative values of the model parameter $\beta$ and $W = -1$ for positive values. The $W = 0$ class included a small black hole branch (SBH) with winding number $w_1 = 1$ and a large black hole branch(LBH) with $w_2 = -1$. The point at which the black hole transitioned from a winding number of $1$ to $-1$ was identified as an annihilation point. In contrast, in general relativity (GR), the topological charge was $W = -1$ for this black hole class.
For $\omega = 1/3$, in the $f(R,T)$ framework, we found that the topological charge was $W = 0$ for all values of the model parameter. However, for $\omega = -2/3$, we encountered an opposite local topology compared to the $\omega = 0$ case. Here, the $W = 0$ topological class comprised an SBH with winding number $w_1 = -1$ and an LBH with $w_2 = 1$. The transition between winding numbers $-1$ and $1$ marked a generation point. Interestingly, the $W = 0$ class appeared only for positive values of $\beta$, in contrast to the $\omega = 0$ case, where it appeared only for negative $\beta$. For negative values of $\beta$, the topological class $W = -1$ emerged. Despite differences in local topology, the global topology remained unchanged, with an overall topological charge of $W = 0$ and $W = 1$. In the GR framework, the only existing topological class for this case was $W = -1$.
For $\omega = -1$, the topology became independent of the model parameter $\beta$, yielding a universal class with $W = 0$, consisting of an SBH with winding number $w_1 = -1$ and an LBH with $w_2 = 1$. The phase transition point here was also identified as a generation point. Finally, for $\omega = -4/3$, we found the topological class $W = 0$ for all values of $\beta$, with the same SBH and LBH configuration in both $f(R,T)$ gravity and the GR framework. A generation point was also observed in this case.
It is noteworthy that the local topology of the black hole solutions changed depending on the sign of $\omega$. For $\omega = 0$ and negative values of $\omega$, an annihilation point was found, while for positive values of $\omega$, a generation point emerged. Therefore, we concluded that the thermodynamic topology of these black holes was significantly affected by the values of the model parameter $\beta$ and the thermodynamic parameter $\omega$. \\
Next we analyzed the thermodynamic topology of black holes in Model II. Here again we identified two distinct topological classes with charges $W = -1$ and $W = 0$. For $\omega = 0$, the black holes exhibits a topological charge of $W = -1$ like Schwarzschild black hole  with no creation or annihilation points. For $\omega = \frac{1}{3}$, $W = 0$ persists across all values of the model parameter $\kappa$. In the case $\omega = -\frac{2}{3}$, $W = 0$ appears for negative $\kappa$, featuring a generation point where small and large black holes (SBH and LBH) transition between winding numbers, while $W = -1$ arises for positive $\kappa$. For $\omega = -1$ and $\omega = -\frac{4}{3}$case, $W = 0$ is observed with same global and local topology as that of Model I. These results highlight that while the model parameter $\kappa$ affects the local topology, the global topology remains unchanged.Hence from the study of thermodynamic topology in both the model,  we can conclude that the introduction of considered  $f(R,T)$ gravity model altered the local topology of these black holes, it did not change their global topology.  \\

In the Table (\ref{tab2}), we have provided the comparison of topological analysis of black hole solution in Model I and Model II). \\

\begin{table}[h!]
    \centering
    \renewcommand{\arraystretch}{1.5}
    \begin{adjustbox}{max width=\textwidth}
    \begin{tabular}{@{}llcccc@{}}
    \toprule
    \textbf{$\omega$ value} & \textbf{Model} & \textbf{Topological Charge ($W$)} & \textbf{SBH Winding ($w_1$)} & \textbf{LBH Winding ($w_2$)} & \textbf{Transition Point} \\
    \midrule
    \multirow{3}{*}{$\omega = 0$}  & Model I & $W = 0$ (negative $\beta$), $W = -1$ (positive $\beta$) & $w_1 = 1$ & $w_2 = -1$ & Annihilation \\
                                    & Model II & $W = -1$ (all $\kappa$) & \multicolumn{2}{c}{No distinct SBH or LBH branches} & None \\
                                    & GR      & $W=-1$ &$w_1=-1$ & $w_1=-1$ &  None  \\
    \midrule
    \multirow{3}{*}{$\omega = 1/3$}  & Model I & $W = 0$ (all $\beta$) & $w_1 = 1$ & $w_2 = -1$ & Annihilation \\
                                    & Model II & $W = 0$ (all $\kappa$) & $w_1=1$&$w_2=-1$ & Annihilation \\
                                    & GR      &  $W = 0$ &$w_1=1$& $w_1=-1$ & Annihilation \\
    \midrule
    \multirow{3}{*}{$\omega = -2/3$} & Model I & $W = 0$ (positive $\beta$), $W = -1$ (negative $\beta$) & $w_1 = -1$ & $w_2 = 1$ & Generation \\
                                    & Model II & $W = 0$ (negative $\kappa$), $W = -1$ (positive $\kappa$) & $w_1 = -1$ & $w_2 = 1$ & Generation \\
                                    & GR      & $W = -1$ (all $\kappa$) & \multicolumn{2}{c}{No distinct SBH or LBH branches} & None \\
    \midrule
    \multirow{3}{*}{$\omega = -1$}   & Model I & $W = 0$ (all $\beta$) & $w_1 = -1$ & $w_2 = 1$ & Generation \\
                                    & Model II & $W = 0$ (all $\kappa$) & $w_1 = -1$ & $w_2 = 1$ & Generation \\
                                    & GR      & $W=0$ & $w_1=-1$ & $w_2=1$ & Generation \\
    \midrule
    \multirow{3}{*}{$\omega = -4/3$} & Model I & $W = 0$ (all $\beta$) & $w_1 = -1$ & $w_2 = 1$ & Generation \\
                                    & Model II & $W = 0$ (all $\kappa$) & $w_1 = -1$ & $w_2 = 1$ & Generation \\
                                    & GR      & $W=0$ & $w_1=-1$& $w_2=1$ &Generation \\
    \bottomrule
    \end{tabular}
    \end{adjustbox}
    \caption{Comparison of topological classes of black hole solutions in $f(R,T)$ gravity Model I, $f(R,T)$ gravity Model II, and General Relativity (GR). SBH refers to Small Black Hole, LBH refers to Large Black Hole. }
    \label{tab2}
\end{table}

We studied the thermodynamic geometry of the black hole using the GTD (Geometrothermodynamics) formalism. We also examined the Ruppeiner metric for the system, but found that the singular points of the Ruppeiner curvature did not align with the system's critical points. As a result, we focused solely on the GTD formalism.
In our analysis, we demonstrated that the singular point, where the GTD scalar curvature diverges, exactly corresponds to the point where the heat capacity changes sign. This is a key indicator of a phase transition. Moreover, the critical point at which the scalar curvature \(R\) diverges depends on the value of the model parameters, meaning that the parameter values influence when the phase transition occurs.\\

In the next part of our analysis, we used black hole shadow data to constrain the model parameter $\beta$. This method relies on the mass-to-distance ratio for Sgr A* and a calibration factor that links the observed shadow radius with the theoretical prediction. The Event Horizon Telescope (EHT) team introduced a parameter, $\delta$, which measures the fractional deviation between the observed shadow radius ($r_s$) and the shadow radius of a Schwarzschild black hole ($r_{sch}$).
Using this approach, we derived constraints on the model parameter $\beta$. For $\omega = 0$, we found that the parameter satisfies the Keck observations when $\beta < -9.00669$. Similarly, for $\omega = 1/3$, the constraint on $\beta$ is $\beta < -2.63325$. For $\omega = -2/3$ and $\omega = -4/3$, the constraints are $\beta < -11.986$ and $\beta < -12.9505$, respectively. \\

Next, we turned our attention to the second model, described by the function $ f(R,T) = f_1(R) + f_2(R) f_3(T)$. In this model, we specifically chose$ f_1(R) = \alpha R$,$f_2(R) = \beta R $, and $ f_3(T) = \gamma T $. We evaluated the solution by examining the strong energy condition (SEC) and the black hole's horizon structure. It was found that while the SEC condition depends on the parameter \(\alpha\), the black hole solution itself is not influenced by this parameter.
The locations of the horizons are determined by solving $N(r) = 0 $, which gives the radii where the metric function vanishes. In the case of $ \omega = 1/3 $, a black hole solution is always present for any value of $\kappa$. For positive $\kappa$, both the Cauchy and event horizons are present, whereas for small negative values of $\kappa$, only the event horizon remains.
For $\omega = -2/3$ and $\omega = -4/3$, the horizon structure becomes more complex. In a certain range of $\kappa$, a degenerate horizon appears, indicating a point where two horizons coincide. Outside this range, the black hole has two distinct horizons. Despite the differences in horizon structure across various cases, an event horizon is always present for all values of $\kappa$ ensuring the existence of a black hole solution.\\

In terms of the thermodynamic behavior, we found similar results for the thermodynamic parameters as in the previous model, with the exception of the $\omega = 0$ case. In the first model, we observed Davies-type phase transitions when the model parameter $\beta$ was negative. However, in the second model, we only encountered a Schwarzschild black hole, which is independent of the model parameters. As a result, no phase transition was observed in the $\omega = 0$ case in Model II.
For $\omega = -2/3$, we found a Davies-type phase transition for negative values of the model parameter $\kappa$, which contrasts with Model I, where the phase transition occurred for positive values of $\beta$.
In the cases of $\omega = 1/3$ and $\omega = -4/3$, the behavior was consistent between both models, with Davies-type transitions present. For $\omega = -1$, as in the previous model, both the GR and $f(R,T)$ gravity frameworks produced identical black hole solutions that were independent of the model parameters, and we again detected Davies-type phase transitions.\\

Next, we investigated the thermodynamic topology and geometry of these black holes. As in the first model, we identified two distinct topological classes, characterized by the topological charges $W = -1$ and $W = 0$. However, for $\omega = 0$ in the $f(R,T)$ gravity framework, we observed that only the $W = -1$ topological charge exists. In contrast to Model I, where both topological classes appeared for $\omega = 0$, the $W = 0$ class disappeared in Model II. Despite this difference, we observed similar topological behavior across the models for other values of $\omega$.
Therefore, we concluded that the thermodynamic topology of these black holes is influenced by the choice of the $f(R,T)$ gravity model. Additionally, the thermodynamic geometry of the black hole also varies between the two models. For $\omega = 0$, we did not observe any singularity in the GTD scalar curvature. However, apart from this specific case, the GTD scalar curvature behaved similarly in both models for the other $\omega$ values. The points where the GTD scalar curvature diverged matched the critical points in Model II as well.\\

The parameter $\kappa$ in Model II was also constrained using black hole shadow data. For $\omega = 1/3$, as the value of $\kappa$ increases, the shadow radius of the black hole grows larger. In contrast, for $\omega = -2/3$ and $\omega = -4/3$, the shadow radius decreases as $\kappa$ increases. The specific constraints on the $\kappa$ parameter are as follows: for $\omega = 1/3$, the model aligns with Keck observations when $5.74018 < \kappa < 9.80015$. Similarly, for $\omega = -2/3$, $\kappa$ is constrained to the range $256.03 < \kappa < 341.818$, while for $\omega = -4/3$, the constraint is $\kappa < 334.827$.\\

As $f(R,T)$ gravity is a relatively new theory that accounts for both matter and geometric aspects, it opens up a vast area of research in the context of black hole physics. In this work, we have considered two models with linear dependencies, as these represent the simplest forms of the theory. However, obtaining exact black hole solutions with more complex functional forms for $f(R,T)$ and exploring solutions beyond those akin to the Kiselev black holes remain largely unexamined. We intend to pursue these avenues in our future endeavours.\\

\section{Acknowledgments}
BH would like to thank DST-INSPIRE, Ministry of Science and Technology fellowship program, Govt. of India for awarding the DST/INSPIRE Fellowship[IF220255] for financial support. Special thanks are extended to Pranjal Sarmah for his valuable suggestions regarding the fundamental concepts discussed in this manuscript. BH would also like to thank R. Karmakar and R. Bora for their fruitful discussions during the drafting of this manuscript.

	\end{document}